\documentclass[12pt,a4paper]{article}

\usepackage{ucs}
\usepackage[usenames,dvipsnames]{xcolor}
\usepackage{tikz}
\usepackage{tkz-tab}
\usepackage{caption}
\usepackage{latexsym}
\usepackage{amssymb}
\usepackage{amsmath}
\usepackage{subcaption}
\newtheorem{theorem}{Theorem}[section]
\newtheorem{corollary}[theorem]{Corollary}
\newtheorem{lemma}[theorem]{Lemma}

\newtheorem{proposition}{Proposition}[section]
\newtheorem{definition}{Definition}[section]
\newtheorem{remark}{Remark}[section]

\usepackage{amsfonts,amssymb,eucal,amsmath}
\title{\bf Finiteness of Stationary Configurations of the Planar Four-vortex Problem. II}

\author{{ Xiang Yu\footnote{Email: xiang.zhiy@foxmail.com, xiang.zhiy@gmail.com}} \\
\small \it School of Economic and Mathematics, Southwestern
University of Finance and Economics, \\
\small \it Chengdu 611130, China}

\date{}

\begin{document}
\maketitle

\begin{abstract}
In an earlier paper \cite{yu2021Finiteness},  we showed that there are finitely many  stationary configurations (consisting of equilibria, rigidly translating configurations,  relative equilibria and collapse configurations) in the planar  four-vortex problem.  However, we only established finiteness of collapse configurations in the sense of  prescribing a 
 collapse constant.  In this paper, by developing ideas of Albouy-Kaloshin and Hampton-Moeckel to do  an analysis of the singularities, we further show that there really are  finitely many  collapse configurations in the   four-vortex problem. This is an  unexpectedly result, because the  $N$-vortex problem has  infinitely many collapse configurations for $N= 3$ and for $N= 5$.  
 We also provide better upper bounds  for collapse configurations than that in \cite{yu2021Finiteness}.  \\




{\bf 2020AMS Subject Classification:} { 76B47	\and 70F10 \and 70F15 \and 	37Nxx}.

\end{abstract}

\begin{center}
\tableofcontents
\end{center}

\section{Introduction}
\ \ \ \

We consider the motion of $N$  point vortices on a plane, the so-called  $N$-vortex problem  introduced by Helmholtz \cite{helmholtz1858integrale}. The equations of motion of the $N$-vortex problem are written as
\begin{equation}\label{eq:vortex's equation1}
\Gamma_n \dot{\mathbf{r}}_n =J\sum_{1 \leq j \leq N, j \neq n} \frac{\Gamma_n\Gamma_j(\mathbf{r}_j-\mathbf{r}_n)}{|\mathbf{r}_j-\mathbf{r}_n|^2}, ~~~~~~~~~~~~~~~n=1, 2, \cdots, N.
\end{equation}
where $J=\left(
           \begin{array}{cc}
             0 & 1 \\
             -1 & 0 \\
           \end{array}
         \right)$,  $\Gamma_n\in \mathbb{R}^*$ is the vorticity (or vortex strength), $\mathbf{r}_n \in {\mathbb{R}}^2$ is the position, of the $n$-th point vortex;
and  $|\cdot|$  denotes the Euclidean norm in ${\mathbb{R}}^2$.

It is well known that,   the $N$-vortex problem has self-similar
collapsing    solutions,  that is, point vortices simultaneously collide with each other such that the relative shape
remains constant; and the
mechanism of vortex collapse plays an important role to understand fluid phenomena. For more detail please refer to \cite{aref1983integrable,novikov1979vortex,yu2021Finiteness}  and references therein.


Following O'Neil  \cite{o1987stationary}, we will call a configuration
stationary if it leads to a self-similar solution.  In \cite{o1987stationary} it is shown that the only stationary
configurations of vortices are equilibria, rigidly translating configurations, relative equilibria (uniformly rotating configurations)
 and collapse configurations. Certainly, collapse configurations lead to self-similar collapsing    solutions.

In this paper, we are interested in the  number of stationary
configurations, especially, the  number of collapse
configurations, for giving vorticities.

As everyone knows, all stationary configurations   for  two and
three point vortices are known  explicitly \cite[etc]{grobli1877specielle,synge1949motion,novikov1979vortex,aref1979motion}. 
However, aside from special cases with certain symmetries \cite{aref2005vortex},
the only general work on stationary
configurations, especially on relative equilibria
 and collapse configurations,  is for  four point vortices.

For the four-vortex problem, O'Neil  \cite{o2007relative}  and Hampton and Moeckel  \cite{hampton2009finiteness}  independently proved that for almost
every  choice of vorticities, there are finite equilibria, rigidly translating configurations and relative equilibria. In \cite{yu2021Finiteness} it is shown that there are finitely many  stationary configurations for
every  choice of vorticities,
however, here finiteness on collapse configurations is only established  for  fixed  collapse constants.

More specifically, in \cite{yu2021Finiteness} we proved
\begin{theorem}\label{Main1}
If all the vorticities of the  four-vortex problem  are nonzero, then there are finitely many  stationary configurations for any given $\Lambda\in \mathbb{C}^*$.
\end{theorem}
\begin{corollary}\label{upperbounds1}
If all the vorticities of the  four-vortex problem  are nonzero, then there are:
\begin{itemize}
  \item[I.] exactly 2 equilibria when the necessary condition $L = 0$ holds
  \item[II.] at most 6 rigidly translating configurations the necessary condition $\Gamma = 0$
holds
  \item[III.] at most 12 collinear relative equilibria, more precisely, \begin{itemize}
                                                   \item[i.] at most 12 collinear relative equilibria when $L \neq 0$
                                                   \item[ii.] at most 10 collinear relative equilibria when $L = 0$
                                                   \item[iii.] at most 6 collinear relative equilibria when $L \neq 0$ and $\Gamma = 0$
                                                 \end{itemize}
  \item[IV.] at most 74 strictly planar  relative equilibria,  furthermore, at most 14 strictly  planar relative equilibria when $L \neq 0$ and $\Gamma = 0$
  \item[V.] at most 130  collapse configurations for any given  $\Lambda\in \mathbb{C}\setminus \mathbb{R}$ when the necessary condition $L = 0$ holds.
\end{itemize}
\end{corollary}
Here please see Definition \ref{definitionstationaryconfigurations1} and  Definition \ref{definitionstationaryconfigurations2} for the meaning of the {collapse constant} $\Lambda$.
We remark that a configuration is called strictly
planar if it is planar but not collinear, and planar configurations include collinear and strictly
planar configurations.

The main purpose of the
present study is to show that there are finitely many  collapse configurations in the four-vortex problem, as a result, we have
\begin{theorem}\label{Main}
If all the vorticities of the  four-vortex problem  are nonzero, then there are finitely many  stationary configurations.
\end{theorem}
The result is an amazing fact. Recall that Synge \cite{{synge1949motion}} proved that the three-vortex problem has a continuum of
collapse configurations. 
Therefore, it is generally  believed that there are infinitely many  collapse configurations in the  $N$-vortex problem. Indeed, if one considers the parameter $\Lambda$ in equations of collapse configurations (see  (\ref{stationaryconfiguration2}), \eqref{stationaryconfigurationmain} or (\ref{stationaryconfigurationmainLambda}) below) as an unknown quantity, then the number of unknown quantities  exceeds the number of equations and the system of equations is indefinite in form. As an example, we find a continuum of
collapse configurations in the  five-vortex problem.

The proof of Theorem \ref{Main} is based upon  an extension of   the elegant method of  Albouy and  Kaloshin for celestial mechanics \cite{Albouy2012Finiteness} by absorbing some advantages of  the method of BKK theory of Hampton and Moeckel  \cite{hampton2009finiteness}.  The principle of
the method is still to follow a possible continuum of collapse configurations in the
complex domain and to study its possible singularities there. But we find that {one can combine ideas of Albouy-Kaloshin and Hampton-Moeckel to do  an analysis of the singularities with more subtlety than before}. Roughly speaking, the method of  Albouy and  Kaloshin does not require any difficult computation, but it can only analyze leading coefficients of a polynomial system; on the other hand, the method of Hampton and Moeckel can analyze more coefficients of a polynomial system, but it requires  difficult computations by employing Computer.

Here we still embed equations of  collapse configurations into a polynomial system (see \cite{yu2021Finiteness} and the following (\ref{stationaryconfigurationmainLambda})),  then  a continuum of  collapse configurations is excluded by an analysis of the singularities. Indeed, if  the polynomial system has infinitely many solutions, then there are Puiseux series solutions. By  analyzing leading coefficients of  Puiseux series solutions,  the
original polynomial system  is replaced by many simpler ``reduced systems" for
which one would like to show that the vorticities must satisfy some constraints. 
Roughly speaking, by showing these constraints on the vorticities are conflict with each other, one shows that  the
original polynomial system  has finitely many solutions. By the way, if constraints on the vorticities  obtained by  analyzing leading coefficients of  Puiseux series solutions are consistent with each other, one can further analyze second coefficients 
and so on to get more constraints on the vorticities for reducing to absurdity.

Once  finiteness  of the number of  collapse configurations  is proved, an explicit upper bound on the number of
 collapse configurations is obtained by a direct application of   B\'{e}zout Theorems.   However, such a bound is probably still far from sharp. Nevertheless, we also provide upper bounds in the following   summary  result:
\begin{corollary}\label{upperbounds}
If all the vorticities of the four-vortex problem are nonzero,  then, besides cases I, II, III and IV are same as that in Corollary \ref{upperbounds},  we have the following  result to replace the case V in Corollary \ref{upperbounds}:
\begin{itemize}
  \item[V.] there are at most 98  collapse configurations when the necessary condition $L = 0$ holds; more precisely, \begin{itemize}
                                                   \item[i.] there are at most 108 central configurations  when $L = 0$
                                                   \item[ii.] there are at most 98  collapse configurations, thus the number of $\Lambda\in \mathbb{S}$ associated with collapse configurations is also no more than 98
                                                   \item[iii.] there are at most 49  collapse configurations for every $\Lambda\in \mathbb{S}\backslash\{\pm1,\pm \mathbf{i}\}$; there are at most 98  collapse configurations for every $\Lambda\in \{\pm \mathbf{i}\}$.
                                                 \end{itemize}
\end{itemize}
\end{corollary}

The paper is structured as follows. In \textbf{Section 2}, we recall some notations and definitions given in \cite{yu2021Finiteness}. In particular, following Hampton and Moeckel  \cite{hampton2009finiteness}, we introduce Puiseux series solutions of  a
polynomial  system.  In \textbf{Section 3}, we  discuss  some tools to classify leading terms of  Puiseux series solutions. In \textbf{Section 4} and \textbf{Section 5}, we   study all possibilities of leading terms of Puiseux series solutions and reduce the problem to the seven diagrams in Figure \ref{fig:Problematicdiagramsnew} and in Figure \ref{fig:Problematicdiagrams2}; in particular, we get constraints on the vorticities corresponding to each of the seven diagrams. 
In \textbf{Section 6}, based upon the prior work, we prove the main result on finiteness. In \textbf{Section 7}, we investigate  upper bounds on the number of
 collapse configurations. Finally, in \textbf{Section 8}, we conclude  the paper by showing the existence of a continuum of
collapse configurations in the  five-vortex problem.

\section{Preliminaries}
\label{Preliminaries}
\indent\par
In this section we recall some notations and definitions that will be needed later. For more detail please refer to \cite{yu2021Finiteness}.

\subsection{Stationary  configurations}
\label{Stationary configurations}
\indent\par
First,   let us consider vortex positions $\mathbf{r}_n\in \mathbb{R}^2$ as complex
numbers $z_n\in \mathbb{C}$, then
  \eqref{eq:vortex's equation1} becomes $\dot{z}_n =-\textbf{i}V_n$, where
\begin{equation}\label{vectorfield}
    V_n= \sum_{1 \leq j \leq N, j \neq n} \frac{\Gamma_j z_{jn}}{r_{jn}^2}= \sum_{ j \neq n} \frac{\Gamma_j }{{\overline{z}_{jn}}},
\end{equation}
$z_{jn}=z_{n}-z_{j}$, $r_{jn}=|z_{jn}|=\sqrt{z_{jn}{\overline{z}_{jn}}}$, $\textbf{i}=\sqrt{-1}$ and the overbar denotes complex conjugation.

Let $\mathbb{C}^N= \{ z = (z_1,  \cdots, z_N):z_j \in \mathbb{C}, j = 1,  \cdots, N \}$ denote the space of configurations for $N$ point vortex, and
let $\mathbb{C}^N \backslash \Delta$ denote the space of collision-free configurations.

\begin{definition}The following quantities are defined:
\begin{center}
$\begin{array}{cc}
  \text{Total vorticity} & \Gamma =\sum_{j=1}^{N}\Gamma_j  \\
  \text{Total vortex angular momentum} & L =\sum_{1\leq j<k\leq N}\Gamma_j\Gamma_k  \\
 \text{ Moment of vorticity }& M =\sum_{j=1}^{N}\Gamma_j z_j \\
 \text{ Angular impulse }& I =\sum_{j=1}^{N}\Gamma_j |z_j|^2=\sum_{j=1}^{N}\Gamma_j z_j{\overline{z}_j}
\end{array}$
\end{center}

\end{definition}
and
\begin{equation}\label{GammaIr}
\Gamma I-M \overline{M} =\sum_{1\leq j<k\leq N}\Gamma_j\Gamma_k z_{jk}{\overline{z}_{jk}}=\sum_{1\leq j<k\leq N}\Gamma_j\Gamma_k r_{jk}^2\triangleq S,
\end{equation}
\begin{equation}\label{GammaL}
\Gamma^2- 2L >0.
\end{equation}
\begin{definition}\label{definitionstationaryconfigurations1}
A configuration $z \in \mathbb{C}^N \backslash \Delta$ is stationary if there exists a constant $\Lambda\in {\mathbb{C}}$ such that
\begin{equation}\label{stationaryconfiguration}
V_j-V_k=\Lambda(z_j-z_k), ~~~~~~~~~~ 1\leq j, k\leq N.
\end{equation}
\end{definition}
\begin{definition}\label{definitionstationaryconfigurations2}
\begin{itemize}
  \item[i.] $z \in \mathbb{C}^N \backslash \Delta$ is an \emph{equilibrium} if $V_1=\cdots=V_N=0$.
  \item[ii.] $z \in \mathbb{C}^N \backslash \Delta$ is \emph{rigidly translating} if $V_1=\cdots=V_N=V$ for some $V\in \mathbb{C}^*$. (The
vortices are said to move with common velocity $V$.)
  \item[iii.] $z \in \mathbb{C}^N \backslash \Delta$ is  a \emph{relative equilibrium} if there exist constants $\lambda\in \mathbb{R}^*,z_0\in \mathbb{C}$ such that $V_n=\lambda(z_n-z_0),~~~~~~~~~~ 1\leq n\leq N$.
  \item[iv.] $z \in \mathbb{C}^N \backslash \Delta$ is a \emph{collapse configuration} if there exist constants $\Lambda,z_0\in \mathbb{C}$ with $\emph{Im}(\Lambda)\neq0$ such that $V_n=\Lambda(z_n-z_0),~~~~~~~~~~ 1\leq n\leq N$.\\
\end{itemize}

Where
~~~$\mathbb{R}^*=\mathbb{R}\backslash\{0\}$, $\mathbb{C}^*=\mathbb{C}\backslash\{0\}$.

\end{definition}
In \cite{o1987stationary}, it is shown that the only stationary
configurations of vortices are
equilibria, rigidly translating configurations, relative equilibria (uniformly rotating configurations) and collapse configurations;
in particular,  $L  =0$ is a necessary condition
for the existence of equilibria, and  $\Gamma  =0$ is a necessary condition
for the existence of configurations.

\begin{definition}
A configuration $z$ is equivalent to a configuration $z'$ if for some $a,b\in \mathbb{C}$ with $b\neq 0$, $z'_n=b(z_n+a),~~~ 1\leq n\leq N$.

 $z\in \mathbb{C}^N \backslash \Delta$ is a translation-normalized configuration  if $M=0$; $z\in \mathbb{C}^N \backslash \Delta$ is a rotation-normalized configuration  if $z_{12}\in\mathbb{R}$.  Fixing the scale of a configuration, we can give the definition of dilation-normalized configuration, however, we do not specify the scale here.

 A  configuration, which is translation-normalized, rotation-normalized and dilation-normalized, is called a \textbf{normalized configuration}.
\end{definition}
\begin{definition}
Relative equilibria and collapse configurations are both called \textbf{central configurations}.
\end{definition}

Recall that equations of relative equilibria and collapse configurations can be unified into the following form
\begin{equation}\label{stationaryconfiguration2}
\Lambda z_n= V_n,~~~~~~~~~~ 1\leq n\leq N,
\end{equation}
and solutions of  equations (\ref{stationaryconfiguration2}) satisfy
\begin{equation}\label{center0}
M=0,
\end{equation}
\begin{equation}\label{LI}
\Lambda I= L.
\end{equation}
 Following Albouy and  Kaloshin \cite{Albouy2012Finiteness}  we introduce
\begin{definition}\label{positivenormalizedcentralconfiguration}
A real normalized central configuration of the planar
$N$-vortex problem is a solution of (\ref{stationaryconfiguration2}) satisfying $z_{12}\in\mathbb{R}$ and $|\Lambda|=1$.
\end{definition}
Note that  real normalized central configurations come in a pair, that is, central configurations is determined up to a common factor $\pm 1$ by normalizing here.
Thus we count  the total central configurations up to a common factor $\pm 1$ below.

\begin{proposition}\label{Iis0}
Collapse configurations satisfy $\Gamma\neq 0$ and \begin{equation}\label{LI0whole}
S=I= L=0.
\end{equation}

For relative equilibria we have
\begin{center}
$S=0 ~~~ \Longleftrightarrow ~~~  \left\{
             \begin{array}{lr}
             \Gamma\neq 0   &\\
             I= L=0 &
             \end{array}
\right.  \text{or} ~~~\left\{
             \begin{array}{lr}
             \Gamma= 0   &\\
             I\neq 0, L\neq 0 &
             \end{array}
\right.$
\end{center}

\end{proposition}

In this paper we mainly study  collapse configurations. Thus we assume that the total vortex angular momentum is trivial in the following, i.e., \eqref{Iis0} holds.

\subsection{Complex central configurations}

\indent\par

We embed (\ref{stationaryconfiguration2}) into
\begin{equation}\label{stationaryconfiguration3}
\begin{array}{c}
  \Lambda z_n=\sum_{ j \neq n} \frac{\Gamma_j }{{w_{jn}}},~~~~~~~~~~ 1\leq n\leq N, \\
 \overline{\Lambda} w_n=\sum_{ j \neq n} \frac{\Gamma_j }{{z_{jn}}},~~~~~~~~~~ 1\leq n\leq N,
\end{array}
\end{equation}
where $z_{jn}=z_{n}-z_{j}$ and $w_{jn}=w_{n}-w_{j}$.
The condition $z_{12}\in\mathbb{R}$ we proposed to remove the rotation freedom
becomes $z_{12}=w_{12}$.

Note that $I, S$ become
$$I=\sum_{j=1}^{N}\Gamma_j z_jw_j, ~~~~~~~S=\sum_{1\leq j<k\leq N}\Gamma_j\Gamma_k r_{jk}^2=\sum_{1\leq j<k\leq N}\Gamma_j\Gamma_k z_{jk}w_{jk}.$$

To the variables $z_n,w_n\in \mathbb{C}$ we add the variables $Z_{jk},W_{jk}\in \mathbb{C}$ $(1\leq j< k\leq N)$ such that
$Z_{jk}=\frac{1}{\Lambda w_{jk}}, W_{jk}=\frac{\Lambda}{z_{jk}}$. For $1\leq k< j\leq N$ we set $Z_{jk}=-Z_{kj}, W_{jk}=-W_{kj}$. Then equations (\ref{stationaryconfiguration2}) together with the condition $z_{12}\in\mathbb{R}$ and $|\Lambda|=1$ becomes
\begin{equation}\label{stationaryconfigurationmain}
\begin{array}{cc}
   z_n=\sum_{ j \neq n} \Gamma_j Z_{jn},&1\leq n\leq N, \\
   w_n=\sum_{ j \neq n} \Gamma_j W_{jn},& 1\leq n\leq N, \\
  \Lambda Z_{jk} w_{jk}=1,&1\leq j< k\leq N, \\
  W_{jk} z_{jk}=\Lambda,&1\leq j< k\leq N, \\
  z_{jk}=z_k-z_j,~~~  w_{jk}=w_k-w_j,&1\leq j, k\leq N, \\
  Z_{jk}=-Z_{kj},~~~ W_{jk}=-W_{kj},&1\leq k< j\leq N, \\
  z_{12}=w_{12}.
\end{array}
\end{equation}
This is a polynomial system in the  variables $\mathcal{Q}=(\mathcal{Z},\mathcal{W})\in(\mathbb{C}^{N}\times\mathbb{C}^{N(N-1)/2})^2$, here
\begin{center}
$\mathcal{Z}=(z_1,z_2,\cdots,z_N,Z_{12},Z_{13},\cdots,Z_{(N-1)N})$, $\mathcal{W}=(w_1,w_2,\cdots,w_N,W_{12},W_{13},\cdots,W_{(N-1)N}).$
\end{center}

It is easy to see that a  real normalized central configuration   of (\ref{stationaryconfiguration2}) is a  solution $\mathcal{Q}=(\mathcal{Z},\mathcal{W})$ of (\ref{stationaryconfigurationmain}) such that $z_n={\overline{w}}_n$   and vice versa.

Following Albouy and  Kaloshin \cite{Albouy2012Finiteness}  we introduce
\begin{definition}[Normalized central configuration]\label{normalizedcentralconfiguration}
 A normalized central configuration is a solution $\mathcal{Q}=(\mathcal{Z},\mathcal{W})$ of (\ref{stationaryconfigurationmain}). A real
normalized central configuration is a normalized central configuration such that
$z_n={\overline{w}}_n$ for any $n=1,2,\cdots,N$. 

A (real)
normalized relative equilibrium (resp. collapse configuration) is (real) a normalized central configuration with $\Lambda=\pm 1$ (resp. $\Lambda\in \mathbb{S}\setminus \{\pm 1\}$).

Where $\mathbb{S}$ is the unit circle in $\mathbb{C}$.
\end{definition}
Definition \ref{normalizedcentralconfiguration} of a real normalized central configuration coincides with
Definition \ref{positivenormalizedcentralconfiguration}.


Note that solutions $\mathcal{Q}=(\mathcal{Z},\mathcal{W})$ of (\ref{stationaryconfigurationmain}) come in a pair: $(-\mathcal{Z},-\mathcal{W})\longmapsto(\mathcal{Z},\mathcal{W})$ sends
solution on solution, that is, a solution of (\ref{stationaryconfigurationmain}) is determined up to a common factor $\pm 1$.

\subsection{Elimination theory}

\indent\par
Recall that, a closed algebraic subset of the affine space $\mathbb{C}^m$ is a set of common zeroes
of a system of polynomials on $\mathbb{C}^m$.
The polynomial system (\ref{stationaryconfigurationmain}) defines a closed algebraic subset. For the planar four-vortex problem, we will prove that  this subset is finite, then the number of  real normalized central configurations is finite.
To distinguish the two possibilities, finitely many or infinitely many points, we
will only use the following results  from elimination theory.
\begin{lemma}\label{Eliminationtheory}\emph{(\cite{Albouy2012Finiteness})}
Let $\mathcal{X}$ be a closed algebraic subset of $\mathbb{C}^m$ and $f:\mathbb{C}^m\rightarrow \mathbb{C}$ be a
polynomial. Either the image $f(\mathcal{X})\subset\mathbb{ C}$ is a finite set, or it is the complement
of a finite set. In the second case one says that $f$ is dominating.
\end{lemma}

In particular, we remark that the polynomial functions   $I$ and $S$, on the closed algebraic subset $\mathcal{A}\subset(\mathbb{C}^{N}\times\mathbb{C}^{N(N-1)/2})^2$ defined by the system (\ref{stationaryconfigurationmain}), are two real constants.

\begin{lemma}\label{Eliminationtheory2}\emph{(\cite{hampton2006finiteness})}
Suppose that a system of $n$ polynomial equations $f_k$ defines an infinite variety $\mathcal{X}\subset \mathbb{C}^m$. Then there is a nonzero rational vector $\alpha=(\alpha_1,   \cdots,  \alpha_m)$,
$m$ nonzero numbers $a_j$, and Puiseux series $x_j(t) = a_j t^{\alpha_j} +\cdots$, $j = 1,\cdots, m$, convergent in some punctured neighborhood $U$ of $t=0$, such that $f_k\left(x_1(t),\cdots,x_m(t)\right)=0$ in $U$, $k = 1,\cdots, n$. Moreover, let $g:\mathbb{C}^m\rightarrow \mathbb{C}$ be a
polynomial, if $g$ is dominating, there exists such a series solution with $g(t) = t$ and
another with $g(t) = t^{-1}$; for example, if the projection from $\mathcal{X}$
onto the $x_i$-axis is dominant, there exists such a series solution with $x_i(t) = t$ and
another with $x_i(t) = t^{-1}$.
\end{lemma}

Lemma \ref{Eliminationtheory2} is a slight variation of Proposition 1 in \cite{hampton2006finiteness}. Since the proof of Lemma \ref{Eliminationtheory2} is  quite similar to that given  by Hampton and Moeckel in \cite{hampton2006finiteness}, no proof will be given here.

\subsection{B\'{e}zout Theorem}

\indent\par
After finiteness  of the number of  central configurations  is established, we mainly  make use  of a refined version of B\'{e}zout Theorem to discuss upper  bounds  of the number of real normalized central configurations.

\begin{lemma}\label{Bezoutrefine}\emph{(\cite{patil1983remarks})}
Let $\mathcal{V}_1, \cdots,\mathcal{V}_m$ be pure dimensional  subvarieties of $\mathbb{P}^\mathcal{N}$.
 Let $\mathcal{U}_1, \cdots,\mathcal{U}_n$ be  the
irreducible components   of $\mathcal{X}\triangleq\mathcal{V}_1\bigcap\cdots \bigcap\mathcal{V}_m$. Then
\begin{equation}\label{Bezoutformnew}
    \sum_{j=1}^{n}l(\mathcal{X};\mathcal{U}_j)deg(\mathcal{U}_j)\leq \prod_{j=1}^{m}deg(\mathcal{V}_j),
\end{equation}
where $l(\mathcal{X};\mathcal{U}_j)$ is the length of well-defined primary ideals, i.e.,  the multiplicity  of $\mathcal{X}$ along $\mathcal{U}_j$.
\end{lemma}
\begin{definition}\emph{(See \cite{eisenbud20163264})} The multiplicity $l(\mathcal{X};P)$ of $\mathcal{X}$ at a point $P\in \mathcal{X}$ is  the degree of the projectivized tangent cone $\mathbb{T}C_P\mathcal{X}$.
The multiplicity $l(\mathcal{X};\mathcal{U})$ of a scheme $\mathcal{X}$  along an
irreducible component $\mathcal{U}$,  is equal to the multiplicity of $\mathcal{X}$ at a
general point of $\mathcal{U}$.
\end{definition}
\begin{lemma}\label{Multiplicity}\emph{(\cite{fulton2013intersection})}
Let $\mathcal{V}_1, \cdots,\mathcal{V}_m$ be  pure-dimensional subschemes    of $\mathbb{P}^\mathcal{N}$, with
$$\sum_{j=1}^{m}dim(\mathcal{V}_j)=(m-1)\mathcal{N}.$$
Assume $P$ is an isolated point of $\mathcal{X}\triangleq\mathcal{V}_1\bigcap\cdots \bigcap\mathcal{V}_m$. Then
\begin{equation}\label{multiplicity}
   l(\mathcal{X};P)\geq \prod_{j=1}^{m}l(\mathcal{V}_j;\mathcal{U})+\sum_{j=1}^{n}deg(\mathcal{U}_j),
\end{equation}
where $\mathcal{U}_1, \cdots,\mathcal{U}_n$ are  the
irreducible components   of $\mathbb{T}C_P\mathcal{V}_1\bigcap\cdots \bigcap\mathbb{T}C_P\mathcal{V}_m$. In particular,
\begin{equation}\label{multiplicity1}
   l(\mathcal{X};P)\geq \prod_{j=1}^{m}l(\mathcal{V}_j;\mathcal{U})
\end{equation}
with equality if and only if $\mathbb{T}C_P\mathcal{V}_1\bigcap\cdots \bigcap\mathbb{T}C_P\mathcal{V}_m=\emptyset$.
\end{lemma}

\section{Puiseux series
solutions and colored diagram}\label{Puiseuxseriessolutionscoloreddiagram}

\indent\par
Consider the following equations for variables $\Lambda, z_n,w_n,Z_{jk},W_{jk}\in \mathbb{C}$, $n=1,\cdots,N$, $1\leq j< k\leq N$:
\begin{equation}\label{stationaryconfigurationmainLambda}
\begin{array}{cc}
   z_n=\sum_{ j \neq n} \Gamma_j Z_{jn},&1\leq n\leq N, \\
   w_n=\sum_{ j \neq n} \Gamma_j W_{jn},& 1\leq n\leq N, \\
  \Lambda Z_{jk} w_{jk}=1,&1\leq j< k\leq N, \\
  W_{jk} z_{jk}=\Lambda,&1\leq j< k\leq N, \\
  z_{jk}=z_k-z_j,~~~  w_{jk}=w_k-w_j,&1\leq j, k\leq N, \\
  Z_{jk}=-Z_{kj},~~~ W_{jk}=-W_{kj},&1\leq k< j\leq N, \\
  z_{12}=w_{12}.
\end{array}
\end{equation}

It is easy to see that, if there are infinitely many $\Lambda\in \mathbb{S}$ such that the equations (\ref{stationaryconfigurationmain})  have a solution, then the system of equations (\ref{stationaryconfigurationmainLambda})  defines an infinite variety $\mathcal{X}\subset \mathbb{C}\times (\mathbb{C}^{N}\times\mathbb{C}^{N(N-1)/2})^2$, and the projection from $\mathcal{X}$
onto the $\Lambda$-axis is dominant. By Lemma \ref{Eliminationtheory2}, it follows that there is a nonzero rational vector
\begin{equation*}
    (\alpha,\beta,\gamma)=\left(\alpha_1,   \cdots,  \alpha_N,\alpha_{12},\cdots,\alpha_{(N-1)N},\beta_1,   \cdots,  \beta_N,\beta_{12},\cdots,\beta_{(N-1)N},\gamma\right),
\end{equation*}
a vector
\begin{equation*}
(a,b,c)=\left(a_1,   \cdots,  a_N,A_{12},\cdots,A_{(N-1)N},b_1,   \cdots,  b_N,B_{12},\cdots,B_{(N-1)N},c\right)\in \mathbb{C^*}^{N(N+1)+1},
\end{equation*}
 and Puiseux series
 \begin{equation}\label{Puiseuxseries0}
   \begin{array}{llr}
     z_n(t) = a_n t^{\alpha_n} +\cdots,&w_n(t) = b_n t^{\beta_n} +\cdots, & n= 1,\cdots, N \\
     Z_{jk}(t)=\frac{1}{\Lambda(t) z_{jk}(t)} = A_{jk} t^{\alpha_{jk}} +\cdots,&W_{jk}(t) = \frac{\Lambda(t)}{ w_{jk}(t)} =B_{jk} t^{\beta_{jk}} +\cdots, & 1\leq j< k\leq N\\
     \Lambda(t) = c t^{\gamma} +\cdots,&& 
   \end{array}
\end{equation}
 convergent in some punctured neighborhood $U$ of $t=0$, such that the system of equations (\ref{stationaryconfigurationmainLambda}) holds for any $t\in U$.

Note that, by the projection from $\mathcal{X}$
onto the $\Lambda$-axis is dominant, there are Puiseux series {as}  (\ref{Puiseuxseries0})  with $\Lambda(t) = t$ or with $\Lambda(t) = t^{-1}$.

Following Hampton and Moeckel \cite{hampton2009finiteness}, a vector of nonzero Puiseux series
\begin{center}
$\mathcal{Q}(t)=\left(\mathcal{Z}(t),\mathcal{W}(t)\right)$
\end{center}
 as (\ref{Puiseuxseries0})
will be said to have order $(\alpha,\beta)$. Here recall that, \begin{center}
$\mathcal{Z}=(z_1,z_2,\cdots,z_N,Z_{12},Z_{13},\cdots,Z_{(N-1)N})$, $\mathcal{W}=(w_1,w_2,\cdots,w_N,W_{12},W_{13},\cdots,W_{(N-1)N}).$
\end{center}

Therefore, to prove the finiteness of $\Lambda\in \mathbb{S}$, it suffices to show that for every nonzero rational vector $(\alpha,\beta)$ there is no Puiseux series
solution of order $(\alpha,\beta)$. To this end, one can apply the method of BKK theory by Hampton and Moeckel  \cite{hampton2009finiteness}. However, we will not utilize this method in this paper, because  it has a large amount of calculation. Instead, we will  utilize a method similar as in \cite{yu2021Finiteness} to show that Puiseux series solutions of a given order
do not exist.



\subsection{Notations and definitions}

\indent\par
\begin{definition}[Notations on Puiseux series]\
For a given Puiseux series $x(t) = a t^{q} +\cdots$ with the leading term $a t^{q}$,   let $q= d(x(t))$ denote the degree of  $x(t)$, and set
\begin{equation*}
    a t^{q}=:L(x(t)).
\end{equation*}

For a given vector of nonzero Puiseux series
\begin{center}
$\mathcal{Q}(t)=\left(\mathcal{Z}(t),\mathcal{W}(t)\right)$
\end{center}
of order $(\alpha,\beta)$ as (\ref{Puiseuxseries0}), set
 \begin{equation*}
  |\mathcal{Z}(t)|=  |\alpha|:=\min\{\alpha_1,   \cdots,  \alpha_N,\alpha_{12},\cdots,\alpha_{(N-1)N}\},
 \end{equation*}
 \begin{equation*}
  |\mathcal{W}(t)|=  |\beta|:=\min\{\beta_1,   \cdots,  \beta_N,\beta_{12},\cdots,\beta_{(N-1)N}\},
 \end{equation*}
 \begin{equation*}
  |\mathcal{Q}(t)|:=\min\{|\mathcal{Z}(t)|,   |\mathcal{W}(t)|\},
 \end{equation*}
 which denote the degrees of $\mathcal{Z}(t),\mathcal{W}(t)$ and $\mathcal{Q}(t)$, respectively.

\end{definition}

\begin{definition}[Notations of asymptotic estimates] \
Given two Puiseux series $x,y$, we have:
\begin{description}
  \item[$x\sim y$]  means $L(x)=L(y)$;
  \item[$x\prec y$]   means $d(x)>d(y)$;
  \item[$x\preceq y$]  means $d(x)\geq d(y)$;
  \item[$x\approx y$]  means $d(x)= d(y)$.
\end{description}

\end{definition}

\begin{definition}[Strokes and circles.]
We pick a Puiseux series $\mathcal{Q}$.
We
write the indices of the bodies in a figure and use two colors for edges and
vertices.

The first color, the $z$-color, is used to mark the minimal degree components
of
\begin{center}
$\mathcal{Z}=(z_1,z_2,\cdots,z_N,Z_{12},Z_{13},\cdots,Z_{(N-1)N})$.
\end{center}
 They correspond to the components that whose degrees are equal to $|\mathcal{Z}|$. We draw a circle around
the name of vertex $\textbf{n}$ if the term $z_n$ is of minimal degree among all the components of $\mathcal{Z}$. We draw a stroke between the names $\textbf{j}$ and $\textbf{k}$ if the term
$Z_{jk}$ is of minimal degree among all the components of $\mathcal{Z}$.
\end{definition}


\indent\par

Similar as in   \cite{yu2021Finiteness}, we have
the following rules mainly concern $z$-diagram, but they apply as well to
the $w$-diagram.

\begin{description}
  \item[{Rule I}]
 There is something at each end of any $z$-stroke: another $z$-stroke
or/and a $z$-circle drawn around the name of the vertex. A $z$-circle cannot be isolated; there must be a $z$-stroke emanating from it. There is at least one
$z$-stroke in the $z$-diagram.
\end{description}

\begin{definition}[$z$-close]

 Consider a Puiseux series. We say that bodies $\textbf{k}$ and $\textbf{l}$
are close in $z$-coordinate, or $z$-close, or that $z_k$ and $z_l$ are close, if $z_{kl}\prec \mathcal{Z}$.
\end{definition}

\begin{description}
  \item[{Rule II}] If bodies $\textbf{k}$ and $\textbf{l}$
are  $z$-close, they are both $z$-circled or both not
$z$-circled.
\end{description}

\begin{definition}[Isolated component]\label{isolatedcomponent2}
An isolated component of the $z$-diagram is a subset of vertices
such that no $z$-stroke is joining a vertex of this subset to a vertex of the
complement.
\end{definition}

\begin{description}
  \item[{Rule III}]  The moment of vorticity of a set of bodies forming an isolated component of the $z$-diagram is $z$-close to the origin.
\end{description}

\begin{description}
  \item[{Rule IV}]  Consider the $z$-diagram or an isolated component of it. If there
is a $z$-circled vertex, there is another one. The $z$-circled vertices can all be
$z$-close together only if the total vorticity of these vertices is zero.
\end{description}

\begin{definition}[Maximal $z$-stroke]
Consider a $z$-stroke from vertex   $\textbf{k}$ to vertex $\textbf{l}$. We say it is
a maximal $z$-stroke if $\textbf{k}$ and $\textbf{l}$ are not $z$-close.
\end{definition}

\begin{description}
  \item[{Rule V}]  There is at least one $z$-circle at certain end of any maximal $z$-stroke. As a result,
if an isolated component of the $z$-diagram has no $z$-circled vertex,
then it has no maximal $z$-stroke.
\end{description}

 On the same diagram we also draw
$w$-strokes and $w$-circles. Graphically we use another color. The previous rules
and definitions apply to $w$-strokes and $w$-circles. What we will call simply the
diagram is the superposition of the $z$-diagram and the $w$-diagram. We will,
for example, adapt Definition \ref{isolatedcomponent2} of an isolated component: a subset of bodies
forms an isolated component of the diagram if and only if it forms an isolated
component of the $z$-diagram and an isolated component of the $w$-diagram.

\begin{definition}[Edges and strokes]
There is an edge between vertex $\textbf{k}$ and vertex $\textbf{l}$ if there
is either a $z$-stroke, or a $w$-stroke, or both. There are three types of edges,
$z$-edges, $w$-edges and $zw$-edges, and only two types of strokes, represented with
two different colors.
\end{definition}
\begin{figure}[!h]
	\centering
	\begin{tikzpicture}
		\vspace*{0cm}\hspace*{0cm} 
	\draw  (-4,  0)  node [black]{$1$}; 
	\draw  (-2,  0)  node [black]{$2$}; 
	\draw [red,very thick] (-3.75,0.05)--(-2.25,0.05); 

\vspace*{0cm}\hspace*{0cm} 
	\draw  (0,  0)  node [black]{$1$}; 
	\draw  (2,  0)  node [black]{$2$}; 
	\draw [red,very thick] (0.25,0.1)--(1.75,0.1); 
		\draw [blue, dashed, thick] (0.2,-0.05)--(1.8,-0.05); 

\vspace*{0cm}\hspace*{0cm} 
	\draw  (4,  0)  node [black]{$1$}; 
	\draw  (6,  0)  node [black]{$2$}; 
		\draw [blue, dashed, thick] (4.2,-0.05)--(5.8,-0.05); 
	\end{tikzpicture}
\caption{A $z$-stroke, a $z$-stroke plus a $z$-stroke, a $w$-stroke,
forming respectively a z-edge, a $zw$-edge, a $w$-edge.  }
	\label{fig:C=0edgestrokenew}
\end{figure}
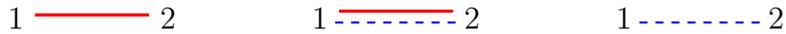

\subsection{New normalization. Main estimates.}

\indent\par

 One does not change a central configuration by multiplying the $z$ coordinates by $t^p$ and the $w$ coordinates
by $t^{-p}$ for any $p\in \mathbb{Q}$. Our diagram is invariant by such an operation, as it considers the
$z$-coordinates and the $z$-coordinates separately.

We used the normalization $z_{12}=w_{12}$ in the previous considerations. In
the following we will normalize instead with $|\mathcal{Z}|=|\mathcal{W}|$. We start with a central configuration normalized with the condition $z_{12}=w_{12}$, then multiply the $z$-coordinates by $t^p$, the $w$-coordinates by $t^{-p}$, in such a way that $|\mathcal{Z}|=|\mathcal{W}|$.

Set $|\mathcal{Z}|=|\mathcal{W}|=q$, here $q\in\mathbb{Q}$. 
Similar as in \cite{yu2021Finiteness}, we have

\begin{proposition}[Estimate 2]\label{Estimate2}
For any $(k,l)$, $1\leq k<l\leq N$, we have $t^{\gamma-q}\preceq z_{kl}\preceq t^{q}$ and $t^{-\gamma-q}\preceq w_{kl}\preceq t^{q}$; thus  $q\leq -\frac{|\gamma|}{2}$.

There is a $z$-stroke between $\textbf{k}$ and $\textbf{l}$ if and only if $w_{kl}\approx t^{-\gamma-q}$; there is a $w$-stroke between $\textbf{k}$ and $\textbf{l}$ if and only if $z_{kl}\approx t^{\gamma-q}$.

 There is a maximal $z$-stroke between $\textbf{k}$ and $\textbf{l}$ if and only if $z_{kl}\approx t^{q}, w_{kl}\approx t^{-\gamma-q}$; there is a maximal $w$-stroke between $\textbf{k}$ and $\textbf{l}$ if and only if $z_{kl}\approx t^{\gamma-q}, w_{kl}\approx t^{q}$.

  There is a $z$-edge between $\textbf{k}$ and $\textbf{l}$ if and only if $z_{kl}\succ t^{\gamma-q},w_{kl}\approx t^{-\gamma-q}$; there is a $w$-edge between $\textbf{k}$ and $\textbf{l}$ if and only if $z_{kl}\approx t^{\gamma-q},w_{kl}\succ t^{-\gamma-q}$

There is a maximal $z$-edge between $\textbf{k}$ and $\textbf{l}$ if and only if $z_{kl}\approx t^{q},w_{kl}\approx t^{-\gamma-q}$; there is a maximal $w$-edge between $\textbf{k}$ and $\textbf{l}$ if and only if $z_{kl}\approx t^{\gamma-q},w_{kl}\approx t^{q}$.

There is a $zw$-edge between $\textbf{k}$ and $\textbf{l}$ if and only if $z_{kl}\approx t^{\gamma-q},w_{kl}\approx t^{-\gamma-q}$.
\end{proposition}

\begin{description}
  \item[{Rule VI}]
  If there are two consecutive $z$-stroke, there is a third $z$-stroke closing the triangle.
\end{description}

 Similar to \cite{yu2021Finiteness}, we  classify all possible diagrams by these rules of coloring diagrams in the following two sections. We divide discussions   into two cases according to
$q< -\frac{|\gamma|}{2}$ or $q=-\frac{|\gamma|}{2}$. Note that we only consider the case that  $q< 0$ below, since we will only face this case when we practically construct  Puiseux series.

\section{Possible diagrams in the case $q< -{|\gamma|}/{2}$}\label{Exclusionofthecaselessgamma12}
\indent\par
In this section, we consider the case that  $q< -\frac{|\gamma|}{2}$.
Then it is easy to see that,  by  estimates in Proposition \ref{Estimate2}, we have
\begin{remark}
The strokes in a $zw$-edge are not maximal. A maximal $z$-stroke (resp. $w$-stroke) is exactly a maximal $z$-edge (resp. $w$-edge). In particular, if vertices $\textbf{k}$ and $\textbf{l}$  form a $z$-stroke (resp. $w$-stroke), then they are   $w$-close (resp. $z$-close).
\end{remark}

Consequently,
we remark  that the rules for coloring diagrams in this section are exactly same as that in Section 3 in \cite{yu2021Finiteness}.

\subsection{Possible diagrams}
\indent\par
Based upon the results in previous sections, it is easy to see that the only possible diagrams are  exactly same as that in Subsection 4.1 in \cite{yu2021Finiteness}, i.e., Figures  2--10 in \cite{yu2021Finiteness}. For convenience, these Figures are presented below.
\begin{figure}[!h]
	\centering
	\begin{tikzpicture}[scale=1.3]
		\vspace*{0cm}\hspace*{0cm} 
	\draw [red,very thick] (-1,  0) circle (0.25) ; 
	\draw [red,very thick] (1,  0) circle (0.25) ; 
	\draw [red,very thick] (-0.75,0.05)--(0.75,0.05); 
	\draw [blue, dashed, thick] (-1,  -1.5) circle (0.2) ; 
	\draw [blue, dashed, thick] (1,  -1.5) circle (0.2); 
		\draw [blue, dashed, thick] (-0.8,-1.55)--(0.8,-1.55); 
\draw	  (-1,0)    node {\large\textbf{1}};
\draw		(1,0) node {\large\textbf{2}};
\draw		(1,-1.5) node {\large\textbf{3}};
\draw		(-1,-1.5) node {\large\textbf{4}};
	\end{tikzpicture}
\caption{  }
	\label{fig:C=0}
\end{figure}

\begin{figure}[!h]
	\centering
	\begin{tikzpicture}[scale=1.3]
		\vspace*{0cm}\hspace*{0cm} 
	\draw [red,very thick] (-5,  0) circle (0.3) ; 
	\draw [red,very thick] (-3,  0) circle (0.3) ; 
\draw [blue, dashed, thick] (-5, 0) circle (0.2); 
	\draw [blue, dashed, thick] (-3, 0) circle (0.2) ; 
	\draw [red,very thick] (-4.7,0.05)--(-3.3,0.05); 
\draw [blue, dashed, thick] (-4.8,-0.05)--(-3.2,-0.05); 
\draw  (-5,  -1.5) ; 
	\draw  (-3,  -1.5) ; 
\draw	  (-5,0)    node {\large\textbf{1}};
\draw		(-3,0) node {\large\textbf{2}};
\draw		(-3,-1.5) node {\large\textbf{3}};
\draw		(-5,-1.5) node {\large\textbf{4}};
			
\vspace*{0cm}\hspace*{0cm} 
		\draw [red,very thick] (-1,  0) circle (0.3) ; 
	\draw [red,very thick] (1,  0) circle (0.3) ; 
\draw [blue, dashed, thick] (-1, 0) circle (0.2); 
	\draw [blue, dashed, thick] (1, 0) circle (0.2) ; 
	\draw [red,very thick] (-0.7,0.05)--(0.7,0.05); 
\draw [blue, dashed, thick] (-0.8,-0.05)--(0.8,-0.05); 
\draw [red,very thick] (-1,  -1.5) circle (0.3) ; 
	\draw [red,very thick] (1,  -1.5) circle (0.3); 
	\draw [red,very thick] (-0.7,-1.55)--(0.7,-1.55); 
\draw	  (-1,0)    node {\large\textbf{1}};
\draw		(1,0) node {\large\textbf{2}};
\draw		(1,-1.5) node {\large\textbf{3}};
\draw		(-1,-1.5) node {\large\textbf{4}};

	\vspace*{0cm}\hspace*{0cm} 
	\draw [red,very thick] (3,  0) circle (0.3) ; 
	\draw [red,very thick] (5,  0) circle (0.3) ; 
\draw [blue, dashed, thick] (3, 0) circle (0.2); 
	\draw [blue, dashed, thick] (5, 0) circle (0.2) ; 
	\draw [red,very thick] (3.3,0.05)--(4.7,0.05); 
\draw [blue, dashed, thick] (3.2,-0.05)--(4.8,-0.05); 
\draw [red,very thick] (3,  -1.5) circle (0.3) ; 
	\draw [red,very thick] (5,  -1.5) circle (0.3) ; 
\draw [blue, dashed, thick] (3, -1.5) circle (0.2); 
	\draw [blue, dashed, thick] (5, -1.5) circle (0.2) ; 
	\draw [red,very thick] (3.3,-1.55)--(4.7,-1.55); 
\draw [blue, dashed, thick] (3.2,-1.45)--(4.8,-1.45); 

\draw	  (3,0)    node {\large\textbf{1}};
\draw		(5,0) node {\large\textbf{2}};
\draw		(5,-1.5) node {\large\textbf{3}};
\draw		(3,-1.5) node {\large\textbf{4}};
	\end{tikzpicture}
\caption{}
	\label{fig:C=21}
\end{figure}

\begin{figure}[!h]
	\centering
	\begin{tikzpicture}[scale=1.3]
		\vspace*{0cm}\hspace*{0cm} 
	\draw [blue, dashed, thick] (0,  0) circle (0.2) ;
	\draw [red,very thick] (0,  0) circle (0.3);
	\draw [blue, dashed, thick] (2,  0) circle (0.2) ;
	\draw [red,very thick] (2,  0) circle (0.3);
	\draw [blue, dashed, thick] (0.2,0)--(1.8,0);
	\draw [blue, dashed, thick] (0.2,-1.5)--(1.8,-1.5);
	\draw [blue, dashed, thick] (0,  -1.5) circle (0.2);
	\draw [red,very thick] (0,  -1.5) circle (0.3);
	\draw[blue, dashed, thick] (2,  -1.5) circle (0.2);
	\draw [red,very thick] (2,  -1.5) circle (0.3);
	\draw [red,very thick] (0, -1.2)--(0,-0.3);
	\draw [red,very thick] (2, -1.2)--(2,-0.3);

\draw	  (0,0)    node {\large\textbf{1}};
\draw		(2,0) node {\large\textbf{2}};
\draw		(2,-1.5) node {\large\textbf{3}};
\draw		(0,-1.5) node {\large\textbf{4}};
	\end{tikzpicture}
\caption{  }
	\label{fig:C=22}
\end{figure}

\begin{figure}[!h]
	\centering
	\begin{tikzpicture}[scale=1.3]
		\vspace*{0cm}\hspace*{0cm} 
	\draw [blue, dashed, thick] (-3,  0) circle (0.2);
	\draw [blue, dashed, thick] (-1,0) circle (0.2);
	\draw [blue, dashed, thick] (-2,-1.5) circle (0.2) ;
	\draw [blue, dashed, thick] (-4,-1.5) circle (0.2);
	
	\draw	  (-3,0)    node {\large\textbf{1}};
\draw		(-1,0) node {\large\textbf{2}};
\draw		(-2,-1.5) node {\large\textbf{3}};
\draw		(-4,-1.5) node {\large\textbf{4}};

	\draw [blue, dashed, thick] (-2.8,-0.05)--(-1.2,-0.05);
	\draw [blue, dashed, thick] (-3.8,-1.5)--(-2.2,-1.5);
		\draw [red,very thick] (-2.7,0.05)--(-1.3,0.05);
	\draw  [red,very thick]  (-1.18,-0.25)--(-1.75,-1.32);
	\draw  [red,very thick]  (-2.82, -0.25)--(-2.25,-1.32);

\vspace*{0cm}\hspace*{0cm} 
	\draw [blue, dashed, thick] (1,  0) circle (0.2) ;
	\draw [blue, dashed, thick] (3,0) circle (0.2) ;
	\draw [blue, dashed, thick] (4,-1.5) circle (0.2) ;
	\draw [blue, dashed, thick] (2,-1.5) circle (0.2) ;
	\draw [red,very thick] (1,  0) circle (0.3) ;
	\draw [red,very thick] (3,0) circle (0.3) ;
	
\draw	  (1,0)    node {\large\textbf{1}};
\draw		(3,0) node {\large\textbf{2}};
\draw		(2,-1.5) node {\large\textbf{3}};
\draw		(4,-1.5) node {\large\textbf{4}};

	\draw [blue, dashed, thick] (2.8,-0.05)--(1.2,-0.05);
	\draw [blue, dashed, thick] (3.8,-1.5)--(2.2,-1.5);
		\draw [red,very thick] (2.7,0.05)--(1.3,0.05);
	\draw  [red,very thick]  (1.18,-0.25)--(1.75,-1.32);
	\draw  [red,very thick]  (2.82, -0.25)--(2.25,-1.32);

	\end{tikzpicture}
\caption{ }
	\label{fig:C=31}
\end{figure}

\begin{figure}[!h]
	\centering
	\begin{tikzpicture}
	\vspace*{0cm}\hspace*{-4cm}
\draw	  (-3/2,0)    node {\large\textbf{1}};
\draw		(3/2,0) node {\large\textbf{2}};
\draw		(0,-3/2*1.732) node {\large\textbf{3}};
\draw		(0,-1) node {\large\textbf{4}};

\draw [red,very thick] (-3/2+.35,0)--(3/2-.35,0);
\draw [blue, dashed,thick] (-3/2+0.35,-.15)--(3/2-.35,-0.15);

\draw [red,very thick] (-3/2+.2,-0.2*1.732)--(-.2,-3/2*1.732+.2*1.732);
\draw [blue, dashed,thick] (-3/2+0.15+.2, -0.2*1.732)--(0.15-.2,-3/2*1.732+0.2*1.732);

\draw [red,very thick] (3/2-0.2,-0.2*1.732)--(0.2,-3/2*1.732+0.2*1.732);
\draw [blue, dashed,thick] (3/2-0.15-0.2,-.2*1.732)--(-0.15+0.2,-3/2*1.732+.2*1.732);

	\vspace*{0cm}\hspace*{4cm}

\draw	[red,very thick]  (-3/2,0)  circle (0.35);
\draw  (-3/2,0)  node {\large\textbf{1}};

\draw	[red,very thick]  (3/2,0)  circle (0.35);
\draw		(3/2,0) node {\large\textbf{2}};

\draw	[red,very thick] (0,-3/2*1.732) circle (0.35);
\draw		(0,-3/2*1.732) node {\large\textbf{3}};

\draw		(0,-1) node {\large\textbf{4}};

\draw [red,very thick] (-3/2+.35,0)--(3/2-.35,0);
\draw [blue, dashed,thick] (-3/2+0.35,-.15)--(3/2-.35,-0.15);

\draw [red,very thick] (-3/2+.2,-0.2*1.732)--(-.2,-3/2*1.732+.2*1.732);
\draw [blue, dashed,thick] (-3/2+0.15+.2, -0.2*1.732)--(0.15-.2,-3/2*1.732+0.2*1.732);

\draw [red,very thick] (3/2-0.2,-0.2*1.732)--(0.2,-3/2*1.732+0.2*1.732);
\draw [blue, dashed,thick] (3/2-0.15-0.2,-.2*1.732)--(-0.15+0.2,-3/2*1.732+.2*1.732);

	\vspace*{0cm}\hspace*{4cm}
	\draw	[blue,dashed, thick]  (-3/2,0)  circle (0.25);
\draw	[red,very thick]  (-3/2,0)  circle (0.35);
\draw  (-3/2,0)  node {\large\textbf{1}};

	\draw	[blue,dashed, thick]  (3/2,0)  circle (0.25);
\draw	[red,very thick]  (3/2,0)  circle (0.35);
\draw		(3/2,0) node {\large\textbf{2}};

	\draw	[blue,dashed, thick]  (0,-3/2*1.732) circle (0.25);
\draw	[red,very thick] (0,-3/2*1.732) circle (0.35);
\draw		(0,-3/2*1.732) node {\large\textbf{3}};

\draw		(0,-1) node {\large\textbf{4}};

\draw [red,very thick] (-3/2+.35,0)--(3/2-.35,0);
\draw [blue, dashed,thick] (-3/2+0.35,-.15)--(3/2-.35,-0.15);

\draw [red,very thick] (-3/2+.2,-0.2*1.732)--(-.2,-3/2*1.732+.2*1.732);
\draw [blue, dashed,thick] (-3/2+0.15+.2, -0.2*1.732)--(0.15-.2,-3/2*1.732+0.2*1.732);

\draw [red,very thick] (3/2-0.2,-0.2*1.732)--(0.2,-3/2*1.732+0.2*1.732);
\draw [blue, dashed,thick] (3/2-0.15-0.2,-.2*1.732)--(-0.15+0.2,-3/2*1.732+.2*1.732);

	\end{tikzpicture}
	\caption{}
\label{fig:C=41}
\end{figure}

\begin{figure}[!h]
	\centering
	\begin{tikzpicture}
	\vspace*{0cm}\hspace*{-4cm}
	\draw	[blue,dashed,thick]  (-3/2,0)  circle (0.35);
	\draw	  (-3/2,0)    node {\large\textbf{1}};
	
		\draw	[blue,dashed,thick]  (3/2,0)  circle (0.35);
	\draw		(3/2,0) node {\large\textbf{2}};
	
		\draw	[blue,dashed,thick]  (0,-3/2*1.732)   circle (0.35);
	\draw		(0,-3/2*1.732) node {\large\textbf{3}};
	
		\draw	[blue,dashed,thick]  (0,-1)  circle (0.35);
	\draw		(0,-1) node {\large\textbf{4}};

	\draw [red,very thick] (-3/2+.35,0.15)--(3/2-.35,0.15);
\draw [blue, dashed,thick] (-3/2+0.35,0)--(3/2-.35,0);
	
	\draw [red,very thick] (-3/2+.2,-0.2*1.732)--(-.2,-3/2*1.732+.2*1.732);
	
	\draw [red,very thick] (3/2-0.2,-0.2*1.732)--(0.2,-3/2*1.732+0.2*1.732);

		\draw [red,very thick]  (-3/2+0.25+0.15,-0.25*2/3)--(0-0.4+0.15,-1+0.4*2/3);  
		
		\draw [red,very thick]  (3/2-0.25-.15,-.25*2/3)--(0.45 -.15,-1+.45*2/3);  

		\draw [red,very thick] (-.07,-3/2*1.732+.4)--(-.07,-1-.4);
			\draw [blue,dashed,thick]  (.07,-3/2*1.732+0.4)--(.07,-1-0.4);

	\vspace*{0cm}\hspace*{4.5cm}
	
	\draw	[blue,dashed, thick]  (-3/2,0)  circle (0.25);
\draw	[red,very thick]  (-3/2,0)  circle (0.35);
\draw  (-3/2,0)  node {\large\textbf{1}};

\draw	[blue,dashed, thick]  (3/2,0)  circle (0.25);
\draw	[red,very thick]  (3/2,0)  circle (0.35);
\draw		(3/2,0) node {\large\textbf{2}};
	
		\draw	[blue,dashed,thick]  (0,-3/2*1.732)   circle (0.35);
\draw		(0,-3/2*1.732) node {\large\textbf{3}};

\draw	[blue,dashed,thick]  (0,-1)  circle (0.35);
\draw		(0,-1) node {\large\textbf{4}};

\draw [red,very thick] (-3/2+.35,0.15)--(3/2-.35,0.15);
\draw [blue, dashed,thick] (-3/2+0.35,0)--(3/2-.35,0);

\draw [red,very thick] (-3/2+.2,-0.2*1.732)--(-.2,-3/2*1.732+.2*1.732);

\draw [red,very thick] (3/2-0.2,-0.2*1.732)--(0.2,-3/2*1.732+0.2*1.732);

\draw [red,very thick]  (-3/2+0.25+0.15,-0.25*2/3)--(0-0.4+0.15,-1+0.4*2/3);  

\draw [red,very thick]  (3/2-0.25-.15,-.25*2/3)--(0.45 -.15,-1+.45*2/3);  

\draw [red,very thick] (-.07,-3/2*1.732+.4)--(-.07,-1-.4);
\draw [blue,dashed,thick]  (.07,-3/2*1.732+0.4)--(.07,-1-0.4);

	\vspace*{0cm}\hspace*{4.5cm}
	\draw	[blue,dashed, thick]  (-3/2,0)  circle (0.25);
	\draw	[red,very thick]  (-3/2,0)  circle (0.35);
	\draw  (-3/2,0)  node {\large\textbf{1}};
	
	\draw	[blue,dashed, thick]  (3/2,0)  circle (0.25);
	\draw	[red,very thick]  (3/2,0)  circle (0.35);
	\draw		(3/2,0) node {\large\textbf{2}};
	
	\draw	[blue,dashed, thick]  (0,-3/2*1.732) circle (0.25);
	\draw	[red,very thick] (0,-3/2*1.732) circle (0.35);
	\draw		(0,-3/2*1.732) node {\large\textbf{3}};
	
		\draw	[blue,dashed, thick]  (0,-1) circle (0.25);
		\draw	[red,very thick](0,-1) circle (0.35);
	\draw		(0,-1) node {\large\textbf{4}};

\draw [red,very thick] (-3/2+.35,0.15)--(3/2-.35,0.15);
\draw [blue, dashed,thick] (-3/2+0.35,0)--(3/2-.35,0);

\draw [red,very thick] (-3/2+.2,-0.2*1.732)--(-.2,-3/2*1.732+.2*1.732);

\draw [red,very thick] (3/2-0.2,-0.2*1.732)--(0.2,-3/2*1.732+0.2*1.732);

\draw [red,very thick]  (-3/2+0.25+0.15,-0.25*2/3)--(0-0.4+0.15,-1+0.4*2/3);  

\draw [red,very thick]  (3/2-0.25-.15,-.25*2/3)--(0.45 -.15,-1+.45*2/3);  

\draw [red,very thick] (-.07,-3/2*1.732+.4)--(-.07,-1-.4);
\draw [blue,dashed,thick]  (.07,-3/2*1.732+0.4)--(.07,-1-0.4);

	\end{tikzpicture}
	\caption{}
\label{fig:C=42}
\end{figure}
\begin{figure}[!h]
	\centering
	\begin{tikzpicture}

	\vspace*{0cm}\hspace*{0cm}

	\draw  (-3/2,0)  node {\large\textbf{1}};

	\draw		(3/2,0) node {\large\textbf{2}};

	\draw		(0,-3/2*1.732) node {\large\textbf{3}};

	\draw		(0,-1) node {\large\textbf{4}};
	
		\draw [red,very thick] (-3/2+.35,0.15)--(3/2-.35,0.15);
	\draw [blue, dashed,thick] (-3/2+0.35,0)--(3/2-.35,0);
	
	\draw [red,very thick] (-3/2+.2,-0.2*1.732)--(-.2,-3/2*1.732+.2*1.732);
	
	\draw [red,very thick] (3/2-0.2,-0.2*1.732)--(0.2,-3/2*1.732+0.2*1.732);

	\draw [blue,dashed,thick]  (-3/2+0.3+0.15,-0.3*2/3)--(0-0.35+0.15,-1+0.35*2/3);  
	
	\draw [blue,dashed,thick]  (3/2-0.25-.15,-.25*2/3)--(0.4 -.15,-1+.4*2/3);  


	\end{tikzpicture}
	\caption{}
\label{fig:C=43}
\end{figure}

\begin{figure}[!h]
	\centering
	\begin{tikzpicture}
	\vspace*{0cm}\hspace*{-4cm}
	\draw	  (-3/2,0)    node {\large\textbf{1}};
	
	\draw		(3/2,0) node {\large\textbf{2}};
	
	\draw		(0,-3/2*1.732) node {\large\textbf{3}};
	
	\draw		(0,-1) node {\large\textbf{4}};

	\draw [red,very thick] (-3/2+.35,0.15)--(3/2-.35,0.15);
\draw [blue, dashed,thick] (-3/2+0.35,0)--(3/2-.35,0);

\draw [blue, dashed,thick] (-3/2+.2,-0.2*1.732)--(-.2,-3/2*1.732+.2*1.732);
\draw [red,very thick] (-3/2-0.15+.25, -0.25*1.732)--(-0.15-.25,-3/2*1.732+0.25*1.732);

\draw [blue, dashed,thick] (3/2-0.2,-0.2*1.732)--(0.2,-3/2*1.732+0.2*1.732);
\draw [red,very thick] (3/2+0.15-0.25,-0.25*1.732)--(0.15+0.25,-3/2*1.732+.25*1.732);

\draw [red,very thick]  (-3/2+0.3+0.15,-0.3*2/3)--(0-0.35+0.15,-1+0.35*2/3);  

\draw [red,very thick]  (3/2-0.25-.15,-.25*2/3)--(0.4 -.15,-1+.4*2/3);  

\draw [red,very thick] (0,-3/2*1.732+.4)--(0,-1-.4);

	\vspace*{0cm}\hspace*{4.5cm}
	
	\draw	[red,very thick]  (-3/2,0)  circle (0.35);
	\draw  (-3/2,0)  node {\large\textbf{1}};
	
	\draw	[red,very thick]  (3/2,0)  circle (0.35);
	\draw		(3/2,0) node {\large\textbf{2}};
	
	\draw	[red,very thick]  (0,-3/2*1.732)   circle (0.35);
	\draw		(0,-3/2*1.732) node {\large\textbf{3}};
	
	\draw		(0,-1) node {\large\textbf{4}};
	
	\draw [red,very thick] (-3/2+.35,0.15)--(3/2-.35,0.15);
\draw [blue, dashed,thick] (-3/2+0.35,0)--(3/2-.35,0);

\draw [blue, dashed,thick] (-3/2+.2,-0.2*1.732)--(-.2,-3/2*1.732+.2*1.732);
\draw [red,very thick] (-3/2-0.15+.25, -0.25*1.732)--(-0.15-.25,-3/2*1.732+0.25*1.732);

\draw [blue, dashed,thick] (3/2-0.2,-0.2*1.732)--(0.2,-3/2*1.732+0.2*1.732);
\draw [red,very thick] (3/2+0.15-0.25,-0.25*1.732)--(0.15+0.25,-3/2*1.732+.25*1.732);

\draw [red,very thick]  (-3/2+0.3+0.15,-0.3*2/3)--(0-0.35+0.15,-1+0.35*2/3);  

\draw [red,very thick]  (3/2-0.25-.15,-.25*2/3)--(0.4 -.15,-1+.4*2/3);  

\draw [red,very thick] (0,-3/2*1.732+.4)--(0,-1-.4);

	\vspace*{0cm}\hspace*{4.5cm}
	\draw	[red,very thick]  (-3/2,0)  circle (0.35);
	\draw  (-3/2,0)  node {\large\textbf{1}};
	
	\draw	[red,very thick]  (3/2,0)  circle (0.35);
	\draw		(3/2,0) node {\large\textbf{2}};
	
	\draw	[red,very thick] (0,-3/2*1.732) circle (0.35);
	\draw		(0,-3/2*1.732) node {\large\textbf{3}};
	
	\draw	[red,very thick](0,-1) circle (0.35);
	\draw		(0,-1) node {\large\textbf{4}};
	
	\draw [red,very thick] (-3/2+.35,0.15)--(3/2-.35,0.15);
	\draw [blue, dashed,thick] (-3/2+0.35,0)--(3/2-.35,0);
	
	\draw [blue, dashed,thick] (-3/2+.2,-0.2*1.732)--(-.2,-3/2*1.732+.2*1.732);
	\draw [red,very thick] (-3/2-0.15+.25, -0.25*1.732)--(-0.15-.25,-3/2*1.732+0.25*1.732);
	
	\draw [blue, dashed,thick] (3/2-0.2,-0.2*1.732)--(0.2,-3/2*1.732+0.2*1.732);
	\draw [red,very thick] (3/2+0.15-0.25,-0.25*1.732)--(0.15+0.25,-3/2*1.732+.25*1.732);

	\draw [red,very thick]  (-3/2+0.3+0.15,-0.3*2/3)--(0-0.4+0.15,-1+0.4*2/3);  
	
	\draw [red,very thick]  (3/2-0.25-.15,-.25*2/3)--(0.4 -.15,-1+.4*2/3);  

	\draw [red,very thick] (0,-3/2*1.732+.4)--(0,-1-.4);

	\end{tikzpicture}
	\caption{}
\label{fig:C=51}
\end{figure}

\begin{figure}[!h]
	\centering
	\begin{tikzpicture}
	\vspace*{0cm}\hspace*{-4cm}
	\draw	  (-3/2,0)    node {\large\textbf{1}};
	
	\draw		(3/2,0) node {\large\textbf{2}};
	
	\draw		(0,-3/2*1.732) node {\large\textbf{3}};
	
	\draw		(0,-1) node {\large\textbf{4}};

	\draw [red,very thick] (-3/2+.35,0.15)--(3/2-.35,0.15);
	\draw [blue, dashed,thick] (-3/2+0.35,0)--(3/2-.35,0);
	
	\draw [blue, dashed,thick] (-3/2+.2,-0.2*1.732)--(-.2,-3/2*1.732+.2*1.732);
	\draw [red,very thick] (-3/2-0.15+.25, -0.25*1.732)--(-0.15-.25,-3/2*1.732+0.25*1.732);
	
	\draw [blue, dashed,thick] (3/2-0.2,-0.2*1.732)--(0.2,-3/2*1.732+0.2*1.732);
	\draw [red,very thick] (3/2+0.15-0.25,-0.25*1.732)--(0.15+0.25,-3/2*1.732+.25*1.732);

	\draw [blue,dashed,thick]  (-3/2+0.3+0.15,-0.3*2/3)--(0-0.35+0.15,-1+0.35*2/3);  
	\draw [red,very thick]  (-3/2+0.4-0.1,-0.4*2/3)--(0-0.3-0.1,-1+0.3*2/3);  
	
	\draw [red,very thick]  (3/2-0.25-.15,-.25*2/3)--(0.4 -.15,-1+.4*2/3);  
	\draw [blue,dashed,thick]   (3/2-0.4+.1,-.4*2/3)--(0.3 +.1,-1+.3*2/3);  

	\draw [red,very thick] (-.07,-3/2*1.732+.4)--(-.07,-1-.4);
	\draw [blue,dashed,thick]  (.07,-3/2*1.732+0.4)--(.07,-1-0.4);

	\vspace*{0cm}\hspace*{4.5cm}
	
	\draw	[red,very thick]  (-3/2,0)  circle (0.35);
	\draw  (-3/2,0)  node {\large\textbf{1}};
	
	\draw	[red,very thick]  (3/2,0)  circle (0.35);
	\draw		(3/2,0) node {\large\textbf{2}};
	
	\draw	[red,very thick]  (0,-3/2*1.732)   circle (0.35);
	\draw		(0,-3/2*1.732) node {\large\textbf{3}};
	
	\draw	[red,very thick]  (0,-1)  circle (0.35);
	\draw		(0,-1) node {\large\textbf{4}};
	
	\draw [red,very thick] (-3/2+.35,0.15)--(3/2-.35,0.15);
	\draw [blue, dashed,thick] (-3/2+0.35,0)--(3/2-.35,0);
	
	\draw [blue, dashed,thick] (-3/2+.2,-0.2*1.732)--(-.2,-3/2*1.732+.2*1.732);
	\draw [red,very thick] (-3/2-0.15+.25, -0.25*1.732)--(-0.15-.25,-3/2*1.732+0.25*1.732);
	
	\draw [blue, dashed,thick] (3/2-0.2,-0.2*1.732)--(0.2,-3/2*1.732+0.2*1.732);
	\draw [red,very thick] (3/2+0.15-0.25,-0.25*1.732)--(0.15+0.25,-3/2*1.732+.25*1.732);

	\draw [blue,dashed,thick]  (-3/2+0.3+0.15,-0.3*2/3)--(0-0.35+0.15,-1+0.35*2/3);  
	\draw [red,very thick]  (-3/2+0.4-0.1,-0.4*2/3)--(0-0.3-0.1,-1+0.3*2/3);  
	
	\draw [red,very thick]  (3/2-0.25-.15,-.25*2/3)--(0.4 -.15,-1+.4*2/3);  
	\draw [blue,dashed,thick]   (3/2-0.4+.1,-.4*2/3)--(0.3 +.1,-1+.3*2/3);  

	\draw [red,very thick] (-.07,-3/2*1.732+.4)--(-.07,-1-.4);
	\draw [blue,dashed,thick]  (.07,-3/2*1.732+0.4)--(.07,-1-0.4);

	\vspace*{0cm}\hspace*{4.5cm}
	\draw	[blue,dashed, thick]  (-3/2,0)  circle (0.25);
	\draw	[red,very thick]  (-3/2,0)  circle (0.35);
	\draw  (-3/2,0)  node {\large\textbf{1}};
	
	\draw	[blue,dashed, thick]  (3/2,0)  circle (0.25);
	\draw	[red,very thick]  (3/2,0)  circle (0.35);
	\draw		(3/2,0) node {\large\textbf{2}};
	
	\draw	[blue,dashed, thick]  (0,-3/2*1.732) circle (0.25);
	\draw	[red,very thick] (0,-3/2*1.732) circle (0.35);
	\draw		(0,-3/2*1.732) node {\large\textbf{3}};
	
	\draw	[blue,dashed, thick]  (0,-1) circle (0.25);
	\draw	[red,very thick](0,-1) circle (0.35);
	\draw		(0,-1) node {\large\textbf{4}};
	
	\draw [red,very thick] (-3/2+.35,0.15)--(3/2-.35,0.15);
	\draw [blue, dashed,thick] (-3/2+0.35,0)--(3/2-.35,0);
	
	\draw [blue, dashed,thick] (-3/2+.2,-0.2*1.732)--(-.2,-3/2*1.732+.2*1.732);
	\draw [red,very thick] (-3/2-0.15+.25, -0.25*1.732)--(-0.15-.25,-3/2*1.732+0.25*1.732);
	
	\draw [blue, dashed,thick] (3/2-0.2,-0.2*1.732)--(0.2,-3/2*1.732+0.2*1.732);
	\draw [red,very thick] (3/2+0.15-0.25,-0.25*1.732)--(0.15+0.25,-3/2*1.732+.25*1.732);

	\draw [blue,dashed,thick]  (-3/2+0.3+0.15,-0.3*2/3)--(0-0.35+0.15,-1+0.35*2/3);  
	\draw [red,very thick]  (-3/2+0.4-0.1,-0.4*2/3)--(0-0.3-0.1,-1+0.3*2/3);  
	
	\draw [red,very thick]  (3/2-0.25-.15,-.25*2/3)--(0.4 -.15,-1+.4*2/3);  
	\draw [blue,dashed,thick]   (3/2-0.4+.1,-.4*2/3)--(0.3 +.1,-1+.3*2/3);  

	\draw [red,very thick] (-.07,-3/2*1.732+.4)--(-.07,-1-.4);
	\draw [blue,dashed,thick]  (.07,-3/2*1.732+0.4)--(.07,-1-0.4);

	\end{tikzpicture}
	\caption{}
\label{fig:C=61}
\end{figure}

Proposition 4.1 in Subsection 4.2 in  \cite{yu2021Finiteness}  may be wrong generally, here we propose the following  proposition to replace Proposition 4.1 in  \cite{yu2021Finiteness} .
\begin{proposition}\label{relationstwoverticesnew}
Suppose a diagram has two $z$-circled vertices (say $\textbf{1}$ and $\textbf{2}$) without other $z$-circled vertices. If vertices $\textbf{1}$ and $\textbf{2}$  form a $z$-stroke,  then $\Gamma_1+\Gamma_2\neq 0$.
\end{proposition}
{\bf Proof.} 

Without loss of generality, assume that
\begin{equation*}
z_1\sim -\Gamma_2 at^{q},~~~~~~~  z_2\sim \Gamma_1at^{q}, ~~~~~~~ w_{12}\sim \frac{1}{ac} t^{-\gamma-q},
\end{equation*}
here recall that $\Lambda\sim c t^{\gamma}$.

If $\Gamma_1+\Gamma_2= 0$, by $L=0$,  it is easy to see that $\sum_{ j > 2} \Gamma_j\neq 0$.

By \begin{equation*}
     w_{12}=(\Gamma_1+\Gamma_2)W_{12}+ \sum_{ j > 2} \Gamma_j (W_{j2}-W_{j1}),
\end{equation*}
it follows that
\begin{equation*}
   \frac{1}{ac} t^{-\gamma-q}\sim  w_{12}=- \sum_{ j > 2} \frac{\Gamma_j \Lambda z_{12}}{z_{1j}z_{2j}}\preceq -\frac{\sum_{ j > 2} \Gamma_j c t^{\gamma} z_{12}}{(\Gamma_1at^{q})^2}\prec-\frac{\sum_{ j > 2} \Gamma_j  c t^{\gamma} t^{q}}{(\Gamma_1at^{q})^2}.
\end{equation*}
Then it is easy to see that
\begin{equation*}
   \gamma<0,~~~~~~~~~~~~~~z_{12}\sim \frac{a\Gamma_1^2}{c^2\sum_{ j > 2} \Gamma_j } t^{q-2\gamma}.
\end{equation*}
By \begin{equation*}
     z_{12}=(\Gamma_1+\Gamma_2)Z_{12}+ \sum_{ j > 2} \Gamma_j  (Z_{j2}-Z_{j1}),
\end{equation*}
it follows that
\begin{equation*}
     \frac{\Lambda z_{12}}{w_{12}}= -\sum_{ j > 2}  \frac{\Gamma_j}{ w_{1j}w_{2j}}.
\end{equation*}
However, we have
\begin{equation*}
  \frac{\Lambda z_{12}}{w_{12}}\approx t^{2q}\succ t^{-2q}\succeq  -\sum_{ j > 2}  \frac{\Gamma_j}{ w_{1j}w_{2j}} ,
\end{equation*}
this leads to a contradiction.

As a result, $\Gamma_1+\Gamma_2\neq 0$ holds.

$~~~~~~~~~~~~~~~~~~~~~~~~~~~~~~~~~~~~~~~~~~~~~~~~~~~~~~~~~~~~~~~~~~~~~~~~~~~~~~~~~~~~~~~~~~~~~~~~~~~~~~~~~~~~~~~~~~~~~~~~~~~~~~~~~~~~~~~~~~~~~~~~~~~~\Box$\\

On the other hand,  it is noteworthy that   Propositions  4.2, 4.3 and 4.4 in Subsection 4.2 in  \cite{yu2021Finiteness} are still correct. For convenience, these Figures are presented below. Their proofs are similar as that in  \cite{yu2021Finiteness},  proofs will be omitted.
\begin{proposition}\label{relationsonvorticities1}
Suppose a diagram has an isolated $z$-stroke, then vertices of it  are both $z$-circled; if the two vertices are $z$-close (for example, provided the two vertices are connected by $w$-stroke), then the total vorticity of them is zero.
\end{proposition}
\begin{proposition}\label{relationsonvorticities2}
Suppose a diagram has an isolated $z$-color triangle, and none of vertices (say $\textbf{1,2,3}$) of it  are $z$-circled, then $\frac{1}{\Gamma_1}+\frac{1}{\Gamma_2}+\frac{1}{\Gamma_3}=0$ or $\Gamma_1\Gamma_2+\Gamma_2 \Gamma_3+ \Gamma_3 \Gamma_1=0$.
\end{proposition}
 \begin{proposition}\label{relationsonvorticities3}
Suppose a fully $z$-stroked sub-diagram with four vertices exists in isolation  in a diagram, and none of vertices (say $\textbf{1,2,3,4}$) of it  are $z$-circled, then
\begin{equation}\label{L1234}
  L_{1234}= \Gamma_1\Gamma_2+\Gamma_2 \Gamma_3+ \Gamma_3 \Gamma_1+\Gamma_4 (\Gamma_1+\Gamma_2+\Gamma_3)=0.
\end{equation}
\end{proposition}

{By these Propositions, an argument  similar to the one used  in Subsection 4.2 in  \cite{yu2021Finiteness} shows that problematic diagrams also consist of the list in Figure 11 in \cite{yu2021Finiteness}. Moreover,  since we are focusing on four-vortex collapse configurations, and we assume that \eqref{Iis0} holds, $L =\sum_{1\leq j<k\leq 4}\Gamma_j\Gamma_k =0$ is a necessary condition, one can further exclude the  diagrams in Figure 11 in \cite{yu2021Finiteness}.} 

\subsection{Exclusion of  diagrams}
\indent\par
Now,  for the sake of completeness, we further exclude diagrams in Figures  2--10 above by a case-by-case analysis.

\paragraph{Figure \ref{fig:C=21}:}
\indent\par
For  diagrams in Figure \ref{fig:C=21}, by Proposition \ref{relationsonvorticities1}, it  is easy to see that $\Gamma_1+\Gamma_2=0$ and the third diagram is  impossible. Moreover, by Proposition \ref{relationstwoverticesnew}, it follows that $\Gamma_1+\Gamma_2\neq0$ for the first two diagrams, this leads to a contradiction. As a result,  diagrams in Figure \ref{fig:C=21} are impossible.

\paragraph{Figure \ref{fig:C=31}:}
\indent\par
For the fist diagram in Figure \ref{fig:C=31}, by Proposition \ref{relationsonvorticities2}, it  is easy to see that $\frac{1}{\Gamma_1}+\frac{1}{\Gamma_2}+\frac{1}{\Gamma_3}=0$, therefore, the fist diagram is  impossible. For the second diagram in Figure \ref{fig:C=31}, by Proposition \ref{relationsonvorticities1}, it  follows that $\Gamma_1+\Gamma_2=0$. However, by Proposition \ref{relationstwoverticesnew}, it follows  that $\Gamma_1+\Gamma_2\neq0$. This leads to a contradiction. As a result,  diagrams in Figure \ref{fig:C=31} are impossible.

\paragraph{Figure \ref{fig:C=41}:}
\indent\par
For  diagrams in Figure \ref{fig:C=41}, it  is easy to see that vorticities satisfy $\frac{1}{\Gamma_1}+\frac{1}{\Gamma_2}+\frac{1}{\Gamma_3}=0$ or $\Gamma_1+\Gamma_2+\Gamma_3=0$, but this is in conflict with  $L =0$. As a result,  diagrams in Figure \ref{fig:C=41} are impossible.

\paragraph{Figure \ref{fig:C=42}:}
\indent\par
For  diagrams in Figure \ref{fig:C=42}, it  is easy to see that vorticities satisfy ${\Gamma_1}+{\Gamma_2}=0$ and $\Gamma_3+\Gamma_4=0$, but these relations are in conflict with  $L =0$. As a result,  diagrams in Figure \ref{fig:C=42} are impossible.

\paragraph{Figure \ref{fig:C=43}:}
\indent\par
For  the diagram in Figure \ref{fig:C=43}, it  is easy to see that vorticities satisfy $\frac{1}{\Gamma_1}+\frac{1}{\Gamma_2}+\frac{1}{\Gamma_3}=0$ and $\frac{1}{\Gamma_1}+\frac{1}{\Gamma_2}+\frac{1}{\Gamma_4}=0$, but these relations are in conflict with  $L =0$. As a result,  diagrams in Figure \ref{fig:C=43} are impossible.

\paragraph{Figure \ref{fig:C=51}:}
\indent\par
For  diagrams in Figure \ref{fig:C=51}, it  is easy to see that vorticities satisfy $\frac{1}{\Gamma_1}+\frac{1}{\Gamma_2}+\frac{1}{\Gamma_3}=0$, but the relation is in conflict with  $L =0$. As a result,  diagrams in Figure \ref{fig:C=51} are impossible.

\paragraph{Figure \ref{fig:C=61}:}
\indent\par
For the last two diagrams in Figure \ref{fig:C=61}, it  is easy to see that vorticities satisfy  $\Gamma_1+\Gamma_2+\Gamma_3+\Gamma_4=0$, but this is in conflict with  $L =0$. As a result,  the last two diagrams in Figure \ref{fig:C=61} are impossible.\\

\begin{figure}
	\centering
	
		\begin{subfigure}[b]{0.2\textwidth}
\centering
		\resizebox{\linewidth}{!}{
		\begin{tikzpicture}
			\hspace{0cm}
	\draw  (-3/2,  0) node {\large\textbf{1}};
\draw (3/2,  0) node{\large\textbf{2}};

\draw  (-3/2,  -2)  node {\large\textbf{4}};
\draw  (3/2,  -2) node {\large\textbf{3}}; 

	\draw [red,very thick] (-3/2,  0) circle (0.35);
	\draw [red,very thick] (3/2,  0) circle (0.35);

	\draw [blue,dashed, thick] (-3/2,  -2) circle (0.35);
	\draw [blue,dashed, thick]  (3/2,  -2) circle (0.35);

		\draw [red,very thick] (-1.2,0)--(1.2,0); 
	\draw [blue,dashed, thick] (-1.2,-2)--(1.2,-2); 
	\end{tikzpicture}
}
	Diagram I
	\end{subfigure}
		\begin{subfigure}[b]{0.2\textwidth}
\centering
		\resizebox{\linewidth}{!}{
		\begin{tikzpicture}
		\draw	  (-3/2,0)    node {\large\textbf{1}};
		
		\draw		(3/2,0) node {\large\textbf{2}};
		
		\draw		(0,-3/2*1.732) node {\large\textbf{3}};
		
		\draw		(0,-1) node {\large\textbf{4}};

		\draw [red,very thick] (-3/2+.35,0.15)--(3/2-.35,0.15);
		\draw [blue, dashed,thick] (-3/2+0.35,0)--(3/2-.35,0);
		
		\draw [blue, dashed,thick] (-3/2+.2,-0.2*1.732)--(-.2,-3/2*1.732+.2*1.732);
		\draw [red,very thick] (-3/2-0.15+.25, -0.25*1.732)--(-0.15-.25,-3/2*1.732+0.25*1.732);
		
		\draw [blue, dashed,thick] (3/2-0.2,-0.2*1.732)--(0.2,-3/2*1.732+0.2*1.732);
		\draw [red,very thick] (3/2+0.15-0.25,-0.25*1.732)--(0.15+0.25,-3/2*1.732+.25*1.732);

		\draw [blue,dashed,thick]  (-3/2+0.3+0.15,-0.3*2/3)--(0-0.35+0.15,-1+0.35*2/3);  
		\draw [red,very thick]  (-3/2+0.4-0.1,-0.4*2/3)--(0-0.3-0.1,-1+0.3*2/3);  
		
		\draw [red,very thick]  (3/2-0.25-.15,-.25*2/3)--(0.4 -.15,-1+.4*2/3);  
		\draw [blue,dashed,thick]   (3/2-0.4+.1,-.4*2/3)--(0.3 +.1,-1+.3*2/3);  

		\draw [red,very thick] (-.07,-3/2*1.732+.4)--(-.07,-1-.4);
		\draw [blue,dashed,thick]  (.07,-3/2*1.732+0.4)--(.07,-1-0.4);
			\end{tikzpicture}
}
		Diagram  II
	\end{subfigure}
\begin{subfigure}[b]{0.2\textwidth}
		\centering
		\resizebox{\linewidth}{!}{
	\begin{tikzpicture}
	
		\draw  (-3/2,  0) node {\large\textbf{1}};
	\draw (3/2,  0) node{\large\textbf{2}};
	
	\draw  (-3/2,  -2)  node {\large\textbf{4}};
	\draw  (3/2,  -2) node {\large\textbf{3}}; 

		\draw [blue,dashed, thick] (-3/2,  0) circle (0.25);
	\draw [red,very thick] (-3/2,  0) circle (0.35);
	
		\draw [blue,dashed, thick] (3/2,  0) circle (0.25);
\draw [red,very thick] (3/2,  0) circle (0.35);

\draw [blue,dashed, thick] (-3/2,  -2) circle (0.25);
\draw [red,very thick](-3/2,  -2) circle (0.35);

\draw [blue,dashed, thick] (3/2,  -2) circle (0.25);
\draw[red,very thick]  (3/2,  -2) circle (0.35);

\draw [red,very thick] (-1.2,0)--(1.2,0); 
\draw [red,very thick] (-1.2,-2)--(1.2,-2);

\draw [blue,dashed, thick] (-3/2,-.3)--(-3/2,-1.7);
\draw [blue,dashed, thick] (3/2,-.3)--(3/2,-1.7);

	\end{tikzpicture}
}
	Diagram III
\end{subfigure}

\caption{Problematic diagrams for $q< -{|\gamma|}/{2}$}
\label{fig:Problematicdiagramsnew}	
\end{figure}
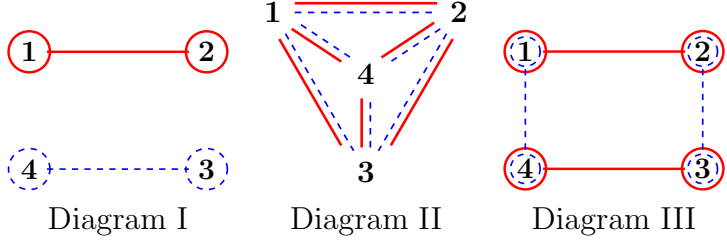

In conclusion, we have derived a list of problematic diagrams consisting of  Diagram I, Diagram II and Diagram III in Figure \ref{fig:Problematicdiagramsnew}. By the way, these diagrams are same as the first three diagrams 
in Figure 11 in \cite{yu2021Finiteness}. 

\subsection{Problematic diagrams}
\indent\par
In general, we could not eliminate Diagram I, Diagram II or Diagram III. 
In this subsection   we obtain   constraints on the
vorticities corresponding to each of these three diagrams.

\subsubsection{Diagram I}\label{DiagramInew}
\indent\par

For the Diagram I,  assume that
\begin{equation*}
    \begin{array}{lll}
      z_1\sim -\Gamma_2 at^{q},~~~~~~~ & z_2\sim \Gamma_1at^{q}, ~~~~~~~& w_{12}\sim \frac{1}{ac} t^{-\gamma-q},\\
      w_3\sim -\Gamma_4 bt^{q},~~~~~~~ & w_4\sim \Gamma_3bt^{q}, ~~~~~~~& z_{34}\sim \frac{c}{b} t^{\gamma-q}.
    \end{array}
\end{equation*}

First,  by Proposition \ref{relationstwoverticesnew}, it follows   that
\begin{center}
$\Gamma_1+\Gamma_2\neq 0$ and $\Gamma_3+\Gamma_4\neq 0$.
\end{center}

Then,   by \begin{equation}\label{w12equ}
     w_{12}=(\Gamma_1+\Gamma_2)W_{12}+ \Gamma_3 (W_{32}-W_{31})+\Gamma_4 (W_{42}-W_{41})\nonumber
\end{equation}
and
\begin{equation}\label{z34equ}
     z_{34}=(\Gamma_3+\Gamma_4)Z_{34}+ \Gamma_1 (Z_{14}-Z_{13})+\Gamma_2 (Z_{24}-Z_{23}),\nonumber
\end{equation}
it follows that
\begin{center}
$\gamma=0$
\end{center}
 and
\begin{equation*}\normalsize
    \begin{array}{c}
       \frac{1}{c^2} =1+(\Gamma_3 +\Gamma_4) (\frac{1}{\Gamma_1}+\frac{1}{\Gamma_2}),\\[8pt]
       {c^2} =1+(\Gamma_1 +\Gamma_2) (\frac{1}{\Gamma_3}+\frac{1}{\Gamma_4}),
     \end{array}
\end{equation*}
or
\begin{equation}
   c^2=-\frac{\Gamma_1\Gamma_2}{\Gamma_3\Gamma_4}.\nonumber
\end{equation}

Note that, when Diagram I is possible, we have
\begin{equation}\label{Ir1}
 r_{12}\approx r_{34}\approx t^{0},  ~~~~~~~~~~~~~~~~~r_{13}\approx r_{14}\approx r_{23}\approx r_{24}\approx t^{q},
\end{equation}
in particular,
\begin{equation}\label{Ir2}
 \frac{r_{ij}}{r_{kl}} \approx t^{0},
\end{equation}
where $(ijkl)$ is any permutation of $\{1,2,3,4\}$, i.e., $\{i,j,k,l\}=\{1,2,3,4\}$.

To summarize, the following relations on the four vorticities should be satisfied if  Diagram I
is possible by choosing a Puiseux series:
\begin{equation}\label{Lambdagamma}
  \begin{array}{ll}
    r_{12}^2\sim \frac{\Gamma_1+\Gamma_2}{c} t^{0}, &r_{34}^2\sim {(\Gamma_3+\Gamma_4)}{c} t^{0};\\[6pt]
    r_{13}^2\sim -\Gamma_2\Gamma_4{ab} t^{2q}, &r_{24}^2\sim -\Gamma_1\Gamma_3{ab} t^{2q};\\[6pt]
    r_{14}^2\sim \Gamma_2\Gamma_3{ab} t^{2q}, &r_{23}^2\sim \Gamma_1\Gamma_4{ab} t^{2q};\\[6pt]
    (\Gamma_1+\Gamma_2)(\Gamma_3+\Gamma_4)\neq 0,
    ~~~~~~~~~~~~&c^2=-\frac{\Gamma_1\Gamma_2}{\Gamma_3\Gamma_4}.
  \end{array}
\end{equation}

\subsubsection{Diagram II}\label{DiagramIInew}
\indent\par
For Diagram II, it is obvious that $z_{kl}\approx t^{\gamma-q},w_{kl}\approx t^{-\gamma-q}$ for any $(k,l)$, $1\leq k<l\leq 4$.
Thus we can assume that
\begin{equation*}
   z_{kl} \sim a_{kl} t^{\gamma-q}, ~~~~~~~~~~~~~~~~~ w_{kl} \sim b_{kl} t^{-\gamma-q}.
\end{equation*}

By \begin{equation}
     \begin{array}{ccc}
       z_n=\sum_{ j \neq n} \Gamma_j Z_{jn},~~~~~ & w_n=\sum_{ j \neq n} \Gamma_j W_{jn},& ~~~~~~n=1,2,3,
     \end{array}
     \nonumber
\end{equation}it follows that
\begin{equation*}\large
  \begin{array}{c}
  \frac {\Gamma_ 2} {a_2 - a_1} + \frac {\Gamma_ 3} {a_3 -
     a_1} + \frac {\Gamma_ 4} {a_4 - a_1} = 0, \\[5pt]
    \frac {\Gamma_ 1} {a_ 1 - a_ 2} + \frac {\Gamma_ 3} {a_3 -
     a_2} + \frac {\Gamma_ 4} {a_ 4 - a_ 2} = 0, \\[5pt]
    \frac {\Gamma_ 1} {a_1 - a_3} + \frac {\Gamma_ 2} {a_2 -
     a_3} + \frac {\Gamma_ 4} {a_ 4 - a_ 3} = 0;\\[5pt]
    \frac{\Gamma_2}{b_2 -b_1 }+\frac{\Gamma_3}{b_3 -b_1}+\frac{\Gamma_4}{b_4-b_1 }=0, \\[5pt]
    \frac{\Gamma_1}{b_1 -b_2}+\frac{\Gamma_3}{b_3 -b_2}+\frac{\Gamma_4}{ b_4-b_2 }=0, \\[5pt]
    \frac{\Gamma_1}{b_1 -b_3}+\frac{\Gamma_2}{b_2 -b_3}+\frac{\Gamma_4}{ b_4-b_3 }=0.
  \end{array}
\end{equation*}

On the other hand, by $S=0$ it follows that
\begin{equation*}
    \sum_{1\leq j<k\leq 4}\Gamma_j\Gamma_k a_{jk}b_{jk}=0.
\end{equation*}

A  straightforward computation shows that there is no solution for the above equations except the case that 
  \begin{center}
   $\Gamma_i =\Gamma_j=\Gamma_k =-\Gamma_l$,
  \end{center}
{where $(ijkl)$ is a certain permutation of $\{1,2,3,4\}$.}

Therefore, Diagram II is impossible except perhaps  for the case that {three of vorticities are equal  and the fourth vorticity is opposite of them}.

Note that, when Diagram II is possible, we have \begin{equation}\label{IIr1}
r_{kl}\approx t^{-q},  ~~~~~~~~~~~~~~~~~1\leq k<l\leq 4,
\end{equation}
in particular,
\begin{equation}\label{IIr2}
 \frac{r_{ij}}{r_{kl}} \approx t^{0},
\end{equation}
{where $(ijkl)$ is any permutation of $\{1,2,3,4\}$.}

\subsubsection{Diagram III}\label{DiagramIIInew}
\indent\par
For Diagram III we can assume that
\begin{equation*}
     \begin{array}{cc}
                                                                                            z_1\sim z_4\sim -\Gamma_2 a t^{q}, & z_2\sim z_3\sim \Gamma_1 a t^{q}, \\
                                                                                           w_1\sim w_2\sim -\Gamma_4 b t^{q}, & w_3\sim w_4\sim \Gamma_1 b t^{q}.
                                                                                         \end{array}
\end{equation*}
It follows that
\begin{equation}
    \Gamma_1 \Gamma_3 =\Gamma_2 \Gamma_4.\nonumber
\end{equation}
Moreover, similar to Subsection 5.3 in \cite{yu2021Finiteness}, it is easy to see that \begin{center}
$(\Gamma_1+\Gamma_2) (\Gamma_2+\Gamma_3)  (\Gamma_3+\Gamma_4)(\Gamma_1+\Gamma_4)\neq 0$.
\end{center}
By $L =0$, it follows that\begin{equation}
    \Gamma_1 \Gamma_3 =\Gamma_2 \Gamma_4<0.\nonumber
\end{equation}

Obviously, \begin{equation}
z_{12} \sim (\Gamma_1+\Gamma_2){a}t^{q}.\nonumber
\end{equation}
By $z_{2}\sim \Gamma_1 \frac{1}{\Lambda w_{12}}$,
it follows that\begin{equation}
w_{12} \sim \frac{1}{ac}t^{-\gamma-q}.\nonumber
\end{equation}
Similarly, \begin{equation}
\begin{array}{lr}
  z_{13} \sim (\Gamma_1+\Gamma_2){a}t^{q}, & w_{13} \sim (\Gamma_1+\Gamma_4){b}t^{q}, \\[5pt]
 z_{14} \sim \frac{c}{b}t^{\gamma-q}, & w_{14} \sim (\Gamma_1+\Gamma_4){b}t^{q}, \\[5pt]
  z_{23} \sim \frac{c\Gamma_3}{b\Gamma_4}t^{\gamma-q}, & w_{23} \sim(\Gamma_1+\Gamma_4){b}t^{q}, \\[5pt]
 z_{24} \sim -(\Gamma_1+\Gamma_2){a}t^{q}, & w_{24} \sim (\Gamma_1+\Gamma_4){b}t^{q}, \\[5pt]
  z_{34} \sim -(\Gamma_1+\Gamma_2){a}t^{q}, & w_{34} \sim -\frac{\Gamma_3}{ac\Gamma_2}t^{-\gamma-q}.
\end{array}
\nonumber
\end{equation}

Note that, when Diagram III is possible, we have \begin{equation}\label{IIIr1}
r_{12}\approx r_{34}\approx t^{-\frac{\gamma}{2}},  ~~~~~~~~~r_{13}\approx r_{24}\approx t^{q},  ~~~~~~~~~~r_{23}\approx r_{14}\approx t^{\frac{\gamma}{2}},
\end{equation}
in particular,
\begin{equation}\label{IIIr2}
 \frac{r_{ij}}{r_{kl}} \approx t^{0},
\end{equation}
{where $(ijkl)$ is any permutation of $\{1,2,3,4\}$.}

\section{Possible diagrams in the case $q= -{|\gamma|}/{2}$}\label{Exclusionofthecaseisgamma12}
\indent\par
In this section, we consider the case that  $q= -\frac{|\gamma|}{2}$.
Without loss of generality, we only consider the case that  $q= -\frac{\gamma}{2}$.

If $q= -\frac{\gamma}{2}$, then for any $(k,l)$, $1\leq k<l\leq 4$, we have $w_{kl}\approx t^{-\frac{\gamma}{2}}$. Hence $Z_{kl}\approx t^{-\frac{\gamma}{2}}$ also holds.
Note that, for any $(k,l)$, vertices $\textbf{k}$ and $\textbf{l}$  form a $z$-stroke, but they are never  $w$-close.

It is easy to see  that
\begin{center}
$w_{n}\preceq t^{-\frac{\gamma}{2}}$ for  $n=1,2,3,4$.
\end{center}

\subsection{The case that some $w_{n}\prec t^{-\frac{\gamma}{2}}$}\label{Exclusionofthecaseisgamma121}
\indent\par
If there is one $w_{n}$, say $ w_{4}$, such that $w_{4}\prec t^{-\frac{\gamma}{2}}$. Then, obviously,
\begin{center}
$w_{1}\approx w_{2}\approx w_{3}\approx t^{-\frac{\gamma}{2}}$.
\end{center}

Let us consider possible diagrams by  Rules in Section \ref{Puiseuxseriessolutionscoloreddiagram}.

\subsubsection{Possible  diagrams}
\indent\par
Obviously, all the possible diagrams contain three $w$-circles at vertices $\textbf{1,2,3}$ and six $z$-strokes.

{\bfseries{Case 1: When there is no any $w$-stroke connected  vertex $\textbf{4}$. }}

Then it is easy to see that vertices  $\textbf{1,2,3}$ formed a $w$-color triangle by Rule I and Rule VI. Next, we  classify all possible diagrams as follows:
\begin{itemize}
  \item  there is no any $z$-circle;
  \item $z$-circle exists but there is  no $z$-circle  at vertex $\textbf{4}$, then there are three $z$-circles at vertices $\textbf{1,2,3}$ by Rule II and Estimate 2 in Proposition \ref{Estimate2};
  \item there is  a $z$-circle  at vertex $\textbf{4}$, then there are also three $z$-circles at vertices $\textbf{1,2,3}$ by Rule IV and Estimate 2 in Proposition \ref{Estimate2}.
\end{itemize}
 Thus
the only possible diagrams are  those   in the following Figure \ref{fig:Case11}.
\begin{figure}[!h]
\centering
	\begin{subfigure}[b]{0.2\textwidth}

		\centering
		\resizebox{\linewidth}{!}{
	\begin{tikzpicture}
\draw	[blue,dashed, thick]  (-3/2,0)  circle (0.25);\draw	[blue,dashed, thick]  (3/2,0)  circle (0.25);\draw	[blue,dashed, thick]  (0,-3/2*1.732)  circle (0.25);
\draw	  (-3/2,0)    node {\large\textbf{1}};
\draw		(3/2,0) node {\large\textbf{2}};
\draw		(0,-3/2*1.732) node {\large\textbf{3}};
\draw		(0,-1) node {\large\textbf{4}};

\draw [red,very thick]  (-3/2+0.3+0.15,-0.3*2/3)--(0-0.35+0.15,-1+0.35*2/3);\draw [red,very thick]  (3/2-0.25-.15,-.25*2/3)--(0.4 -.15,-1+.4*2/3); \draw [red,very thick] (0,-3/2*1.732+.4)--(0,-1-.4);

\draw [red,very thick] (-3/2+.35,0)--(3/2-.35,0);
\draw [blue, dashed,thick] (-3/2+0.35,-.15)--(3/2-.35,-0.15);

\draw [red,very thick] (-3/2+.2,-0.2*1.732)--(-.2,-3/2*1.732+.2*1.732);
\draw [blue, dashed,thick] (-3/2+0.15+.2, -0.2*1.732)--(0.15-.2,-3/2*1.732+0.2*1.732);

\draw [red,very thick] (3/2-0.2,-0.2*1.732)--(0.2,-3/2*1.732+0.2*1.732);
\draw [blue, dashed,thick] (3/2-0.15-0.2,-.2*1.732)--(-0.15+0.2,-3/2*1.732+.2*1.732);
		\end{tikzpicture}
}
	Diagram i
\end{subfigure}
	\begin{subfigure}[b]{0.2\textwidth}

		\centering
		\resizebox{\linewidth}{!}{
	\begin{tikzpicture}
\draw	[blue,dashed, thick]  (-3/2,0)  circle (0.25);\draw	[blue,dashed, thick]  (3/2,0)  circle (0.25);\draw	[blue,dashed, thick]  (0,-3/2*1.732)  circle (0.25);
\draw	[red,very thick]  (-3/2,0)  circle (0.35);
\draw  (-3/2,0)  node {\large\textbf{1}};

\draw	[red,very thick]  (3/2,0)  circle (0.35);
\draw		(3/2,0) node {\large\textbf{2}};

\draw	[red,very thick] (0,-3/2*1.732) circle (0.35);
\draw		(0,-3/2*1.732) node {\large\textbf{3}};

\draw		(0,-1) node {\large\textbf{4}};

\draw [red,very thick]  (-3/2+0.3+0.15,-0.3*2/3)--(0-0.35+0.15,-1+0.35*2/3);\draw [red,very thick]  (3/2-0.25-.15,-.25*2/3)--(0.4 -.15,-1+.4*2/3); \draw [red,very thick] (0,-3/2*1.732+.4)--(0,-1-.4);
\draw [red,very thick] (-3/2+.35,0)--(3/2-.35,0);
\draw [blue, dashed,thick] (-3/2+0.35,-.15)--(3/2-.35,-0.15);

\draw [red,very thick] (-3/2+.2,-0.2*1.732)--(-.2,-3/2*1.732+.2*1.732);
\draw [blue, dashed,thick] (-3/2+0.15+.2, -0.2*1.732)--(0.15-.2,-3/2*1.732+0.2*1.732);

\draw [red,very thick] (3/2-0.2,-0.2*1.732)--(0.2,-3/2*1.732+0.2*1.732);
\draw [blue, dashed,thick] (3/2-0.15-0.2,-.2*1.732)--(-0.15+0.2,-3/2*1.732+.2*1.732);

\end{tikzpicture}
}
	Diagram ii
\end{subfigure}
	\begin{subfigure}[b]{0.2\textwidth}

		\centering
		\resizebox{\linewidth}{!}{
	\begin{tikzpicture}
\draw	[blue,dashed, thick]  (-3/2,0)  circle (0.25);\draw	[blue,dashed, thick]  (3/2,0)  circle (0.25);\draw	[blue,dashed, thick]  (0,-3/2*1.732)  circle (0.25);
	\draw	[blue,dashed, thick]  (-3/2,0)  circle (0.25);
\draw	[red,very thick]  (-3/2,0)  circle (0.35);
\draw  (-3/2,0)  node {\large\textbf{1}};

	\draw	[blue,dashed, thick]  (3/2,0)  circle (0.25);
\draw	[red,very thick]  (3/2,0)  circle (0.35);
\draw		(3/2,0) node {\large\textbf{2}};

	\draw	[blue,dashed, thick]  (0,-3/2*1.732) circle (0.25);
\draw	[red,very thick] (0,-3/2*1.732) circle (0.35);
\draw		(0,-3/2*1.732) node {\large\textbf{3}};

\draw		(0,-1) node {\large\textbf{4}}; \draw	[red,very thick](0,-1) circle (0.35);

\draw [red,very thick]  (-3/2+0.3+0.15,-0.3*2/3)--(0-0.35+0.15,-1+0.35*2/3);\draw [red,very thick]  (3/2-0.25-.15,-.25*2/3)--(0.4 -.15,-1+.4*2/3); \draw [red,very thick] (0,-3/2*1.732+.4)--(0,-1-.4);
\draw [red,very thick] (-3/2+.35,0)--(3/2-.35,0);
\draw [blue, dashed,thick] (-3/2+0.35,-.15)--(3/2-.35,-0.15);

\draw [red,very thick] (-3/2+.2,-0.2*1.732)--(-.2,-3/2*1.732+.2*1.732);
\draw [blue, dashed,thick] (-3/2+0.15+.2, -0.2*1.732)--(0.15-.2,-3/2*1.732+0.2*1.732);

\draw [red,very thick] (3/2-0.2,-0.2*1.732)--(0.2,-3/2*1.732+0.2*1.732);
\draw [blue, dashed,thick] (3/2-0.15-0.2,-.2*1.732)--(-0.15+0.2,-3/2*1.732+.2*1.732);

	\end{tikzpicture}
}
	Diagram iii
\end{subfigure}
	\caption{}
\label{fig:Case11}
\end{figure}

{\bfseries{Case 2: When there is some $w$-stroke connected  vertex $\textbf{4}$. }}

Then it is easy to see that six $w$-strokes by  Rule VI. Next, we  classify all possible diagrams as follows:
\begin{itemize}
  \item  there is no any $z$-circle;
  \item $z$-circle exists, then there are four $z$-circles at vertices $\textbf{1,2,3,4}$ by Rule II and Estimate 2 in Proposition \ref{Estimate2}.
\end{itemize}
 Thus
the only possible diagrams are  those   in the following Figure \ref{fig:Case12}.
\begin{figure}[!h]
\centering
	\begin{subfigure}[b]{0.2\textwidth}

		\centering
		\resizebox{\linewidth}{!}{
	\begin{tikzpicture}
\draw	[blue,dashed, thick]  (-3/2,0)  circle (0.25);
	\draw  (-3/2,0)  node {\large\textbf{1}};
	
	\draw	[blue,dashed, thick]  (3/2,0)  circle (0.25);
	\draw		(3/2,0) node {\large\textbf{2}};
	
	\draw	[blue,dashed, thick]  (0,-3/2*1.732) circle (0.25);
	\draw		(0,-3/2*1.732) node {\large\textbf{3}};
	
	\draw		(0,-1) node {\large\textbf{4}};
	
	\draw [red,very thick] (-3/2+.35,0.15)--(3/2-.35,0.15);
	\draw [blue, dashed,thick] (-3/2+0.35,0)--(3/2-.35,0);
	
	\draw [blue, dashed,thick] (-3/2+.2,-0.2*1.732)--(-.2,-3/2*1.732+.2*1.732);
	\draw [red,very thick] (-3/2-0.15+.25, -0.25*1.732)--(-0.15-.25,-3/2*1.732+0.25*1.732);
	
	\draw [blue, dashed,thick] (3/2-0.2,-0.2*1.732)--(0.2,-3/2*1.732+0.2*1.732);
	\draw [red,very thick] (3/2+0.15-0.25,-0.25*1.732)--(0.15+0.25,-3/2*1.732+.25*1.732);

	\draw [blue,dashed,thick]  (-3/2+0.3+0.15,-0.3*2/3)--(0-0.35+0.15,-1+0.35*2/3);  
	\draw [red,very thick]  (-3/2+0.4-0.1,-0.4*2/3)--(0-0.3-0.1,-1+0.3*2/3);  
	
	\draw [red,very thick]  (3/2-0.25-.15,-.25*2/3)--(0.4 -.15,-1+.4*2/3);  
	\draw [blue,dashed,thick]   (3/2-0.4+.1,-.4*2/3)--(0.3 +.1,-1+.3*2/3);  

	\draw [red,very thick] (-.07,-3/2*1.732+.4)--(-.07,-1-.4);
	\draw [blue,dashed,thick]  (.07,-3/2*1.732+0.4)--(.07,-1-0.4);

\end{tikzpicture}
}
	Diagram iv
\end{subfigure}
	\begin{subfigure}[b]{0.2\textwidth}

		\centering
		\resizebox{\linewidth}{!}{
	\begin{tikzpicture}
\draw	[blue,dashed, thick]  (-3/2,0)  circle (0.25);
	\draw	[red,very thick]  (-3/2,0)  circle (0.35);
	\draw  (-3/2,0)  node {\large\textbf{1}};
	
	\draw	[blue,dashed, thick]  (3/2,0)  circle (0.25);
	\draw	[red,very thick]  (3/2,0)  circle (0.35);
	\draw		(3/2,0) node {\large\textbf{2}};
	
	\draw	[blue,dashed, thick]  (0,-3/2*1.732) circle (0.25);
	\draw	[red,very thick] (0,-3/2*1.732) circle (0.35);
	\draw		(0,-3/2*1.732) node {\large\textbf{3}};
	
	\draw	[blue,dashed, thick]  (0,-1) circle (0.25);
	\draw	[red,very thick](0,-1) circle (0.35);
	\draw		(0,-1) node {\large\textbf{4}};
	
	\draw [red,very thick] (-3/2+.35,0.15)--(3/2-.35,0.15);
	\draw [blue, dashed,thick] (-3/2+0.35,0)--(3/2-.35,0);
	
	\draw [blue, dashed,thick] (-3/2+.2,-0.2*1.732)--(-.2,-3/2*1.732+.2*1.732);
	\draw [red,very thick] (-3/2-0.15+.25, -0.25*1.732)--(-0.15-.25,-3/2*1.732+0.25*1.732);
	
	\draw [blue, dashed,thick] (3/2-0.2,-0.2*1.732)--(0.2,-3/2*1.732+0.2*1.732);
	\draw [red,very thick] (3/2+0.15-0.25,-0.25*1.732)--(0.15+0.25,-3/2*1.732+.25*1.732);

	\draw [blue,dashed,thick]  (-3/2+0.3+0.15,-0.3*2/3)--(0-0.35+0.15,-1+0.35*2/3);  
	\draw [red,very thick]  (-3/2+0.4-0.1,-0.4*2/3)--(0-0.3-0.1,-1+0.3*2/3);  
	
	\draw [red,very thick]  (3/2-0.25-.15,-.25*2/3)--(0.4 -.15,-1+.4*2/3);  
	\draw [blue,dashed,thick]   (3/2-0.4+.1,-.4*2/3)--(0.3 +.1,-1+.3*2/3);  

	\draw [red,very thick] (-.07,-3/2*1.732+.4)--(-.07,-1-.4);
	\draw [blue,dashed,thick]  (.07,-3/2*1.732+0.4)--(.07,-1-0.4);

	\end{tikzpicture}
}
	Diagram v
\end{subfigure}
	\caption{}
\label{fig:Case12}
\end{figure}

\subsubsection{Exclusion of  diagrams}
\indent\par

{\bfseries{Case i:  }} For Diagram i we can assume that
\begin{equation*}
w_4\prec w_n\sim b_n t^{-\frac{\gamma}{2}}, ~~~~~~~~n=1,2,3,
\end{equation*}
where $b_n$ are three unequal nonzero numbers.

By \begin{equation}
     z_n=\sum_{ j \neq n} \Gamma_j Z_{jn},~~~~~~~n=1,2,3,\nonumber
\end{equation}it follows that
\begin{equation*}\large
  \begin{array}{c}
    \frac{\Gamma_2}{b_2 -b_1 }+\frac{\Gamma_3}{b_3 -b_1}+\frac{\Gamma_4}{-b_1 }=0, \\
    \frac{\Gamma_1}{b_1 -b_2}+\frac{\Gamma_3}{b_3 -b_2}+\frac{\Gamma_4}{ -b_2 }=0, \\
    \frac{\Gamma_1}{b_1 -b_3}+\frac{\Gamma_2}{b_2 -b_3}+\frac{\Gamma_4}{ -b_3 }=0.
  \end{array}
\end{equation*}
A  straightforward computation shows that there is no solution for the above equations except the case that $\Gamma_1 =\Gamma_2=\Gamma_3 =-\Gamma_4$.

As a result,  Diagram i is impossible except perhaps  for the case that
\begin{center}
$\Gamma_1 =\Gamma_2=\Gamma_3 =-\Gamma_4$.
\end{center}

Note that, when Diagram i is possible, we have \begin{equation}\label{ir1}
\begin{array}{lr}
  r_{kl}\approx t^{\frac{\gamma}{2}}=t^{-q}, & ~~~~~~~~~~~1\leq k<l\leq 3,\\
  t^{-q}\prec r_{j4}\prec t^{q}, & ~~~~~~~~~~~~~1\leq j\leq 3;
\end{array}
\end{equation}in particular,
\begin{equation}\label{ir2}
 \frac{r_{kl}}{r_{j4}} \prec t^{0}\prec \frac{r_{j4}}{r_{kl}},
\end{equation}
{where $(jkl)$ is any permutation of $\{1,2,3\}$.}
\\

{\bfseries{Case ii:  }} For Diagram ii it is easy to see that\begin{equation*}
z_4\prec z_1\sim z_2\sim z_3.
\end{equation*}
By
\begin{equation*}
    \sum_{j=1}^4\Gamma_jz_j=0,
\end{equation*}
it follows that\begin{equation*}
    \Gamma_1+\Gamma_2+\Gamma_3=0.
\end{equation*}
However, this relation is in conflict with  $L =0$.

As a result,  Diagram ii is impossible.\\

{\bfseries{Case iii:  }} For Diagram iii, without loss of generality, , we can assume that
\begin{equation*}
z_1\sim z_2\sim z_3\sim -\Gamma_4 t^{-\frac{\gamma}{2}},  ~~~~~~~~z_4\sim (\Gamma_1+\Gamma_2+\Gamma_3) t^{-\frac{\gamma}{2}}
\end{equation*}

By \begin{equation}
     w_4=\sum_{ k \neq 4} \Gamma_k Z_{k4},\nonumber
\end{equation}it follows that
\begin{equation*}
\frac{\Gamma_1}{\Gamma }+\frac{\Gamma_2}{\Gamma}+\frac{\Gamma_3}{\Gamma}=0,
\end{equation*}
or
\begin{equation*}
    \Gamma_1+\Gamma_2+\Gamma_3=0.
\end{equation*}
However, this relation is in conflict with  $L =0$.

As a result,  Diagram iii is impossible.\\

{\bfseries{Case iv:  }} For Diagram iv it is easy to see that the result is similar to  Diagram i, i.e., Diagram iv is impossible except perhaps  for the case that $\Gamma_1 =\Gamma_2=\Gamma_3 =-\Gamma_4$.

Note that, when Diagram iv is possible, we have \begin{equation}\label{ivr1}
 r_{kl}\approx t^{\frac{\gamma}{2}}=t^{-q},  ~~~~~~~~~~~~~~1\leq k<l\leq 4;
\end{equation}in particular,
\begin{equation}\label{ivr2}
 \frac{r_{ij}}{r_{kl}} \approx t^{0},
\end{equation}
{where $(ijkl)$ is any permutation of $\{1,2,3,4\}$.}\\

{\bfseries{Case v:  }} For Diagram v it is easy to see that
\begin{equation*}
 z_1\sim z_2\sim z_3\sim z_4,
\end{equation*}
it follows that
\begin{equation*}
    \Gamma_1+\Gamma_2+\Gamma_3+\Gamma_4=0.
\end{equation*}
However, this relation is in conflict with  $L =0$.

As a result,  Diagram v is impossible.\\

\subsection{The case that  all $w_{n}\approx t^{-\frac{\gamma}{2}}$}\label{Exclusionofthecaseis122}
\indent\par
If  $w_{1}\approx w_{2}\approx w_{3}\approx w_{4}\approx t^{-\frac{\gamma}{2}}$.

Let us consider possible diagrams by  Rules in Section \ref{Puiseuxseriessolutionscoloreddiagram}.

\subsubsection{Possible  diagrams}
\indent\par
Obviously, all the possible diagrams contain four $w$-circles and six $z$-strokes. Then, similar to the above subsection,  it is easy to show that the following five diagrams in Figure \ref{fig:gamma=21} exhaust all the possible diagrams:

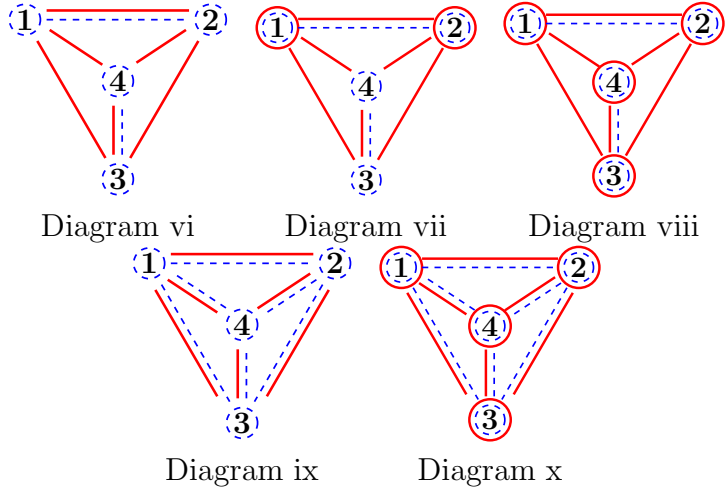
\begin{figure}[!h]
	\centering
	\begin{subfigure}[b]{0.2\textwidth}

		\centering
		\resizebox{\linewidth}{!}{
	\begin{tikzpicture}
	\draw	[blue,dashed,thick]  (-3/2,0)  circle (0.25);
	\draw	  (-3/2,0)    node {\large\textbf{1}};
	
		\draw	[blue,dashed,thick]  (3/2,0)  circle (0.25);
	\draw		(3/2,0) node {\large\textbf{2}};
	
		\draw	[blue,dashed,thick]  (0,-3/2*1.732)   circle (0.25);
	\draw		(0,-3/2*1.732) node {\large\textbf{3}};
	
		\draw	[blue,dashed,thick]  (0,-1)  circle (0.25);
	\draw		(0,-1) node {\large\textbf{4}};

	\draw [red,very thick] (-3/2+.35,0.15)--(3/2-.35,0.15);
\draw [blue, dashed,thick] (-3/2+0.35,0)--(3/2-.35,0);
	
	\draw [red,very thick] (-3/2+.2,-0.2*1.732)--(-.2,-3/2*1.732+.2*1.732);
	
	\draw [red,very thick] (3/2-0.2,-0.2*1.732)--(0.2,-3/2*1.732+0.2*1.732);

		\draw [red,very thick]  (-3/2+0.25+0.15,-0.25*2/3)--(0-0.4+0.15,-1+0.4*2/3);  
		
		\draw [red,very thick]  (3/2-0.25-.15,-.25*2/3)--(0.45 -.15,-1+.45*2/3);  

		\draw [red,very thick] (-.07,-3/2*1.732+.4)--(-.07,-1-.4);
			\draw [blue,dashed,thick]  (.07,-3/2*1.732+0.4)--(.07,-1-0.4);
	\end{tikzpicture}
}
	Diagram vi
\end{subfigure}
	\begin{subfigure}[b]{0.2\textwidth}
\centering
		\resizebox{\linewidth}{!}{
		\begin{tikzpicture}
	
	\draw	[blue,dashed, thick]  (-3/2,0)  circle (0.25);
\draw	[red,very thick]  (-3/2,0)  circle (0.35);
\draw  (-3/2,0)  node {\large\textbf{1}};

\draw	[blue,dashed, thick]  (3/2,0)  circle (0.25);
\draw	[red,very thick]  (3/2,0)  circle (0.35);
\draw		(3/2,0) node {\large\textbf{2}};
	
		\draw	[blue,dashed,thick]  (0,-3/2*1.732)   circle (0.25);
\draw		(0,-3/2*1.732) node {\large\textbf{3}};

\draw	[blue,dashed,thick]  (0,-1)  circle (0.25);
\draw		(0,-1) node {\large\textbf{4}};

\draw [red,very thick] (-3/2+.35,0.15)--(3/2-.35,0.15);
\draw [blue, dashed,thick] (-3/2+0.35,0)--(3/2-.35,0);

\draw [red,very thick] (-3/2+.2,-0.2*1.732)--(-.2,-3/2*1.732+.2*1.732);

\draw [red,very thick] (3/2-0.2,-0.2*1.732)--(0.2,-3/2*1.732+0.2*1.732);

\draw [red,very thick]  (-3/2+0.25+0.15,-0.25*2/3)--(0-0.4+0.15,-1+0.4*2/3);  

\draw [red,very thick]  (3/2-0.25-.15,-.25*2/3)--(0.45 -.15,-1+.45*2/3);  

\draw [red,very thick] (-.07,-3/2*1.732+.4)--(-.07,-1-.4);
\draw [blue,dashed,thick]  (.07,-3/2*1.732+0.4)--(.07,-1-0.4);

	\end{tikzpicture}
}
	Diagram vii
\end{subfigure}
	\begin{subfigure}[b]{0.2\textwidth}
\centering
		\resizebox{\linewidth}{!}{
		\begin{tikzpicture}
	\draw	[blue,dashed, thick]  (-3/2,0)  circle (0.25);
	\draw	[red,very thick]  (-3/2,0)  circle (0.35);
	\draw  (-3/2,0)  node {\large\textbf{1}};
	
	\draw	[blue,dashed, thick]  (3/2,0)  circle (0.25);
	\draw	[red,very thick]  (3/2,0)  circle (0.35);
	\draw		(3/2,0) node {\large\textbf{2}};
	
	\draw	[blue,dashed, thick]  (0,-3/2*1.732) circle (0.25);
	\draw	[red,very thick] (0,-3/2*1.732) circle (0.35);
	\draw		(0,-3/2*1.732) node {\large\textbf{3}};
	
		\draw	[blue,dashed, thick]  (0,-1) circle (0.25);
		\draw	[red,very thick](0,-1) circle (0.35);
	\draw		(0,-1) node {\large\textbf{4}};

\draw [red,very thick] (-3/2+.35,0.15)--(3/2-.35,0.15);
\draw [blue, dashed,thick] (-3/2+0.35,0)--(3/2-.35,0);

\draw [red,very thick] (-3/2+.2,-0.2*1.732)--(-.2,-3/2*1.732+.2*1.732);

\draw [red,very thick] (3/2-0.2,-0.2*1.732)--(0.2,-3/2*1.732+0.2*1.732);

\draw [red,very thick]  (-3/2+0.25+0.15,-0.25*2/3)--(0-0.4+0.15,-1+0.4*2/3);  

\draw [red,very thick]  (3/2-0.25-.15,-.25*2/3)--(0.45 -.15,-1+.45*2/3);  

\draw [red,very thick] (-.07,-3/2*1.732+.4)--(-.07,-1-.4);
\draw [blue,dashed,thick]  (.07,-3/2*1.732+0.4)--(.07,-1-0.4);
	
	\end{tikzpicture}
}
	Diagram viii
\end{subfigure}
	
	\begin{subfigure}[b]{0.2\textwidth}
		\centering
		\resizebox{\linewidth}{!}{
	\begin{tikzpicture}
	
	\draw	[blue,dashed, thick]  (-3/2,0)  circle (0.25);
	\draw  (-3/2,0)  node {\large\textbf{1}};
	
	\draw	[blue,dashed, thick]  (3/2,0)  circle (0.25);
	\draw		(3/2,0) node {\large\textbf{2}};
	\draw	[blue,dashed, thick]  (0,-3/2*1.732)  circle (0.25);
	\draw		(0,-3/2*1.732) node {\large\textbf{3}};
	\draw	[blue,dashed, thick]  (0,-1)  circle (0.25);
	\draw		(0,-1) node {\large\textbf{4}};
	
	\draw [red,very thick] (-3/2+.35,0.15)--(3/2-.35,0.15);
	\draw [blue, dashed,thick] (-3/2+0.35,0)--(3/2-.35,0);
	
	\draw [blue, dashed,thick] (-3/2+.2,-0.2*1.732)--(-.2,-3/2*1.732+.2*1.732);
	\draw [red,very thick] (-3/2-0.15+.25, -0.25*1.732)--(-0.15-.25,-3/2*1.732+0.25*1.732);
	
	\draw [blue, dashed,thick] (3/2-0.2,-0.2*1.732)--(0.2,-3/2*1.732+0.2*1.732);
	\draw [red,very thick] (3/2+0.15-0.25,-0.25*1.732)--(0.15+0.25,-3/2*1.732+.25*1.732);

	\draw [blue,dashed,thick]  (-3/2+0.3+0.15,-0.3*2/3)--(0-0.35+0.15,-1+0.35*2/3);  
	\draw [red,very thick]  (-3/2+0.4-0.1,-0.4*2/3)--(0-0.3-0.1,-1+0.3*2/3);  
	
	\draw [red,very thick]  (3/2-0.25-.15,-.25*2/3)--(0.4 -.15,-1+.4*2/3);  
	\draw [blue,dashed,thick]   (3/2-0.4+.1,-.4*2/3)--(0.3 +.1,-1+.3*2/3);  

	\draw [red,very thick] (-.07,-3/2*1.732+.4)--(-.07,-1-.4);
	\draw [blue,dashed,thick]  (.07,-3/2*1.732+0.4)--(.07,-1-0.4);
	
	\end{tikzpicture}
}
	Diagram ix
\end{subfigure}
\begin{subfigure}[b]{0.2\textwidth}
		\centering
		\resizebox{\linewidth}{!}{
	\begin{tikzpicture}
	\draw	[blue,dashed, thick]  (-3/2,0)  circle (0.25);
	\draw	[red,very thick]  (-3/2,0)  circle (0.35);
	\draw  (-3/2,0)  node {\large\textbf{1}};
	
	\draw	[blue,dashed, thick]  (3/2,0)  circle (0.25);
	\draw	[red,very thick]  (3/2,0)  circle (0.35);
	\draw		(3/2,0) node {\large\textbf{2}};
	
	\draw	[blue,dashed, thick]  (0,-3/2*1.732) circle (0.25);
	\draw	[red,very thick] (0,-3/2*1.732) circle (0.35);
	\draw		(0,-3/2*1.732) node {\large\textbf{3}};
	
	\draw	[blue,dashed, thick]  (0,-1) circle (0.25);
	\draw	[red,very thick](0,-1) circle (0.35);
	\draw		(0,-1) node {\large\textbf{4}};
	
	\draw [red,very thick] (-3/2+.35,0.15)--(3/2-.35,0.15);
	\draw [blue, dashed,thick] (-3/2+0.35,0)--(3/2-.35,0);
	
	\draw [blue, dashed,thick] (-3/2+.2,-0.2*1.732)--(-.2,-3/2*1.732+.2*1.732);
	\draw [red,very thick] (-3/2-0.15+.25, -0.25*1.732)--(-0.15-.25,-3/2*1.732+0.25*1.732);
	
	\draw [blue, dashed,thick] (3/2-0.2,-0.2*1.732)--(0.2,-3/2*1.732+0.2*1.732);
	\draw [red,very thick] (3/2+0.15-0.25,-0.25*1.732)--(0.15+0.25,-3/2*1.732+.25*1.732);

	\draw [blue,dashed,thick]  (-3/2+0.3+0.15,-0.3*2/3)--(0-0.35+0.15,-1+0.35*2/3);  
	\draw [red,very thick]  (-3/2+0.4-0.1,-0.4*2/3)--(0-0.3-0.1,-1+0.3*2/3);  
	
	\draw [red,very thick]  (3/2-0.25-.15,-.25*2/3)--(0.4 -.15,-1+.4*2/3);  
	\draw [blue,dashed,thick]   (3/2-0.4+.1,-.4*2/3)--(0.3 +.1,-1+.3*2/3);  

	\draw [red,very thick] (-.07,-3/2*1.732+.4)--(-.07,-1-.4);
	\draw [blue,dashed,thick]  (.07,-3/2*1.732+0.4)--(.07,-1-0.4);
	
	\end{tikzpicture}
}
	Diagram x
\end{subfigure}

	\caption{Possible diagrams for $q=-\frac{\gamma}{2}$ and all $w_{n}\approx t^{-\frac{\gamma}{2}}$}
\label{fig:gamma=21}
\end{figure}

\subsubsection{Exclusion of  diagrams}
\indent\par

{\bfseries{Case vi:  }} For Diagram vi we can assume that
\begin{equation*}
     \begin{array}{cc}
                                                                                            w_1\sim  -\Gamma_2 a t^{-\frac{\gamma}{2}}, & w_2\sim  \Gamma_1 a t^{-\frac{\gamma}{2}}, \\[5pt]
                                                                                           w_3\sim  -\Gamma_4 b t^{-\frac{\gamma}{2}}, &  w_4\sim \Gamma_3 b t^{-\frac{\gamma}{2}}.

                                                                                         \end{array}
\end{equation*}
By \begin{equation}
     z_n=\sum_{ j \neq n} \Gamma_j Z_{jn},~~~~~~~n=1,2,3,\nonumber
\end{equation}it follows that
\begin{equation*}\large
  \begin{array}{c}
    \frac{\Gamma_2}{-\Gamma_2 a-\Gamma_1 a}+\frac{\Gamma_3}{-\Gamma_2 a+\Gamma_4 b}+\frac{\Gamma_4}{-\Gamma_2 a-\Gamma_3 b }=0, \\
    \frac{\Gamma_1}{\Gamma_1 a+\Gamma_2 a}+\frac{\Gamma_3}{\Gamma_1 a+\Gamma_4 b}+\frac{\Gamma_4}{\Gamma_1 a-\Gamma_3 b }=0, \\
    \frac{\Gamma_1}{-\Gamma_4 b-\Gamma_1 a}+\frac{\Gamma_2}{-\Gamma_4 b-\Gamma_1 a}+\frac{\Gamma_4}{-\Gamma_4 b-\Gamma_3 b }=0.
  \end{array}
\end{equation*}
A  straightforward computation shows that there is no solution for the above equations except the case $\Gamma_3 =\Gamma_4=(\sqrt{3}-2)^{\pm 1}\Gamma_1=(\sqrt{3}-2)^{\pm 1}\Gamma_2$.

Thus Diagram vi is impossible except perhaps  for the case
\begin{center}
$\Gamma_3 =\Gamma_4=(\sqrt{3}-2)^{\pm 1}\Gamma_1=(\sqrt{3}-2)^{\pm 1}\Gamma_2$.
\end{center}

Note that, when Diagram vi is possible, we have \begin{equation}\label{vir}
\begin{array}{c}
  r_{12}\approx r_{34}\approx t^{\frac{\gamma}{2}}=t^{-q},\\ 
  t^{-q}\prec r_{13},r_{24},r_{14},r_{23}\prec t^{q}.
\end{array}
\end{equation}\\

{\bfseries{Case vii:  }} For Diagram vii we can assume that
\begin{equation*}
   w_1\sim  -\Gamma_2 a t^{-\frac{\gamma}{2}},  w_2\sim  \Gamma_1 a t^{-\frac{\gamma}{2}}.
\end{equation*}

However, it is easy to see that
\begin{equation*}
    z_1\sim z_2 \approx t^{-\frac{\gamma}{2}},
\end{equation*}
thus we have $\Gamma_1+\Gamma_2=0$. Then it is obvious that $w_{12}\prec t^{-\frac{\gamma}{2}}$, this is in conflict with $w_{12}\approx t^{-\frac{\gamma}{2}}$. As a result,  Diagram vii is impossible.\\

{\bfseries{Case viii:  }} For Diagram viii we can assume that
\begin{equation*}
     \begin{array}{cc}
                                                                                            w_1\sim  -\Gamma_2 a t^{-\frac{\gamma}{2}}, & w_2\sim  \Gamma_1 a t^{-\frac{\gamma}{2}}, \\[5pt]
                                                                                           w_3\sim  -\Gamma_4 b t^{-\frac{\gamma}{2}}, &  w_4\sim \Gamma_3 b t^{-\frac{\gamma}{2}},\\[5pt]
                                                                                           z_1\sim  z_2 \sim a' t^{-\frac{\gamma}{2}}, & z_3\sim  z_4\sim b' t^{-\frac{\gamma}{2}}. \\
                                                                                         \end{array}
\end{equation*}
Note that 
\begin{equation*}
    (\Gamma_1+\Gamma_2)(\Gamma_3+\Gamma_4)\neq 0,
\end{equation*}
and
\begin{equation*}
    b'=-(\Gamma_1+\Gamma_2)a'/(\Gamma_3+\Gamma_4).
\end{equation*}

By \begin{equation}
     w_{12}=(\Gamma_1+\Gamma_2)W_{12}+ \Gamma_3 (W_{32}-W_{31})+\Gamma_4 (W_{42}-W_{41}),\nonumber
\end{equation}it follows that
\begin{equation*}
    w_{12}=(\Gamma_1+\Gamma_2)\frac{\Lambda}{z_{12}}+O(t^{\frac{7}{2}\gamma}),
\end{equation*}
or\begin{equation*}
    z_{12}=\frac{(\Gamma_1+\Gamma_2)\Lambda}{w_{12}}+O(t^{\frac{11}{2}\gamma})\approx t^{\frac{3}{2}\gamma},
\end{equation*}

By
\begin{equation}
 z_{12}=(\Gamma_1+\Gamma_2)Z_{12}+ \Gamma_3 (Z_{32}-Z_{31})+ \Gamma_4 (Z_{42}-Z_{41}),\nonumber
\end{equation}
it follows that
\begin{equation*}
    z_{12}=\frac{\Gamma_1+\Gamma_2}{\Lambda w_{12}}-\Gamma_3 \frac{w_{12}}{\Lambda w_{13}w_{23}}- \Gamma_4 \frac{w_{12}}{\Lambda w_{41}w_{42}}.
\end{equation*}

Therefore,
\begin{equation*}
    \frac{\Gamma_1+\Gamma_2}{ w_{12}^2}\sim \frac{1}{ w_{31}w_{32}}+\frac{1}{ w_{14}w_{24}}.
\end{equation*}

Similarly, we also have
\begin{equation*}
    \frac{\Gamma_3+\Gamma_4}{ w_{34}^2}\sim \frac{1}{ w_{13}w_{14}}+\frac{1}{ w_{23}w_{24}}.
\end{equation*}

As a result, nonzero numbers $a,b$ satisfy
\begin{equation}\label{abequC}
    \begin{array}{c}
       \frac{\Gamma_1+\Gamma_2}{ a^2(\Gamma_1+\Gamma_2)^2}= \frac{1}{ (-\Gamma_4 b+\Gamma_2 a)(-\Gamma_4 b-\Gamma_1 a)}+\frac{1}{ (\Gamma_3 b+\Gamma_2 a)(\Gamma_3 b-\Gamma_1 a)}, \\[8pt]
       \frac{\Gamma_3+\Gamma_4}{ b^2(\Gamma_3+\Gamma_4)^2}= \frac{1}{ (-\Gamma_4 b+\Gamma_2 a)(\Gamma_3 b+\Gamma_2 a)}+\frac{1}{ (-\Gamma_4 b-\Gamma_1 a)(\Gamma_3 b-\Gamma_1 a)}.
     \end{array}
\end{equation}

On the other hand, by \begin{equation}
     z_n=\sum_{ j \neq n} \Gamma_j Z_{jn},~~~~~~~n=1,2,3,4,\nonumber
\end{equation}it follows that
\begin{equation}\label{ababequC}
  \begin{array}{c}
     c a'=-\frac{\Gamma_2}{(\Gamma_1+\Gamma_2)a}+\frac{\Gamma_3}{\Gamma_4 b-\Gamma_2 a}-  \frac{\Gamma_4}{\Gamma_3 b+\Gamma_2 a},\\[8pt]
    c a'= \frac{\Gamma_1}{(\Gamma_1+\Gamma_2)a}+\frac{\Gamma_3}{\Gamma_4 b+\Gamma_1 a}+  \frac{\Gamma_4}{\Gamma_1 a-\Gamma_3 b},\\[8pt]
     c b'=\frac{\Gamma_1}{\Gamma_2 a-\Gamma_4 b}-\frac{\Gamma_2}{\Gamma_1 a+\Gamma_4 b}-  \frac{\Gamma_4}{\Gamma_3 b+\Gamma_4 b},\\[8pt]
    c b'= \frac{\Gamma_1}{\Gamma_2 a+\Gamma_3 b}+\frac{\Gamma_2}{-\Gamma_1 a+\Gamma_3 b}+  \frac{\Gamma_3}{\Gamma_3 b+\Gamma_4 b}.
  \end{array}
\end{equation}

A  straightforward computation shows that, if there is a solution for the  equations (\ref{abequC}) and (\ref{ababequC}), we have
\begin{equation*}
(\Gamma_1\Gamma_4-\Gamma_2\Gamma_3) (\Gamma_1\Gamma_3-\Gamma_2\Gamma_4)=0.
\end{equation*}

As a matter of fact, if we use an equivalent form of the system  (\ref{stationaryconfiguration3}) (see \cite{yu2021Finiteness} and  (\ref{stationaryconfigurationnew6new}) below), then it is easy to see that, by 
\begin{equation*}
    {F_z}-\Lambda f_z=0 ~~~~~~~\text{and}~~~~~~~
 G_z+\Lambda g_z=0,
\end{equation*}
it follows that
\begin{equation}\label{FfGg}\large
    \begin{array}{c}
     (\Gamma_1+\Gamma_2) \left(\Gamma_3^2+\Gamma_4^2\right)-\left(\Gamma_1^2+\Gamma_2^2\right) (\Gamma_3+\Gamma_4)=0,\\[8pt]
    \Gamma_1 \Gamma_2 (\Gamma_1+\Gamma_2)^2+\Gamma_3 \Gamma_4 (\Gamma_3+\Gamma_4)^2 - (\Gamma_1+\Gamma_2)^2 (\Gamma_3+\Gamma_4)^2=0.
    \end{array}
\end{equation}
However, a  straightforward computation shows that there is no solution for the  equations (\ref{FfGg}) and $L=0$.

As a result, Diagram viii is impossible.\\

{\bfseries{Case ix:  }} For Diagram ix it is easy to see that, for any $(k,l)$, $1\leq k<l\leq 4$, we have
\begin{center}
$z_{kl}\approx t^{\frac{3}{2}\gamma}$,   ~~~~~~~~~~~~~~~~~~~$w_{kl}\approx t^{-\frac{1}{2}\gamma}$.
\end{center}

Therefore, 
\begin{equation}\label{ixr1}
    r_{kl}\approx t^{\frac{\gamma}{2}}=t^{-q},~~~~~~~~~~~~~~~1\leq k<l\leq 4;
\end{equation}
in particular,
\begin{equation}\label{ixr2}
 \frac{r_{ij}}{r_{kl}} \approx t^{0},
\end{equation}
{where $(ijkl)$ is any permutation of $\{1,2,3,4\}$.}\\ \\

{\bfseries{Case x:  }} For Diagram x, it is easy to see that
\begin{equation*}
    z_1\sim z_2 \sim z_3 \sim z_4\approx t^{-\frac{\gamma}{2}}.
\end{equation*}
Then we have
\begin{equation*}
    \Gamma_1+\Gamma_2+\Gamma_3+\Gamma_4=0.
\end{equation*}
However, this relation is in conflict with  $L =0$. Thus Diagram x is impossible.

\subsection{Problematic diagrams}

\indent\par

\begin{figure}[!h]
\centering
	\begin{subfigure}[b]{0.2\textwidth}

		\centering
		\resizebox{\linewidth}{!}{
	\begin{tikzpicture}
\draw	[blue,dashed, thick]  (-3/2,0)  circle (0.25);\draw	[blue,dashed, thick]  (3/2,0)  circle (0.25);\draw	[blue,dashed, thick]  (0,-3/2*1.732)  circle (0.25);
\draw	  (-3/2,0)    node {\large\textbf{1}};
\draw		(3/2,0) node {\large\textbf{2}};
\draw		(0,-3/2*1.732) node {\large\textbf{3}};
\draw		(0,-1) node {\large\textbf{4}};

\draw [red,very thick]  (-3/2+0.3+0.15,-0.3*2/3)--(0-0.35+0.15,-1+0.35*2/3);\draw [red,very thick]  (3/2-0.25-.15,-.25*2/3)--(0.4 -.15,-1+.4*2/3); \draw [red,very thick] (0,-3/2*1.732+.4)--(0,-1-.4);

\draw [red,very thick] (-3/2+.35,0)--(3/2-.35,0);
\draw [blue, dashed,thick] (-3/2+0.35,-.15)--(3/2-.35,-0.15);

\draw [red,very thick] (-3/2+.2,-0.2*1.732)--(-.2,-3/2*1.732+.2*1.732);
\draw [blue, dashed,thick] (-3/2+0.15+.2, -0.2*1.732)--(0.15-.2,-3/2*1.732+0.2*1.732);

\draw [red,very thick] (3/2-0.2,-0.2*1.732)--(0.2,-3/2*1.732+0.2*1.732);
\draw [blue, dashed,thick] (3/2-0.15-0.2,-.2*1.732)--(-0.15+0.2,-3/2*1.732+.2*1.732);
		\end{tikzpicture}
}
	Diagram i
\end{subfigure}
\begin{subfigure}[b]{0.2\textwidth}
		\centering
		\resizebox{\linewidth}{!}{
\begin{tikzpicture}
\draw	[blue,dashed, thick]  (-3/2,0)  circle (0.25);
	\draw  (-3/2,0)  node {\large\textbf{1}};
	
	\draw	[blue,dashed, thick]  (3/2,0)  circle (0.25);
	\draw		(3/2,0) node {\large\textbf{2}};
	
	\draw	[blue,dashed, thick]  (0,-3/2*1.732) circle (0.25);
	\draw		(0,-3/2*1.732) node {\large\textbf{3}};
	
	\draw		(0,-1) node {\large\textbf{4}};
	
	\draw [red,very thick] (-3/2+.35,0.15)--(3/2-.35,0.15);
	\draw [blue, dashed,thick] (-3/2+0.35,0)--(3/2-.35,0);
	
	\draw [blue, dashed,thick] (-3/2+.2,-0.2*1.732)--(-.2,-3/2*1.732+.2*1.732);
	\draw [red,very thick] (-3/2-0.15+.25, -0.25*1.732)--(-0.15-.25,-3/2*1.732+0.25*1.732);
	
	\draw [blue, dashed,thick] (3/2-0.2,-0.2*1.732)--(0.2,-3/2*1.732+0.2*1.732);
	\draw [red,very thick] (3/2+0.15-0.25,-0.25*1.732)--(0.15+0.25,-3/2*1.732+.25*1.732);

	\draw [blue,dashed,thick]  (-3/2+0.3+0.15,-0.3*2/3)--(0-0.35+0.15,-1+0.35*2/3);  
	\draw [red,very thick]  (-3/2+0.4-0.1,-0.4*2/3)--(0-0.3-0.1,-1+0.3*2/3);  
	
	\draw [red,very thick]  (3/2-0.25-.15,-.25*2/3)--(0.4 -.15,-1+.4*2/3);  
	\draw [blue,dashed,thick]   (3/2-0.4+.1,-.4*2/3)--(0.3 +.1,-1+.3*2/3);  

	\draw [red,very thick] (-.07,-3/2*1.732+.4)--(-.07,-1-.4);
	\draw [blue,dashed,thick]  (.07,-3/2*1.732+0.4)--(.07,-1-0.4);

\end{tikzpicture}
}
	Diagram iv
\end{subfigure}
\begin{subfigure}[b]{0.2\textwidth}
		\centering
		\resizebox{\linewidth}{!}{
	\begin{tikzpicture}
	\draw	[blue,dashed,thick]  (-3/2,0)  circle (0.25);
	\draw	  (-3/2,0)    node {\large\textbf{1}};
	
		\draw	[blue,dashed,thick]  (3/2,0)  circle (0.25);
	\draw		(3/2,0) node {\large\textbf{2}};
	
		\draw	[blue,dashed,thick]  (0,-3/2*1.732)   circle (0.25);
	\draw		(0,-3/2*1.732) node {\large\textbf{3}};
	
		\draw	[blue,dashed,thick]  (0,-1)  circle (0.25);
	\draw		(0,-1) node {\large\textbf{4}};

	\draw [red,very thick] (-3/2+.35,0.15)--(3/2-.35,0.15);
\draw [blue, dashed,thick] (-3/2+0.35,0)--(3/2-.35,0);
	
	\draw [red,very thick] (-3/2+.2,-0.2*1.732)--(-.2,-3/2*1.732+.2*1.732);
	
	\draw [red,very thick] (3/2-0.2,-0.2*1.732)--(0.2,-3/2*1.732+0.2*1.732);

		\draw [red,very thick]  (-3/2+0.25+0.15,-0.25*2/3)--(0-0.4+0.15,-1+0.4*2/3);  
		
		\draw [red,very thick]  (3/2-0.25-.15,-.25*2/3)--(0.45 -.15,-1+.45*2/3);  

		\draw [red,very thick] (-.07,-3/2*1.732+.4)--(-.07,-1-.4);
			\draw [blue,dashed,thick]  (.07,-3/2*1.732+0.4)--(.07,-1-0.4);
	\end{tikzpicture}
}
	Diagram vi
\end{subfigure}
\begin{subfigure}[b]{0.2\textwidth}
		\centering
		\resizebox{\linewidth}{!}{
	\begin{tikzpicture}
	
	\draw	[blue,dashed, thick]  (-3/2,0)  circle (0.25);
	\draw  (-3/2,0)  node {\large\textbf{1}};
	
	\draw	[blue,dashed, thick]  (3/2,0)  circle (0.25);
	\draw		(3/2,0) node {\large\textbf{2}};
	\draw	[blue,dashed, thick]  (0,-3/2*1.732)  circle (0.25);
	\draw		(0,-3/2*1.732) node {\large\textbf{3}};
	\draw	[blue,dashed, thick]  (0,-1)  circle (0.25);
	\draw		(0,-1) node {\large\textbf{4}};
	
	\draw [red,very thick] (-3/2+.35,0.15)--(3/2-.35,0.15);
	\draw [blue, dashed,thick] (-3/2+0.35,0)--(3/2-.35,0);
	
	\draw [blue, dashed,thick] (-3/2+.2,-0.2*1.732)--(-.2,-3/2*1.732+.2*1.732);
	\draw [red,very thick] (-3/2-0.15+.25, -0.25*1.732)--(-0.15-.25,-3/2*1.732+0.25*1.732);
	
	\draw [blue, dashed,thick] (3/2-0.2,-0.2*1.732)--(0.2,-3/2*1.732+0.2*1.732);
	\draw [red,very thick] (3/2+0.15-0.25,-0.25*1.732)--(0.15+0.25,-3/2*1.732+.25*1.732);

	\draw [blue,dashed,thick]  (-3/2+0.3+0.15,-0.3*2/3)--(0-0.35+0.15,-1+0.35*2/3);  
	\draw [red,very thick]  (-3/2+0.4-0.1,-0.4*2/3)--(0-0.3-0.1,-1+0.3*2/3);  
	
	\draw [red,very thick]  (3/2-0.25-.15,-.25*2/3)--(0.4 -.15,-1+.4*2/3);  
	\draw [blue,dashed,thick]   (3/2-0.4+.1,-.4*2/3)--(0.3 +.1,-1+.3*2/3);  

	\draw [red,very thick] (-.07,-3/2*1.732+.4)--(-.07,-1-.4);
	\draw [blue,dashed,thick]  (.07,-3/2*1.732+0.4)--(.07,-1-0.4);
	
	\end{tikzpicture}
}
	Diagram ix
\end{subfigure}
	\caption{Problematic diagrams for $q= -{|\gamma|}/{2}$}
\label{fig:Problematicdiagrams2}
\end{figure}
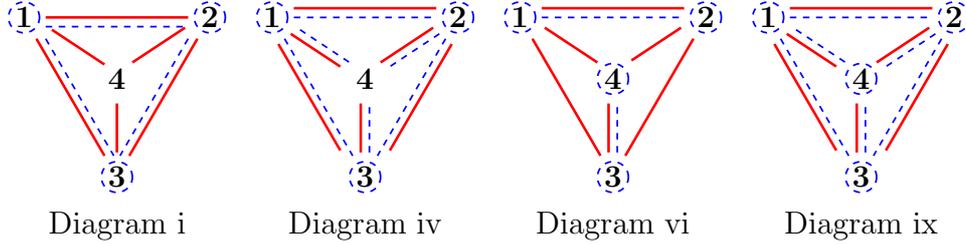

In conclusion, we have derived a list of problematic diagrams consisting of Diagram i, Diagram iv, Diagram vi and Diagram ix in Figure \ref{fig:Problematicdiagrams2}.

\section{Finiteness results of collapse configurations}\label{Finitenessresultnew}
\indent\par
In this section we mainly prove the following result.
\begin{theorem}\label{Main2}
If the vorticities $\Gamma_n$ $(n\in\{1,2,3,4\})$ are nonzero, then the associated four-vortex problem has finitely many  collapse configurations.
\end{theorem}
  Theorem \ref{Main2}   is an obvious inference of the following Theorems \ref{Main2collapse1}, \ref{Main2collapse2} and \ref{Main2collapse3}.  And we  remark that the following results of  finiteness are all on  normalized collapse configurations in  {the} complex domain,  rather  than  real   configurations.

First, we establish the following results.
\begin{lemma}\label{nonadjacentdistancesbj1}
Any ratio $r_{jk}^2/r_{lm}^2$ of two nonadjacent
distances' squares is not dominating on the closed algebraic subset $\mathcal X$ defined by the system (\ref{stationaryconfigurationmainLambda}). 
\end{lemma}
\begin{lemma}\label{nonadjacentdistancesbj2}
Suppose the closed algebraic subset $\mathcal X$ defined by the system (\ref{stationaryconfigurationmainLambda})  consists of   infinitely many points. If all ratios $r_{jk}^2/r_{lm}^2$ of two nonadjacent
distances' squares are not dominating on the closed algebraic subset $\mathcal X$, Diagram I is impossible. 
\end{lemma}

\noindent\emph{Proof of Lemma \ref{nonadjacentdistancesbj1}}:

If system  (\ref{stationaryconfigurationmainLambda}) possesses  infinitely many solutions, without lose of generality, assume that $r_{12}^2/r_{34}^2$ is not dominating on the closed algebraic subset $\mathcal X$. Then some level set $r_{12}^2/r_{34}^2\equiv const\neq 0$, denoted by $\mathcal{X}_{lev}$, also contains  infinitely many points of $\mathcal X$.

On $\mathcal{X}_{lev}$, there is a Puiseux series as  (\ref{Puiseuxseries0})  with $r_{12}^2(t)/r_{34}^2(t) = t$. By considering the Puiseux series, it is obvious to see that only Diagram i  and Diagram vi   are possible.\\

{\bfseries{Case 1: If Diagram i  occurs. }}

Then, it is easy to show that, Diagram vi   is impossible, and we have
\begin{center}
$\Gamma_1 =\Gamma_2=\Gamma_3 =-\Gamma_4$~~~~~~~~~~~ or~~~~~~~~~~~ $\Gamma_1 =\Gamma_2=\Gamma_4 =-\Gamma_3$.
\end{center}

By considering a Puiseux series as  (\ref{Puiseuxseries0})  with $r_{12}^2(t)/r_{34}^2(t) = t^{-1}$, it is easy to show that we have
 \begin{center}
$\Gamma_3 =\Gamma_4=\Gamma_1 =-\Gamma_2$~~~~~~~~~~~ or~~~~~~~~~~~ $\Gamma_3 =\Gamma_4=\Gamma_2 =-\Gamma_1$.
\end{center}

However, the two requirements on vorticities above are not compatible with each other. This leads to a contradiction.\\

{\bfseries{Case 2: If Diagram vi  occurs. }}

Then, it is easy to show that, Diagram i   is impossible, and we have
\begin{center}
$\Gamma_2 =\Gamma_4=(\sqrt{3}-2)^{\pm 1}\Gamma_1=(\sqrt{3}-2)^{\pm 1}\Gamma_3$
\end{center}
or
\begin{center}
$\Gamma_2 =\Gamma_3=(\sqrt{3}-2)^{\pm 1}\Gamma_1=(\sqrt{3}-2)^{\pm 1}\Gamma_4$.
\end{center}
Without loss of generality, assume vorticities satisfy \begin{center}
$\Gamma_2 =\Gamma_4=(\sqrt{3}-2)^{\pm 1}\Gamma_1=(\sqrt{3}-2)^{\pm 1}\Gamma_3$.
\end{center}
Then we are in one of diagrams in the following Figure \ref{fig:Completediagramsvi}.
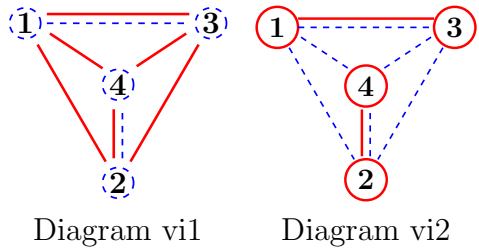
\begin{figure}[!h]
	\centering
	\begin{subfigure}[b]{0.2\textwidth}

		\centering
		\resizebox{\linewidth}{!}{
	\begin{tikzpicture}
	\draw	[blue,dashed,thick]  (-3/2,0)  circle (0.25);
	\draw	  (-3/2,0)    node {\large\textbf{1}};
	
		\draw	[blue,dashed,thick]  (3/2,0)  circle (0.25);
	\draw		(3/2,0) node {\large\textbf{3}};
	
		\draw	[blue,dashed,thick]  (0,-3/2*1.732)   circle (0.25);
	\draw		(0,-3/2*1.732) node {\large\textbf{2}};
	
		\draw	[blue,dashed,thick]  (0,-1)  circle (0.25);
	\draw		(0,-1) node {\large\textbf{4}};

	\draw [red,very thick] (-3/2+.35,0.15)--(3/2-.35,0.15);
\draw [blue, dashed,thick] (-3/2+0.35,0)--(3/2-.35,0);
	
	\draw [red,very thick] (-3/2+.2,-0.2*1.732)--(-.2,-3/2*1.732+.2*1.732);
	
	\draw [red,very thick] (3/2-0.2,-0.2*1.732)--(0.2,-3/2*1.732+0.2*1.732);

		\draw [red,very thick]  (-3/2+0.25+0.15,-0.25*2/3)--(0-0.4+0.15,-1+0.4*2/3);  
		
		\draw [red,very thick]  (3/2-0.25-.15,-.25*2/3)--(0.45 -.15,-1+.45*2/3);  

		\draw [red,very thick] (-.07,-3/2*1.732+.4)--(-.07,-1-.4);
			\draw [blue,dashed,thick]  (.07,-3/2*1.732+0.4)--(.07,-1-0.4);
	\end{tikzpicture}
}
	Diagram vi1
\end{subfigure}
	\begin{subfigure}[b]{0.2\textwidth}

		\centering
		\resizebox{\linewidth}{!}{
	\begin{tikzpicture}
	\draw	[red,very thick]  (-3/2,0)  circle (0.35);
	\draw	  (-3/2,0)    node {\large\textbf{1}};
	
		\draw	[red,very thick]  (3/2,0)  circle (0.35);
	\draw		(3/2,0) node {\large\textbf{3}};
	
		\draw	[red,very thick]  (0,-3/2*1.732)   circle (0.35);
	\draw		(0,-3/2*1.732) node {\large\textbf{2}};
	
		\draw	[red,very thick]  (0,-1)  circle (0.35);
	\draw		(0,-1) node {\large\textbf{4}};

	\draw [red,very thick] (-3/2+.35,0.15)--(3/2-.35,0.15);
\draw [blue, dashed,thick] (-3/2+0.35,0)--(3/2-.35,0);
	
	\draw [blue,dashed,thick] (-3/2+.2,-0.2*1.732)--(-.2,-3/2*1.732+.2*1.732);
	
	\draw [blue,dashed,thick] (3/2-0.2,-0.2*1.732)--(0.2,-3/2*1.732+0.2*1.732);

		\draw [blue,dashed,thick]  (-3/2+0.25+0.15,-0.25*2/3)--(0-0.4+0.15,-1+0.4*2/3);  
		
		\draw [blue,dashed,thick] (3/2-0.25-.15,-.25*2/3)--(0.45 -.15,-1+.45*2/3);  

		\draw [red,very thick] (-.07,-3/2*1.732+.4)--(-.07,-1-.4);
			\draw [blue,dashed,thick]  (.07,-3/2*1.732+0.4)--(.07,-1-0.4);
	\end{tikzpicture}
}
	Diagram vi2
\end{subfigure}

	\caption{Complete problematic diagrams for $\Gamma_2 =\Gamma_4=(\sqrt{3}-2)^{\pm 1}\Gamma_1=(\sqrt{3}-2)^{\pm 1}\Gamma_3$}
\label{fig:Completediagramsvi}
\end{figure}

For Diagram vi1 or Diagram vi2, we have
 \begin{center}
 $\frac{r_{13}^2}{r_{12}^2} \prec t^{0}$,
\end{center}
thus $\frac{r_{13}^2}{r_{12}^2}$ is dominating and there is a Puiseux series as  (\ref{Puiseuxseries0}) with $\frac{r_{13}^2(t)}{r_{12}^2(t)} = t^{-1}$. By considering the Puiseux series, it is obvious to see that  Diagram vi1 and Diagram vi2  are impossible. This leads to a contradiction.\\

This completes the proof of Lemma \ref{nonadjacentdistancesbj1}.

$~~~~~~~~~~~~~~~~~~~~~~~~~~~~~~~~~~~~~~~~~~~~~~~~~~~~~~~~~~~~~~~~~~~~~~~~~~~~~~~~~~~~~~~~~~~~~~~~~~~~~~~~~~~~~~~~~~~~~~~~~~~~~~~~~~~~~~~~~~~~~~~~~~~~\Box$\\

\noindent\emph{Proof of Lemma \ref{nonadjacentdistancesbj2}}:

First, by the condition that all ratios $r_{jk}^2/r_{lm}^2$ of two nonadjacent
distances' squares are not dominating, it is easy to see that    Diagram i  is impossible.

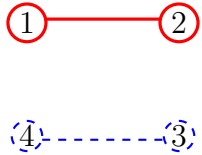
\begin{figure}[!h]
	\centering
	\begin{tikzpicture}
		\vspace*{0cm}\hspace*{0cm} 
	\draw [red,very thick] (-1,  0) circle (0.25) node [black]{$1$}; 
	\draw [red,very thick] (1,  0) circle (0.25) node [black]{$2$}; 
	\draw [red,very thick] (-0.75,0.05)--(0.75,0.05); 
	\draw [blue, dashed, thick] (-1,  -1.5) circle (0.2) node [black]{$4$}; 
	\draw [blue, dashed, thick] (1,  -1.5) circle (0.2) node [black]{$3$}; 
		\draw [blue, dashed, thick] (-0.8,-1.55)--(0.8,-1.55); 
	\end{tikzpicture}
\caption{Diagram I   }
	\label{fig:DiagramI}
\end{figure}

Next, if Diagram I   is possible, without loss of generality, assume that we are in the diagram in Figure \ref{fig:DiagramI}.
Then,
by the condition that all ratios $r_{jk}^2/r_{lm}^2$ of two nonadjacent
distances' squares are not dominating and by \eqref{Lambdagamma}, it follows that
\begin{equation}\label{DiagramI12-34}
    \begin{array}{l}
     r_{34}^2=-\frac{\Gamma_1\Gamma_2(\Gamma_3+\Gamma_4)}{\Gamma_3\Gamma_4(\Gamma_1+\Gamma_2)} r_{12}^2 ,\\[6pt]
    r_{24}^2= \frac{\Gamma_1\Gamma_3}{\Gamma_2\Gamma_4} r_{13}^2,\\[6pt]
    r_{23}^2= \frac{\Gamma_1\Gamma_4}{\Gamma_2\Gamma_3} r_{14}^2.
    \end{array}
\end{equation}
We claim that all ratios $r_{jk}^2/r_{jl}^2$ of two adjacent
distances' squares are  dominating. Here we only prove that $r_{13}^2/r_{14}^2$ is dominating. Otherwise, $r_{13}^2/r_{14}^2\equiv const$, by \eqref{Lambdagamma}, it is simple to see that
\begin{center}
$r_{13}^2/r_{14}^2\equiv -\frac{\Gamma_1\Gamma_4}{\Gamma_1\Gamma_3}$, ~~~~~~~~~~~ or~~~~~~~~~~~ $\Gamma_1\Gamma_3 r_{13}^2+\Gamma_1\Gamma_4 r_{14}^2=0$.
\end{center}
Therefore, by \eqref{DiagramI12-34} and $S=0$, it follows that
\begin{equation*}
    \Gamma_1\Gamma_2 r_{12}^2+\Gamma_3\Gamma_4 r_{34}^2=0,
\end{equation*}
thus $\Gamma_1+\Gamma_2+\Gamma_3+\Gamma_4=0$, but this is in conflict with  $L =0$. As a result, $r_{13}^2/r_{14}^2$ is dominating. Similarly, it is easy to show that all ratios $r_{jk}^2/r_{jl}^2$ of two adjacent
distances' squares are  dominating.

To prove that Diagram I   is impossible, let us further consider a Puiseux series as  (\ref{Puiseuxseries0}) with $\frac{r_{jk}^2(t)}{r_{jl}^2(t)} = t$ or with $\frac{r_{jk}^2(t)}{r_{jl}^2(t)} = t^{-1}$,  it is easy to see that  only Diagram I, Diagram III  and Diagram vi   are possible. Below by showing that these diagrams are impossible,  we infer that   Diagram I   is impossible.

By considering $\frac{r_{12}^2(t)}{r_{13}^2(t)} = t^{-1}$, we are faced the following cases:\\\\
{\bfseries{Case 1: If Diagram I  occurs. }}

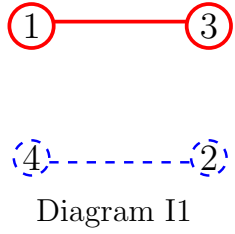
\begin{figure}[!h]
	\centering
\begin{subfigure}[b]{0.2\textwidth}
		\centering
		\resizebox{\linewidth}{!}{
	\begin{tikzpicture}
	\draw [red,very thick] (-1,  0) circle (0.25) node [black]{$1$}; 
	\draw [red,very thick] (1,  0) circle (0.25) node [black]{$3$}; 
	\draw [red,very thick] (-0.75,0.05)--(0.75,0.05); 
	\draw [blue, dashed, thick] (-1,  -1.5) circle (0.2) node [black]{$4$}; 
	\draw [blue, dashed, thick] (1,  -1.5) circle (0.2) node [black]{$2$}; 
		\draw [blue, dashed, thick] (-0.8,-1.55)--(0.8,-1.55); 
	\end{tikzpicture}}

	Diagram I1
\end{subfigure}
\caption{Problematic diagrams for Diagram I with  $r_{12}^2(t)/r_{13}^2(t) = t^{-1}$}
	\label{fig:CompletediagramsI1}
\end{figure}

Then, it is easy to show that, we are in the diagram in  Figure \ref{fig:CompletediagramsI1}.

For Diagram I1, it is easy to see that, vorticities satisfy
\begin{equation}\label{DiagramI13-24}
    \begin{array}{l}
     r_{24}^2=-\frac{\Gamma_1\Gamma_3(\Gamma_2+\Gamma_4)}{\Gamma_2\Gamma_4(\Gamma_1+\Gamma_3)} r_{13}^2 ,\\[6pt]
    r_{34}^2= \frac{\Gamma_1\Gamma_2}{\Gamma_3\Gamma_4} r_{12}^2,\\[6pt]
    r_{23}^2= \frac{\Gamma_1\Gamma_4}{\Gamma_2\Gamma_3} r_{14}^2.
    \end{array}
\end{equation}
Then, by \eqref{DiagramI12-34} and \eqref{DiagramI13-24}, it follows that
  \begin{center}
  $\Gamma_1+\Gamma_2+\Gamma_3+\Gamma_4=0$,
\end{center}
but this is in conflict with  $L =0$.

As a result, Diagram I does not occur.\\

\noindent{\bfseries{Case 2: If Diagram III  occurs. }}

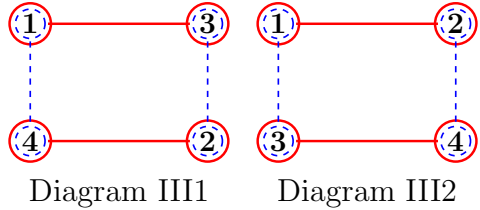
\begin{figure}[!h]
	\centering
\begin{subfigure}[b]{0.2\textwidth}
		\centering
		\resizebox{\linewidth}{!}{
	\begin{tikzpicture}
	
		\draw  (-3/2,  0) node {\large\textbf{1}};
	\draw (3/2,  0) node{\large\textbf{3}};
	
	\draw  (-3/2,  -2)  node {\large\textbf{4}};
	\draw  (3/2,  -2) node {\large\textbf{2}}; 

		\draw [blue,dashed, thick] (-3/2,  0) circle (0.25);
	\draw [red,very thick] (-3/2,  0) circle (0.35);
	
		\draw [blue,dashed, thick] (3/2,  0) circle (0.25);
\draw [red,very thick] (3/2,  0) circle (0.35);

\draw [blue,dashed, thick] (-3/2,  -2) circle (0.25);
\draw [red,very thick](-3/2,  -2) circle (0.35);

\draw [blue,dashed, thick] (3/2,  -2) circle (0.25);
\draw[red,very thick]  (3/2,  -2) circle (0.35);

\draw [red,very thick] (-1.2,0)--(1.2,0); 
\draw [red,very thick] (-1.2,-2)--(1.2,-2);

\draw [blue,dashed, thick] (-3/2,-.3)--(-3/2,-1.7);
\draw [blue,dashed, thick] (3/2,-.3)--(3/2,-1.7);

	\end{tikzpicture}
}
	Diagram III1
\end{subfigure}
\begin{subfigure}[b]{0.2\textwidth}
		\centering
		\resizebox{\linewidth}{!}{
	\begin{tikzpicture}
	
		\draw  (-3/2,  0) node {\large\textbf{1}};
	\draw (3/2,  0) node{\large\textbf{2}};
	
	\draw  (-3/2,  -2)  node {\large\textbf{3}};
	\draw  (3/2,  -2) node {\large\textbf{4}}; 

		\draw [blue,dashed, thick] (-3/2,  0) circle (0.25);
	\draw [red,very thick] (-3/2,  0) circle (0.35);
	
		\draw [blue,dashed, thick] (3/2,  0) circle (0.25);
\draw [red,very thick] (3/2,  0) circle (0.35);

\draw [blue,dashed, thick] (-3/2,  -2) circle (0.25);
\draw [red,very thick](-3/2,  -2) circle (0.35);

\draw [blue,dashed, thick] (3/2,  -2) circle (0.25);
\draw[red,very thick]  (3/2,  -2) circle (0.35);

\draw [red,very thick] (-1.2,0)--(1.2,0); 
\draw [red,very thick] (-1.2,-2)--(1.2,-2);

\draw [blue,dashed, thick] (-3/2,-.3)--(-3/2,-1.7);
\draw [blue,dashed, thick] (3/2,-.3)--(3/2,-1.7);

	\end{tikzpicture}
}
	Diagram III2
\end{subfigure}
\caption{Problematic diagrams for Diagram III with  $r_{12}^2(t)/r_{13}^2(t) = t^{-1}$}
	\label{fig:CompletediagramsIII1}
\end{figure}

Then, it is easy to show that, we are in one of diagrams in  Figure \ref{fig:CompletediagramsIII1}.

For Diagram III1, it is easy to see that, vorticities satisfy
\begin{equation}\label{DiagramIII12-34}
    \begin{array}{l}
    \Gamma_1 \Gamma_2 =\Gamma_3 \Gamma_4, \\[6pt]
     r_{24}^2=\frac{\Gamma_2}{\Gamma_3} r_{13}^2 ,\\[6pt]
    r_{34}^2= - r_{12}^2,\\[6pt]
    r_{23}^2= \frac{\Gamma_2}{\Gamma_4} r_{14}^2.
    \end{array}
\end{equation}
Then, by \eqref{DiagramI12-34} and \eqref{DiagramIII12-34}, it follows that
  \begin{center}
  $\Gamma_1=\Gamma_2=\Gamma_3=\Gamma_4$,
\end{center}
but this is in conflict with  $L =0$.

 Similarly, for Diagram III2, we have
\begin{equation}\label{DiagramIII14-23}
    \begin{array}{l}
    \Gamma_1 \Gamma_4 =\Gamma_2 \Gamma_3, \\[6pt]
     r_{24}^2=\frac{\Gamma_4}{\Gamma_3} r_{13}^2 ,\\[6pt]
    r_{34}^2= \frac{\Gamma_4}{\Gamma_2} r_{12}^2,\\[6pt]
    r_{23}^2= - r_{14}^2.
    \end{array}
\end{equation}
Then,  by \eqref{DiagramI12-34} and \eqref{DiagramIII12-34}, it follows that
  \begin{center}
  $\Gamma_1 \Gamma_4 =\Gamma_2 \Gamma_3$, and $\Gamma_1\Gamma_4=-\Gamma_2\Gamma_3$,
\end{center}
but this leads to a contradiction.

As a result, Diagram III does not occur.\\

\noindent{\bfseries{Case 3: If Diagram vi  occurs. }}

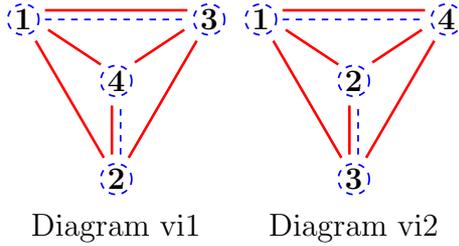
\begin{figure}[!h]
	\centering
\begin{subfigure}[b]{0.2\textwidth}

		\centering
		\resizebox{\linewidth}{!}{
	\begin{tikzpicture}
	\draw	[blue,dashed,thick]  (-3/2,0)  circle (0.25);
	\draw	  (-3/2,0)    node {\large\textbf{1}};
	
		\draw	[blue,dashed,thick]  (3/2,0)  circle (0.25);
	\draw		(3/2,0) node {\large\textbf{3}};
	
		\draw	[blue,dashed,thick]  (0,-3/2*1.732)   circle (0.25);
	\draw		(0,-3/2*1.732) node {\large\textbf{2}};
	
		\draw	[blue,dashed,thick]  (0,-1)  circle (0.25);
	\draw		(0,-1) node {\large\textbf{4}};

	\draw [red,very thick] (-3/2+.35,0.15)--(3/2-.35,0.15);
\draw [blue, dashed,thick] (-3/2+0.35,0)--(3/2-.35,0);
	
	\draw [red,very thick] (-3/2+.2,-0.2*1.732)--(-.2,-3/2*1.732+.2*1.732);
	
	\draw [red,very thick] (3/2-0.2,-0.2*1.732)--(0.2,-3/2*1.732+0.2*1.732);

		\draw [red,very thick]  (-3/2+0.25+0.15,-0.25*2/3)--(0-0.4+0.15,-1+0.4*2/3);  
		
		\draw [red,very thick]  (3/2-0.25-.15,-.25*2/3)--(0.45 -.15,-1+.45*2/3);  

		\draw [red,very thick] (-.07,-3/2*1.732+.4)--(-.07,-1-.4);
			\draw [blue,dashed,thick]  (.07,-3/2*1.732+0.4)--(.07,-1-0.4);
	\end{tikzpicture}
}
	Diagram vi1
\end{subfigure}\begin{subfigure}[b]{0.2\textwidth}

		\centering
		\resizebox{\linewidth}{!}{
	\begin{tikzpicture}
	\draw	[blue,dashed,thick]  (-3/2,0)  circle (0.25);
	\draw	  (-3/2,0)    node {\large\textbf{1}};
	
		\draw	[blue,dashed,thick]  (3/2,0)  circle (0.25);
	\draw		(3/2,0) node {\large\textbf{4}};
	
		\draw	[blue,dashed,thick]  (0,-3/2*1.732)   circle (0.25);
	\draw		(0,-3/2*1.732) node {\large\textbf{3}};
	
		\draw	[blue,dashed,thick]  (0,-1)  circle (0.25);
	\draw		(0,-1) node {\large\textbf{2}};

	\draw [red,very thick] (-3/2+.35,0.15)--(3/2-.35,0.15);
\draw [blue, dashed,thick] (-3/2+0.35,0)--(3/2-.35,0);
	
	\draw [red,very thick] (-3/2+.2,-0.2*1.732)--(-.2,-3/2*1.732+.2*1.732);
	
	\draw [red,very thick] (3/2-0.2,-0.2*1.732)--(0.2,-3/2*1.732+0.2*1.732);

		\draw [red,very thick]  (-3/2+0.25+0.15,-0.25*2/3)--(0-0.4+0.15,-1+0.4*2/3);  
		
		\draw [red,very thick]  (3/2-0.25-.15,-.25*2/3)--(0.45 -.15,-1+.45*2/3);  

		\draw [red,very thick] (-.07,-3/2*1.732+.4)--(-.07,-1-.4);
			\draw [blue,dashed,thick]  (.07,-3/2*1.732+0.4)--(.07,-1-0.4);
	\end{tikzpicture}
}
	Diagram vi2
\end{subfigure}
\caption{Problematic diagrams for Diagram vi with  $r_{12}^2(t)/r_{13}^2(t) = t^{-1}$}
	\label{fig:Completediagramsvi1}
\end{figure}

Then, it is easy to show that, we are in one of diagrams in  Figure \ref{fig:Completediagramsvi1}.

For these diagrams, we have
\begin{center}
$\Gamma_2 =\Gamma_4=(\sqrt{3}-2)^{\pm 1}\Gamma_1=(\sqrt{3}-2)^{\pm 1}\Gamma_3$
\end{center}
or
\begin{center}
$\Gamma_2 =\Gamma_3=(\sqrt{3}-2)^{\pm 1}\Gamma_1=(\sqrt{3}-2)^{\pm 1}\Gamma_4$.
\end{center}
Without loss of generality, assume vorticities satisfy
\begin{equation}\label{vorticitiesvi13-24}
    \Gamma_2 =\Gamma_4=(\sqrt{3}-2)^{\pm 1}\Gamma_1=(\sqrt{3}-2)^{\pm 1}\Gamma_3
\end{equation}
Then we are in Diagram vi1.

By considering a Puiseux series as  (\ref{Puiseuxseries0})  with $r_{13}^2(t)/r_{14}^2(t) = t^{-1}$, once again, we are faced the following subcases:\\\\
\noindent{\bfseries{Subcase 1: If Diagram I  occurs. }}

Then, it is easy to show that, we are in the diagram in  Figure \ref{fig:CompletediagramsI2}.
\begin{figure}[!h]
	\centering
\begin{subfigure}[b]{0.2\textwidth}
		\centering
		\resizebox{\linewidth}{!}{
	\begin{tikzpicture}
	\draw [red,very thick] (-1,  0) circle (0.25) node [black]{$1$}; 
	\draw [red,very thick] (1,  0) circle (0.25) node [black]{$4$}; 
	\draw [red,very thick] (-0.75,0.05)--(0.75,0.05); 
	\draw [blue, dashed, thick] (-1,  -1.5) circle (0.2) node [black]{$2$}; 
	\draw [blue, dashed, thick] (1,  -1.5) circle (0.2) node [black]{$3$}; 
		\draw [blue, dashed, thick] (-0.8,-1.55)--(0.8,-1.55); 
	\end{tikzpicture}}
	Diagram I2
\end{subfigure}
\caption{Problematic diagrams for Diagram I with  $r_{13}^2(t)/r_{14}^2(t) = t^{-1}$}
	\label{fig:CompletediagramsI2}
\end{figure}

For Diagram I2, it is easy to see that, vorticities satisfy
\begin{equation}\label{DiagramI14-23}
    \begin{array}{l}
      r_{23}^2=-\frac{\Gamma_1\Gamma_4(\Gamma_3+\Gamma_2)}{\Gamma_3\Gamma_2(\Gamma_1+\Gamma_4)} r_{14}^2 ,\\[6pt]
    r_{24}^2= \frac{\Gamma_1\Gamma_3}{\Gamma_2\Gamma_4} r_{13}^2,\\[6pt]
    r_{34}^2= \frac{\Gamma_1\Gamma_2}{\Gamma_4\Gamma_3} r_{12}^2.
    \end{array}
\end{equation}
Then, by \eqref{DiagramI12-34} and \eqref{DiagramI14-23}, it also follows that
  \begin{center}
  $\Gamma_1+\Gamma_2+\Gamma_3+\Gamma_4=0$,
\end{center}
but this is in conflict with  $L =0$.

As a result, Diagram I does not occur.\\

\noindent{\bfseries{Subcase 2: If Diagram III  occurs. }}

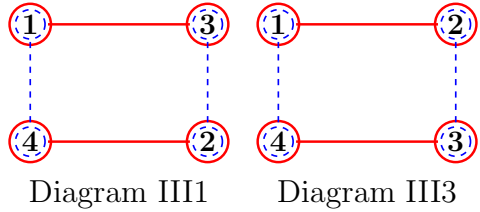
\begin{figure}[!h]
	\centering
\begin{subfigure}[b]{0.2\textwidth}
		\centering
		\resizebox{\linewidth}{!}{
	\begin{tikzpicture}
	
		\draw  (-3/2,  0) node {\large\textbf{1}};
	\draw (3/2,  0) node{\large\textbf{3}};
	
	\draw  (-3/2,  -2)  node {\large\textbf{4}};
	\draw  (3/2,  -2) node {\large\textbf{2}}; 

		\draw [blue,dashed, thick] (-3/2,  0) circle (0.25);
	\draw [red,very thick] (-3/2,  0) circle (0.35);
	
		\draw [blue,dashed, thick] (3/2,  0) circle (0.25);
\draw [red,very thick] (3/2,  0) circle (0.35);

\draw [blue,dashed, thick] (-3/2,  -2) circle (0.25);
\draw [red,very thick](-3/2,  -2) circle (0.35);

\draw [blue,dashed, thick] (3/2,  -2) circle (0.25);
\draw[red,very thick]  (3/2,  -2) circle (0.35);

\draw [red,very thick] (-1.2,0)--(1.2,0); 
\draw [red,very thick] (-1.2,-2)--(1.2,-2);

\draw [blue,dashed, thick] (-3/2,-.3)--(-3/2,-1.7);
\draw [blue,dashed, thick] (3/2,-.3)--(3/2,-1.7);

	\end{tikzpicture}
}
	Diagram III1
\end{subfigure}
\begin{subfigure}[b]{0.2\textwidth}
		\centering
		\resizebox{\linewidth}{!}{
	\begin{tikzpicture}
	
		\draw  (-3/2,  0) node {\large\textbf{1}};
	\draw (3/2,  0) node{\large\textbf{2}};
	
	\draw  (-3/2,  -2)  node {\large\textbf{4}};
	\draw  (3/2,  -2) node {\large\textbf{3}}; 

		\draw [blue,dashed, thick] (-3/2,  0) circle (0.25);
	\draw [red,very thick] (-3/2,  0) circle (0.35);
	
		\draw [blue,dashed, thick] (3/2,  0) circle (0.25);
\draw [red,very thick] (3/2,  0) circle (0.35);

\draw [blue,dashed, thick] (-3/2,  -2) circle (0.25);
\draw [red,very thick](-3/2,  -2) circle (0.35);

\draw [blue,dashed, thick] (3/2,  -2) circle (0.25);
\draw[red,very thick]  (3/2,  -2) circle (0.35);

\draw [red,very thick] (-1.2,0)--(1.2,0); 
\draw [red,very thick] (-1.2,-2)--(1.2,-2);

\draw [blue,dashed, thick] (-3/2,-.3)--(-3/2,-1.7);
\draw [blue,dashed, thick] (3/2,-.3)--(3/2,-1.7);

	\end{tikzpicture}
}
	Diagram III3
\end{subfigure}
\caption{Problematic diagrams for Diagram III with  $r_{13}^2(t)/r_{14}^2(t) = t^{-1}$}
	\label{fig:CompletediagramsIII2}
\end{figure}

Then, it is easy to show that, we are in one of diagrams in  Figure \ref{fig:CompletediagramsIII2}.

For Diagram III1, we have shown  that it is impossible.

For Diagram III3, we have
\begin{equation}\label{DiagramIII13-24}
    \begin{array}{l}
    \Gamma_1 \Gamma_3 =\Gamma_2 \Gamma_4, \\[6pt]
      r_{23}^2=\frac{\Gamma_3}{\Gamma_4} r_{14}^2 ,\\[6pt]
    r_{34}^2= \frac{\Gamma_3}{\Gamma_2} r_{12}^2,\\[6pt]
    r_{24}^2= - r_{13}^2.
    \end{array}
\end{equation}
Then,  by \eqref{DiagramI12-34} and \eqref{DiagramIII13-24}, it follows that
  \begin{center}
  $\Gamma_1 \Gamma_3 =\Gamma_2 \Gamma_4$, and $\Gamma_1\Gamma_3=-\Gamma_2\Gamma_4$,
\end{center}
but this leads to a contradiction.

As a result, Diagram III does not occur.\\

\noindent{\bfseries{Subcase 3: If Diagram vi  occurs. }}

\begin{figure}[!h]
	\centering
\begin{subfigure}[b]{0.2\textwidth}

		\centering
		\resizebox{\linewidth}{!}{
	\begin{tikzpicture}
	\draw	[blue,dashed,thick]  (-3/2,0)  circle (0.25);
	\draw	  (-3/2,0)    node {\large\textbf{1}};
	
		\draw	[blue,dashed,thick]  (3/2,0)  circle (0.25);
	\draw		(3/2,0) node {\large\textbf{2}};
	
		\draw	[blue,dashed,thick]  (0,-3/2*1.732)   circle (0.25);
	\draw		(0,-3/2*1.732) node {\large\textbf{3}};
	
		\draw	[blue,dashed,thick]  (0,-1)  circle (0.25);
	\draw		(0,-1) node {\large\textbf{4}};

	\draw [red,very thick] (-3/2+.35,0.15)--(3/2-.35,0.15);
\draw [blue, dashed,thick] (-3/2+0.35,0)--(3/2-.35,0);
	
	\draw [red,very thick] (-3/2+.2,-0.2*1.732)--(-.2,-3/2*1.732+.2*1.732);
	
	\draw [red,very thick] (3/2-0.2,-0.2*1.732)--(0.2,-3/2*1.732+0.2*1.732);

		\draw [red,very thick]  (-3/2+0.25+0.15,-0.25*2/3)--(0-0.4+0.15,-1+0.4*2/3);  
		
		\draw [red,very thick]  (3/2-0.25-.15,-.25*2/3)--(0.45 -.15,-1+.45*2/3);  

		\draw [red,very thick] (-.07,-3/2*1.732+.4)--(-.07,-1-.4);
			\draw [blue,dashed,thick]  (.07,-3/2*1.732+0.4)--(.07,-1-0.4);
	\end{tikzpicture}
}
	Diagram vi3
\end{subfigure}\begin{subfigure}[b]{0.2\textwidth}

		\centering
		\resizebox{\linewidth}{!}{
	\begin{tikzpicture}
	\draw	[blue,dashed,thick]  (-3/2,0)  circle (0.25);
	\draw	  (-3/2,0)    node {\large\textbf{1}};
	
		\draw	[blue,dashed,thick]  (3/2,0)  circle (0.25);
	\draw		(3/2,0) node {\large\textbf{4}};
	
		\draw	[blue,dashed,thick]  (0,-3/2*1.732)   circle (0.25);
	\draw		(0,-3/2*1.732) node {\large\textbf{2}};
	
		\draw	[blue,dashed,thick]  (0,-1)  circle (0.25);
	\draw		(0,-1) node {\large\textbf{3}};

	\draw [red,very thick] (-3/2+.35,0.15)--(3/2-.35,0.15);
\draw [blue, dashed,thick] (-3/2+0.35,0)--(3/2-.35,0);
	
	\draw [red,very thick] (-3/2+.2,-0.2*1.732)--(-.2,-3/2*1.732+.2*1.732);
	
	\draw [red,very thick] (3/2-0.2,-0.2*1.732)--(0.2,-3/2*1.732+0.2*1.732);

		\draw [red,very thick]  (-3/2+0.25+0.15,-0.25*2/3)--(0-0.4+0.15,-1+0.4*2/3);  
		
		\draw [red,very thick]  (3/2-0.25-.15,-.25*2/3)--(0.45 -.15,-1+.45*2/3);  

		\draw [red,very thick] (-.07,-3/2*1.732+.4)--(-.07,-1-.4);
			\draw [blue,dashed,thick]  (.07,-3/2*1.732+0.4)--(.07,-1-0.4);
	\end{tikzpicture}
}
	Diagram vi4
\end{subfigure}
\caption{Problematic diagrams for Diagram vi with  $r_{13}^2(t)/r_{14}^2(t) = t^{-1}$}
	\label{fig:Completediagramsvi2}
\end{figure}
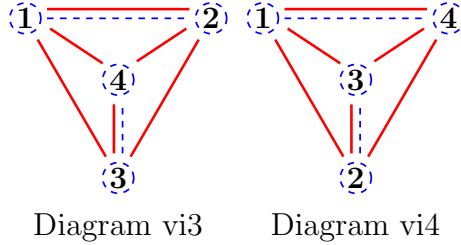

Then, it is easy to show that, we are in one of diagrams in  Figure \ref{fig:Completediagramsvi2}.
For these diagrams, we have
\begin{center}
$\Gamma_3 =\Gamma_4=(\sqrt{3}-2)^{\pm 1}\Gamma_1=(\sqrt{3}-2)^{\pm 1}\Gamma_2$
\end{center}
or
\begin{center}
$\Gamma_2 =\Gamma_3=(\sqrt{3}-2)^{\pm 1}\Gamma_1=(\sqrt{3}-2)^{\pm 1}\Gamma_4$.
\end{center}
However,  the two requirements on vorticities above are both in conflict with \eqref{vorticitiesvi13-24}. This leads to a contradiction.

As a result,  Diagram vi  does not occur in  Case 3.\\

This completes the proof of Lemma \ref{nonadjacentdistancesbj2}.

$~~~~~~~~~~~~~~~~~~~~~~~~~~~~~~~~~~~~~~~~~~~~~~~~~~~~~~~~~~~~~~~~~~~~~~~~~~~~~~~~~~~~~~~~~~~~~~~~~~~~~~~~~~~~~~~~~~~~~~~~~~~~~~~~~~~~~~~~~~~~~~~~~~~~\Box$\\

Therefore, we assume that  all ratios  of two nonadjacent
distances' squares are constants, and Diagram I is impossible from now on. Note that Diagram i  is also impossible by  the condition that all ratios  of two nonadjacent
distances' squares are not dominating.\\

\subsection{Finiteness of collapse configurations except two special cases}
\indent\par
In this subsection we consider the finiteness  problem  of collapse configurations in the case that vorticities do not satisfy the following relations:
 \begin{itemize}
   \item $\Gamma_i =\Gamma_j=\Gamma_k =-\Gamma_l$,
   \item $\Gamma_i =\Gamma_j=(\sqrt{3}-2)\Gamma_k=(\sqrt{3}-2)\Gamma_l$,
 \end{itemize}
where $(ijkl)$ is a certain permutation of $\{1,2,3,4\}$.

In this case only  Diagram III  and Diagram ix   are possible.

\begin{theorem}\label{Main2collapse1}
If the vorticities $\Gamma_n$ $(n\in\{1,2,3,4\})$ are nonzero and do not satisfy any relation above, then the four-vortex problem has finitely many
 collapse configurations.
\end{theorem}

{\bf Proof.}

If there are infinitely many $\Lambda\in \mathbb{S}$ such that  equations (\ref{stationaryconfigurationmainLambda})  have a solution, then there is a Puiseux series as  (\ref{Puiseuxseries0})  with $\Lambda(t) = t$. By considering the Puiseux series, we are faced the following cases:\\\\
\noindent{\bfseries{Case 1: If we are in Diagram ix. }}

Then, by \eqref{ixr1}, it follows that all products $r_{jk}^2 r_{jl}^2$ of two adjacent
distances' squares are  dominating.

By considering a Puiseux series as  (\ref{Puiseuxseries0})  with $r_{12}^2r_{13}^2 = t^{-1}$, we are in Diagram III, indeed,  in one of diagrams in  Figure \ref{fig:CompletediagramsIII2}.

Thus vorticities t satisfy the following relation.
\begin{equation}\label{DiagramIII1213}
    (\Gamma_1 \Gamma_2 -\Gamma_3 \Gamma_4)(\Gamma_1 \Gamma_3 -\Gamma_2 \Gamma_4)=0.
\end{equation}

Similarly, by considering a Puiseux series as  (\ref{Puiseuxseries0})  with $r_{12}^2r_{14}^2 = t^{-1}$, we have
\begin{equation}\label{DiagramIII1214}
    (\Gamma_1 \Gamma_2 -\Gamma_3 \Gamma_4)(\Gamma_1 \Gamma_4 -\Gamma_2 \Gamma_3)=0;
\end{equation}
by considering a Puiseux series as  (\ref{Puiseuxseries0})  with $r_{13}^2r_{14}^2 = t^{-1}$, we have
\begin{equation}\label{DiagramIII1314}
    (\Gamma_1 \Gamma_3 -\Gamma_2 \Gamma_4)(\Gamma_1 \Gamma_4 -\Gamma_2 \Gamma_3)=0.
\end{equation}

Then,  by \eqref{DiagramIII1213}, \eqref{DiagramIII1214}, \eqref{DiagramIII1314} and $L=0$, it follows that
  \begin{center}
  $\Gamma_i =\Gamma_j=(\sqrt{3}-2)\Gamma_k=(\sqrt{3}-2)\Gamma_l$,
  \end{center}
where $(ijkl)$ is a certain permutation of $\{1,2,3,4\}$.
This leads to a contradiction.

As a result,  Diagram ix  is impossible.\\

\noindent{\bfseries{Case 2: We are in Diagram III. }}

Without loss of generality, suppose we are in Diagram III1.
Then, we have
\begin{equation}\label{DiagramIII12}
    \Gamma_1 \Gamma_2 -\Gamma_3 \Gamma_4=0,
\end{equation}
and  products $r_{12}^2 r_{13}^2$ and $r_{12}^2 r_{14}^2$ are  dominating.

\begin{figure}[!h]
	\centering
\begin{subfigure}[b]{0.2\textwidth}
		\centering
		\resizebox{\linewidth}{!}{
	\begin{tikzpicture}
	
		\draw  (-3/2,  0) node {\large\textbf{1}};
	\draw (3/2,  0) node{\large\textbf{2}};
	
	\draw  (-3/2,  -2)  node {\large\textbf{3}};
	\draw  (3/2,  -2) node {\large\textbf{4}}; 
	
		\draw [blue,dashed, thick] (-3/2,  0) circle (0.25);
	\draw [red,very thick] (-3/2,  0) circle (0.35);
	
		\draw [blue,dashed, thick] (3/2,  0) circle (0.25);
\draw [red,very thick] (3/2,  0) circle (0.35);

\draw [blue,dashed, thick] (-3/2,  -2) circle (0.25);
\draw [red,very thick](-3/2,  -2) circle (0.35);

\draw [blue,dashed, thick] (3/2,  -2) circle (0.25);
\draw[red,very thick]  (3/2,  -2) circle (0.35);

\draw [red,very thick] (-1.2,0)--(1.2,0); 
\draw [red,very thick] (-1.2,-2)--(1.2,-2);

\draw [blue,dashed, thick] (-3/2,-.3)--(-3/2,-1.7);
\draw [blue,dashed, thick] (3/2,-.3)--(3/2,-1.7);

	\end{tikzpicture}
}
	
\end{subfigure}
\caption{Problematic diagrams for Diagram III with  $r_{12}^2(t)r_{13}^2(t) = t$}
	\label{fig:CompletediagramsIII3}
\end{figure}
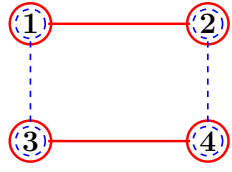

By considering a Puiseux series as  (\ref{Puiseuxseries0})  with $r_{12}^2r_{13}^2 = t$, we are in  the diagram in  Figure \ref{fig:CompletediagramsIII3}. Therefore, we have
\begin{equation}\label{DiagramIII14}
    \Gamma_1 \Gamma_4 -\Gamma_2 \Gamma_3=0.
\end{equation}

Similarly, by considering a Puiseux series as  (\ref{Puiseuxseries0})  with $r_{12}^2r_{14}^2 = t$, we have
\begin{equation}\label{DiagramIII13}
    \Gamma_1 \Gamma_3 -\Gamma_2 \Gamma_4=0.
\end{equation}

However, there is no solution for \eqref{DiagramIII12}, \eqref{DiagramIII14}, \eqref{DiagramIII13} and $L=0$.
This leads to a contradiction.

As a result,  Diagram III  is also impossible.\\

This completes the proof of Theorem \ref{Main2collapse1}.

$~~~~~~~~~~~~~~~~~~~~~~~~~~~~~~~~~~~~~~~~~~~~~~~~~~~~~~~~~~~~~~~~~~~~~~~~~~~~~~~~~~~~~~~~~~~~~~~~~~~~~~~~~~~~~~~~~~~~~~~~~~~~~~~~~~~~~~~~~~~~~~~~~~~~\Box$\\

\subsection{Finiteness of  collapse configurations in case $\Gamma_1 =\Gamma_2=\Gamma_3 =-\Gamma_4$}
\indent\par
In this subsection we consider the finiteness  problem  of collapse configurations in  case that vorticities  satisfy the following relation:
\begin{center}
$\Gamma_i =\Gamma_j=\Gamma_k =-\Gamma_l$,
\end{center}
where $(ijkl)$ is a certain permutation of $\{1,2,3,4\}$.
Without loss of generality, we only consider the case
\begin{center}
$\Gamma_1 =\Gamma_2=\Gamma_3 =-\Gamma_4$.
\end{center}
In this case only  Diagram II,  Diagram iv and Diagram ix     are possible.

\begin{theorem}\label{Main2collapse2}
If the vorticities $\Gamma_n$ $(n\in\{1,2,3,4\})$ are nonzero and  satisfy the relation $\Gamma_1 =\Gamma_2=\Gamma_3 =-\Gamma_4$, then the four-vortex problem has finitely many
 collapse configurations.
\end{theorem}

{\bf Proof.}

If there are infinitely many $\Lambda\in \mathbb{S}$ such that  equations (\ref{stationaryconfigurationmainLambda})  have a solution, then there is a Puiseux series as  (\ref{Puiseuxseries0})  with $\Lambda(t) = t$. By considering the Puiseux series, we are in  Diagram II,  Diagram iv or Diagram ix. However, all the three diagrams yield that \begin{equation*}
r_{kl}\approx t^{-q}\prec t^0,  ~~~~~~~~~~~~~~~~~1\leq k<l\leq 4.
\end{equation*}
It follows that $r_{kl}^2$ are all dominating. Then,  by considering a Puiseux series as  (\ref{Puiseuxseries0})  with $r_{12}^2 = t^{-1}$, it is easy to see that this contradicts the fact that $r_{kl}\prec t^{0}$. Consequently, we arrive at the conclusion   that there are finitely many $\Lambda\in \mathbb{S}$ such that  equations (\ref{stationaryconfigurationmainLambda})  have a solution, the proof of Theorem \ref{Main2collapse2} is now completed.

$~~~~~~~~~~~~~~~~~~~~~~~~~~~~~~~~~~~~~~~~~~~~~~~~~~~~~~~~~~~~~~~~~~~~~~~~~~~~~~~~~~~~~~~~~~~~~~~~~~~~~~~~~~~~~~~~~~~~~~~~~~~~~~~~~~~~~~~~~~~~~~~~~~~~\Box$\\

\subsection{Finiteness of  collapse configurations in case $\Gamma_3 =\Gamma_4=(\sqrt{3}-2)\Gamma_1 =(\sqrt{3}-2)\Gamma_2$}
\indent\par
In this subsection we consider the finiteness  problem  of collapse configurations in  case that vorticities  satisfy the following relation:
\begin{center}
$\Gamma_i =\Gamma_j=(\sqrt{3}-2)\Gamma_k=(\sqrt{3}-2)\Gamma_l$,
\end{center}
where $(ijkl)$ is a certain permutation of $\{1,2,3,4\}$.
Without loss of generality, we only consider the case
\begin{equation}\label{vorticities12-34}
    \Gamma_3 =\Gamma_4=(\sqrt{3}-2)\Gamma_1 =(\sqrt{3}-2)\Gamma_2.
\end{equation}
In this case only  Diagram III,  Diagram vi and Diagram ix     are possible.

\begin{theorem}\label{Main2collapse3}
If the vorticities $\Gamma_n$ $(n\in\{1,2,3,4\})$ are nonzero and  satisfy the relation $\Gamma_3 =\Gamma_4=(\sqrt{3}-2)\Gamma_1 =(\sqrt{3}-2)\Gamma_2$, then the four-vortex problem has finitely many
 collapse configurations.
\end{theorem}

{\bf Proof.}

If there are infinitely many $\Lambda\in \mathbb{S}$ such that  equations (\ref{stationaryconfigurationmainLambda})  have a solution, then there is a Puiseux series as  (\ref{Puiseuxseries0})  with $\Lambda(t) = t$. By considering the Puiseux series, we are in Diagram III,  Diagram vi or Diagram ix.

We divide our proof in the following cases:\\

\noindent{\bfseries{Case 1: If we are in Diagram vi or Diagram ix. }}

Then, by \eqref{vir} and \eqref{ixr1}, it follows that all products   of two adjacent
distances' squares are  dominating.

\begin{figure}[!h]
	\centering
\begin{subfigure}[b]{0.2\textwidth}
		\centering
		\resizebox{\linewidth}{!}{
	\begin{tikzpicture}
	
		\draw  (-3/2,  0) node {\large\textbf{1}};
	\draw (3/2,  0) node{\large\textbf{3}};
	
	\draw  (-3/2,  -2)  node {\large\textbf{4}};
	\draw  (3/2,  -2) node {\large\textbf{2}}; 
	
		\draw [blue,dashed, thick] (-3/2,  0) circle (0.25);
	\draw [red,very thick] (-3/2,  0) circle (0.35);
	
		\draw [blue,dashed, thick] (3/2,  0) circle (0.25);
\draw [red,very thick] (3/2,  0) circle (0.35);

\draw [blue,dashed, thick] (-3/2,  -2) circle (0.25);
\draw [red,very thick](-3/2,  -2) circle (0.35);

\draw [blue,dashed, thick] (3/2,  -2) circle (0.25);
\draw[red,very thick]  (3/2,  -2) circle (0.35);

\draw [red,very thick] (-1.2,0)--(1.2,0); 
\draw [red,very thick] (-1.2,-2)--(1.2,-2);

\draw [blue,dashed, thick] (-3/2,-.3)--(-3/2,-1.7);
\draw [blue,dashed, thick] (3/2,-.3)--(3/2,-1.7);

	\end{tikzpicture}
}
Diagram III1	
\end{subfigure}\begin{subfigure}[b]{0.2\textwidth}
		\centering
		\resizebox{\linewidth}{!}{
	\begin{tikzpicture}
	
		\draw  (-3/2,  0) node {\large\textbf{1}};
	\draw (3/2,  0) node{\large\textbf{2}};
	
	\draw  (-3/2,  -2)  node {\large\textbf{4}};
	\draw  (3/2,  -2) node {\large\textbf{3}}; 
	
		\draw [blue,dashed, thick] (-3/2,  0) circle (0.25);
	\draw [red,very thick] (-3/2,  0) circle (0.35);
	
		\draw [blue,dashed, thick] (3/2,  0) circle (0.25);
\draw [red,very thick] (3/2,  0) circle (0.35);

\draw [blue,dashed, thick] (-3/2,  -2) circle (0.25);
\draw [red,very thick](-3/2,  -2) circle (0.35);

\draw [blue,dashed, thick] (3/2,  -2) circle (0.25);
\draw[red,very thick]  (3/2,  -2) circle (0.35);

\draw [red,very thick] (-1.2,0)--(1.2,0); 
\draw [red,very thick] (-1.2,-2)--(1.2,-2);

\draw [blue,dashed, thick] (-3/2,-.3)--(-3/2,-1.7);
\draw [blue,dashed, thick] (3/2,-.3)--(3/2,-1.7);

	\end{tikzpicture}
}
Diagram III3	
\end{subfigure}
\caption{Problematic diagrams for Diagram III with  $r_{12}^2(t)r_{13}^2(t) = t^{-1}$}
	\label{fig:CompletediagramsIII4}
\end{figure}
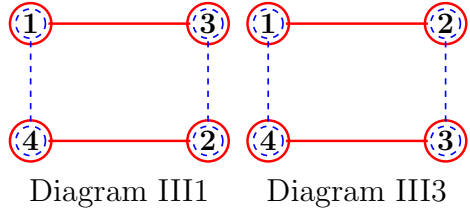

By considering a Puiseux series as  (\ref{Puiseuxseries0})  with $r_{12}^2r_{13}^2 = t^{-1}$, we are in Diagram III, indeed,  in one of diagrams in  Figure \ref{fig:CompletediagramsIII4}.

For Diagram III1, it follows that $\Gamma_1 \Gamma_2 =\Gamma_3 \Gamma_4$. This is in conflict with \eqref{vorticities12-34}. Thus we are only in Diagram III3, it follows that
\begin{equation}\label{DiagramIII13new}
    \begin{array}{c}
       \Gamma_1 \Gamma_3 =\Gamma_2 \Gamma_4, \\[6pt]
      r_{23}^2=\frac{\Gamma_3}{\Gamma_4} r_{14}^2 ,\\[6pt]
    r_{34}^2= \frac{\Gamma_3}{\Gamma_2} r_{12}^2,\\[6pt]
    r_{24}^2= - r_{13}^2.
    \end{array}
\end{equation}

Similarly, by considering a Puiseux series as  (\ref{Puiseuxseries0})  with $r_{12}^2r_{14}^2 = t^{-1}$, we are in Diagram III2 and it follows that
\begin{equation}\label{DiagramIII14new}
    \begin{array}{l}
    \Gamma_1 \Gamma_4 =\Gamma_2 \Gamma_3, \\[6pt]
     r_{24}^2=\frac{\Gamma_4}{\Gamma_3} r_{13}^2 ,\\[6pt]
    r_{34}^2= \frac{\Gamma_4}{\Gamma_2} r_{12}^2,\\[6pt]
    r_{23}^2= - r_{14}^2.
    \end{array}
\end{equation}

It is easy to see that \eqref{DiagramIII13new} and \eqref{DiagramIII14new} are in conflict with each other.

As a result,  Diagram vi and Diagram ix  are impossible. Therefore, by considering any Puiseux series as  (\ref{Puiseuxseries0}), we are only in Diagram III.
\\

\noindent{\bfseries{Case 2: We are in Diagram III. }}

Without loss of generality, suppose we are in Diagram III3.
Then, it follows that \eqref{DiagramIII13new} holds,
and  products $r_{12}^2 r_{13}^2$ and $r_{13}^2 r_{14}^2$ are  dominating.

By considering a Puiseux series as  (\ref{Puiseuxseries0})  with $r_{12}^2r_{13}^2 = t$, we are in   Diagram III2, and it follows that \eqref{DiagramIII14new} holds. However, \eqref{DiagramIII13new} and \eqref{DiagramIII14new} are in conflict with each other.

As a result,  Diagram III is also impossible.\\

Consequently, we arrive at the conclusion   that there are finitely many $\Lambda\in \mathbb{S}$ such that  equations (\ref{stationaryconfigurationmainLambda})  have a solution, the proof of Theorem \ref{Main2collapse3} is now completed.

$~~~~~~~~~~~~~~~~~~~~~~~~~~~~~~~~~~~~~~~~~~~~~~~~~~~~~~~~~~~~~~~~~~~~~~~~~~~~~~~~~~~~~~~~~~~~~~~~~~~~~~~~~~~~~~~~~~~~~~~~~~~~~~~~~~~~~~~~~~~~~~~~~~~~\Box$\\

\section{Upper bounds on the  number of $\Lambda$ and on collapse configurations}
\indent\par
In this section we provide   upper  bounds  for the number of real normalized collapse configurations  and for the number of the correlative $\Lambda$, mainly by making use of B\'{e}zout Theorem.

First, let us focus on the numbers of normalized collapse configurations  and the correlative $\Lambda$ in the complex domain. Recall that,
normalized collapse configurations are characterized by  nonzero solutions of the following equations such that $\Lambda\in \mathbb{S}\backslash\{\pm1\}$.
\begin{equation}\label{stationaryconfigurationnew6new}
\left\{
             \begin{array}{rr}
                M_z=0, &~~~~~~ M_w=0,\\
             L=0,~~~~~~ I=0, &z_2-z_1= w_2-w_1, \\
             {F_z}-\Lambda f_z=0, &
             {\Lambda } {F_w}-f_w=0,\\
 G_z+\Lambda g_z=0,&\Lambda  G_w+g_w=0.
             \end{array}
\right.
\end{equation}
where
\begin{equation}
\begin{array}{c}
  M_z=\sum_{j=1}^4\Gamma_jz_j , ~~~~~~ M_w=\sum_{j=1}^4\Gamma_jw_j,\\
   F_z=\sum_{j=1}^4\Gamma_jz_j^2 w_j, ~~~~~~f_z=\sum_{1\leq j< k\leq 4}\Gamma_j\Gamma_k(z_j+ z_k),\\
    F_w=\sum_{j=1}^4\Gamma_jz_j w_j^2,~~~~~~f_w=\sum_{1\leq j< k\leq 4}\Gamma_j\Gamma_k(w_j+ w_k),\\
G_z=\Gamma_1 w_1 z_2 z_3 z_4+\Gamma_2 w_2 z_1 z_3 z_4+\Gamma_3 w_3 z_1 z_2  z_4+\Gamma_4 w_4 z_1 z_2 z_3,\\
g_z=\sum_{1\leq j< k\leq 4,l<m,\{j,k,l,m\}=\{1,2,3,4\}}\Gamma_j \Gamma_k z_l z_m,\\
G_w=\Gamma_1 z_1 w_2 w_3 w_4+\Gamma_2 z_2 w_1 w_3 w_4+\Gamma_3 z_3 w_1 w_2  w_4+\Gamma_4 z_4 w_1 w_2 w_3,\\
g_w=\sum_{1\leq j< k\leq 4,l<m,\{j,k,l,m\}=\{1,2,3,4\}}\Gamma_j \Gamma_k w_l w_m.
\end{array}\nonumber
\end{equation}
For more detail please refer to Subsection 7.2 in \cite{yu2021Finiteness}.

We now know  that \eqref{stationaryconfigurationnew6new} has finitely many solutions in $\Lambda, z,w$. In the following, we focus on the solutions of \eqref{stationaryconfigurationnew6new} such that $\Lambda\in \mathbb{S}\backslash\{\pm1\}$ and $(z,w)\neq(0,0)$.

For the system (\ref{stationaryconfigurationnew6new}), by considering the map $ (z, w) \mapsto (\overline{z}, \overline{w})$, one can prove that, the  number of solutions  associated with $\Lambda$ is equal to the  number of solutions   associated with $\overline{\Lambda}$ $(=\Lambda^{-1})$. By considering the map $ (z, w) \mapsto (\mathbf{i} z, \mathbf{i}w)$, one can prove that the number of solutions  associated with $\Lambda$ is equal {to} the  number of solutions associated with $-{\Lambda}$.

{Let $\Lambda_j\in \mathbb{S}$ $(j=1,\cdots,n_{\Lambda})$ be all points  in the first quadrant associated with nonzero solutions of the system (\ref{stationaryconfigurationnew6new}), and
$\mathcal{N}_{\Lambda_j}$ be the  number of nonzero solutions (counting with the appropriate mulitiplicity) of the system (\ref{stationaryconfigurationnew6new}) associated with $\Lambda_j$. Notations $\mathcal{N}_{\bar{\Lambda}_j}$, $\mathcal{N}_{-\Lambda_j}$, $\mathcal{N}_{\mathbf{i}}$ and $\mathcal{N}_{-\mathbf{i}}$ have similar meaning; but note that $\mathcal{N}_{\mathbf{i}}$ and $\mathcal{N}_{-\mathbf{i}}$ may be trivial, and yet
\begin{center}
$\mathcal{N}_{\Lambda_j}=\mathcal{N}_{\bar{\Lambda}_j}=\mathcal{N}_{-\Lambda_j}=\mathcal{N}_{-\bar{\Lambda}_j}\geq 2$.
\end{center}
Indeed, by considering the map $ (z, w) \mapsto (- z, -w)$, one can prove that every $\mathcal{N}_{\Lambda_j}$ is an even number, so are $\mathcal{N}_{\mathbf{i}}$ and $\mathcal{N}_{-\mathbf{i}}$.
 We also use notations $\mathcal{N}_{1}$ and $\mathcal{N}_{-1}$ to denote the  number of nonzero solutions  associated with $\Lambda=1$ and $\Lambda=-1$, respectively. We remark  that $\mathcal{N}_{1}$ and $\mathcal{N}_{1}$ are both even numbers, moreover, \begin{center}
$\mathcal{N}_{1}=\mathcal{N}_{-1}\geq 20$,
\end{center}
since there are exactly $20$ collinear nonzero solutions (counting with the appropriate mulitiplicity) when $L=0$, please see Subsubsection 7.2.1 in \cite{yu2021Finiteness} for more detail.}

Let us embed the system (\ref{stationaryconfigurationnew6new})  into a polynomial system in the projective space $\mathbb{P}^{9}_{\mathbb{C}}$:
\begin{equation}\label{stationaryconfigurationnewc1new}
\left\{
             \begin{array}{rr}
                M_z=0, &~~~~~~ M_w=0,\\
             I=0, &z_2-z_1= w_2-w_1, \\
              {F_z}-t{\Lambda}f_z=0, &
             {\Lambda } {F_w}-t^3f_w=0,\\
 G_z+t{\Lambda}g_z=0,&\Lambda  G_w+t^3g_w=0.
             \end{array}
\right.
\end{equation}
It is clear  that the degree of the system (\ref{stationaryconfigurationnewc1new}) is no more than $2\times3\times4\times4\times5=480$.

To obtain an upper bound of the  number of nonzero solutions for the system (\ref{stationaryconfigurationnew6new}), let us estimate a lower bound of degree of the system (\ref{stationaryconfigurationnewc1new}) for three cases:\begin{itemize}
                                                                 \item $t=0$ and $\Lambda\neq0$;
                                                                 \item $t\neq0$ and $\Lambda=0$;
                                                                 \item $t=0$ and $\Lambda=0$.
                                                               \end{itemize}

{\bfseries{Case 1: $t=0$ and $\Lambda\neq0$. }}\\
Then the system (\ref{stationaryconfigurationnewc1new}) reduces to \begin{equation}\label{stationaryconfigurationnewc1new1}
\left\{
             \begin{array}{rr}
                M_z=0, &~~~~~~ M_w=0,\\
             I=0, &z_2-z_1= w_2-w_1, \\
              {F_z}=0, &
              {F_w}=0,\\
 G_z=0,&  G_w=0.
             \end{array}
\right.
\end{equation}
This system is  an algebraic variety in $\mathbb{P}^{7}_{\mathbb{C}}$ which is the same as the system (7.76) in \cite{yu2021Finiteness} in form (note that $L=0$ here).

A straightforward computation similar to the one used for (7.76) in \cite{yu2021Finiteness} shows that the degree of the system (\ref{stationaryconfigurationnewc1new1}) is no less than $6\times2+4\times2=20$.\\

{\bfseries{Case 2: $t\neq0$ and $\Lambda=0$. }}\\
Then the system (\ref{stationaryconfigurationnewc1new}) reduces to \begin{equation}\label{stationaryconfigurationnewc1new2}
\left\{
             \begin{array}{rr}
                M_z=0, &~~~~~~ M_w=0,\\
             I=0, &z_2-z_1= w_2-w_1, \\
              {F_z}=0, &
             f_w=0,\\
 G_z=0,&g_w=0.
             \end{array}
\right.
\end{equation}
This system is also an algebraic variety in $\mathbb{P}^{7}_{\mathbb{C}}$.

A straightforward computation  shows that the  variety (\ref{stationaryconfigurationnewc1new2}) consists of  a  one-dimensional irreducible components (i.e., a one-dimensional lines):
\begin{equation*}
 w_1=w_2=w_3=w_4=0, ~~M_z=0, ~~z_2-z_1=0.
\end{equation*}
Moreover, a straightforward computation shows that the  multiplicity of  this  one-dimensional irreducible components is at least $2$.
It follows that the degree of the variety (\ref{stationaryconfigurationnewc1new2}) is no less than $2$.
\\

{\bfseries{Case 3: $t=0$ and $\Lambda=0$. }}\\
Then the system (\ref{stationaryconfigurationnewc1new}) reduces to \begin{equation}\label{stationaryconfigurationnewc1new3}
\left\{
             \begin{array}{rr}
                M_z=0, &~~~~~~ M_w=0,\\
             I=0, &~~~~~~~~~~~~z_2-z_1= w_2-w_1, \\
              {F_z}=0, &~~~~~~
 G_z=0.
             \end{array}
\right.
\end{equation}
This system is also an algebraic variety in $\mathbb{P}^{7}_{\mathbb{C}}$.

A straightforward computation  shows that the  variety (\ref{stationaryconfigurationnewc1new3}) consists of  thirteen  one-dimensional irreducible components (i.e., thirteen one-dimensional lines); moreover, a straightforward computation shows that  the multiplicity of  the    irreducible component,
\begin{center}
$z_1=z_2=z_3=z_4=0, M_w=0, w_2-w_1=0$,
\end{center}
is at least $6$.

Therefore, the degree of the variety (\ref{stationaryconfigurationnewc1new3}) is no less than $12\times1+1\times6=18$.\\

Since the system (\ref{stationaryconfigurationnewc1new}) is a disjoint union of  (\ref{stationaryconfigurationnew6new}),  (\ref{stationaryconfigurationnewc1new1}), (\ref{stationaryconfigurationnewc1new2}) and (\ref{stationaryconfigurationnewc1new3}), it follows that the degree of the system (\ref{stationaryconfigurationnew6new}) is no more than \begin{center}
$480-20-2-18=440$.
\end{center}
On the other hand, note that we have
\begin{proposition}\label{isolatedtrivialsolutionnew}
Assume $L=0$ and $\Lambda\neq0$. If  one solution of the system (\ref{stationaryconfigurationnew6new}) satisfies $z_j=z_k$ (or $w_j=w_k$)  for some $ j\neq k$, then  the solution is trivial.
\end{proposition}
The Proposition is essentially Proposition  7.1 in \cite{yu2021Finiteness}, so the proof is omitted here. By  Proposition \ref{isolatedtrivialsolutionnew}, it follows that there is  an isolated  trivial solution of system (\ref{stationaryconfigurationnew6new}) for any given $\Lambda\neq0$. Moreover, it is easy to see that the multiplicity of  this trivial solution is  $8$.

As a result, we have
\begin{equation}\label{estnumber0}
    4\sum_{j=1}^{n_{\Lambda}}\mathcal{N}_{\Lambda_j}+\mathcal{N}_{\mathbf{i}}+\mathcal{N}_{-\mathbf{i}}+\mathcal{N}_{1}+\mathcal{N}_{-1}\leq 440-8=432.
\end{equation}

Based on this inequality, we have the following result.
\begin{corollary}\label{upperbounds2}
If the vorticities $\Gamma_n$ $(n\in\{1,2,3,4\})$ are nonzero and satisfy $L = 0$,  then the associated four-vortex problem has:
\begin{itemize}
  \item at most 108 real central configurations
  \item  at most 98 real collapse configurations
  \item  at most 98 $\Lambda$ in $\mathbb{S}$ associated with real collapse configurations
  \item at most 49 real collapse configurations for every $\Lambda\in \mathbb{S}\backslash\{\pm1,\pm \mathbf{i}\}$; at most 98 real collapse configurations for every $\Lambda\in \{\pm \mathbf{i}\}$
  \item at most 216  central configurations in complex domain
  \item  at most 196  collapse configurations in complex domain
  \item  at most {196 $\Lambda$ in $ \mathbb{S}$} associated with collapse configurations in complex domain
  \item at most 49  collapse configurations in complex domain for every $\Lambda\in \mathbb{S}\backslash\{\pm1,\pm \mathbf{i}\}$; at most 98  collapse configurations in complex domain for every $\Lambda\in \{\pm \mathbf{i}\}$.

\end{itemize}
\end{corollary}

{\bf Proof.} 

By \eqref{estnumber1}, it  follows that
\begin{equation}\label{estnumber1}
    4\sum_{j=1}^{n_{\Lambda}}\mathcal{N}_{\Lambda_j}+\mathcal{N}_{\mathbf{i}}+\mathcal{N}_{-\mathbf{i}}\leq 392.
\end{equation}

It is clear that
\begin{equation*}
  n_{\Lambda}\leq~\left\{
             \begin{array}{lr}
             [\frac{392}{8}]=49,  &~~~~~~ \text{if}~~ \mathcal{N}_{\mathbf{i}}=0; \\[5pt]
            [\frac{392-4}{8}]=48.  &~~~~~~ \text{if}~~ \mathcal{N}_{\mathbf{i}}>0.
             \end{array}
\right.
\end{equation*}
Consequently, the number of  $\Lambda\in \mathbb{S}\backslash\{\pm1\}$ such that the system \eqref{stationaryconfigurationnew6new} has  nonzero solutions is no more than $196$.

It is also clear that
\begin{equation}\label{estnumber2}
  \mathcal{N}_{\Lambda_j}\leq [\frac{392}{4}]=98,~~~~~~~~~~~~~~~~~~\mathcal{N}_{\mathbf{i}}=\mathcal{N}_{-\mathbf{i}}\leq [\frac{392}{2}]=196.
\end{equation}

Recall that, two solutions $(\Lambda,z,w)$, $(\Lambda,-z,-w)$ of the system \eqref{stationaryconfigurationnew6new}  correspond to a same central configuration in complex domain.
Therefore, in complex domain,  by \eqref{estnumber0}, there are at most 216  central configurations;
   by \eqref{estnumber1}, there are  at most 196  collapse configurations and
    at most {196 $\Lambda$ in $\mathbb{S}\backslash\{\pm1\}$} associated with collapse configurations;
  by \eqref{estnumber2}, there are  at most 49  collapse configurations for every $\Lambda\in \mathbb{S}\backslash\{\pm1,\pm \mathbf{i}\}$ and at most 98  collapse configurations  for every $\Lambda\in \{\pm \mathbf{i}\}$.

  Moreover, by considering the map $ (z, w) \mapsto (\mathbf{i} z, \mathbf{i}w)$, it is easy to see that the total number of real  collapse configurations for the system (\ref{stationaryconfigurationnew6new}) corresponding to $\Lambda_j$ and that  associated with $-\Lambda_j$ is no more than $[\frac{\mathcal{N}_{\Lambda_j}}{2}]$; the total number associated with $\Lambda=\mathbf{i}$ or $-\mathbf{i}$ is no more than $[\frac{\mathcal{N}_{\mathbf{i}}}{2}]$; the total number associated with $\Lambda=1$ or $-1$ is no more than $[\frac{\mathcal{N}_{1}}{2}]$.

  As a result, by \eqref{estnumber0}, there are at most 108 real central configurations;
   by \eqref{estnumber1}, there are  at most 98 real collapse configurations and
    at most {98 $\Lambda$ in $\mathbb{S}\backslash\{\pm1\}$} associated with real collapse configurations;
  by \eqref{estnumber2}, there are  at most 49 real collapse configurations for every $\Lambda\in \mathbb{S}\backslash\{\pm1,\pm \mathbf{i}\}$ and at most 98 real collapse configurations  for every $\Lambda\in \{\pm \mathbf{i}\}$.

The proof is completed.

$~~~~~~~~~~~~~~~~~~~~~~~~~~~~~~~~~~~~~~~~~~~~~~~~~~~~~~~~~~~~~~~~~~~~~~~~~~~~~~~~~~~~~~~~~~~~~~~~~~~~~~~~~~~~~~~~~~~~~~~~~~~~~~~~~~~~~~~~~~~~~~~~~~~~\Box$\\

\section{Conclusion}
\indent\par
To summarize,  the main  result  Theorem \ref{Main} obviously follows from Theorem \ref{Main1} and Theorem \ref{Main2}; and Corollary \ref{upperbounds} follows from  Corollary \ref{upperbounds1} and Corollary \ref{upperbounds2}.\\

We conclude  the paper by showing that  the five-vortex problem has a continuum of
collapse configurations. To this aim, let us consider one of {four} one-parameter {families} of  real  configurations such that
\begin{equation*}
 z_1=-z_2=a, ~~~~~z_3=-z_4=b+\mathbf{i} c,~~~~~ z_5=0;
\end{equation*}
where\begin{equation*}
    b=\pm\sqrt{\frac{144 a^6-121 a^2}{64 a^4-20}},~~~~~~~~~~ c=\pm\sqrt{2 a^2-\frac{144 a^6-121 a^2}{64 a^4-20}}, 
\end{equation*}{and}\begin{equation*}
    a\in (-\frac{3}{2},-\frac{\sqrt{\frac{11}{3}}}{2})\bigcup (\frac{\sqrt{\frac{11}{3}}}{2},\frac{3}{2}).
\end{equation*}
{Then the vortices $\textbf{1},\textbf{2},\textbf{3},\textbf{4}$ form a  parallelogram, and the vortex $\textbf{5}$ is located at the central position (see Figure \ref{fig:continuumcollapseconfigurations}).}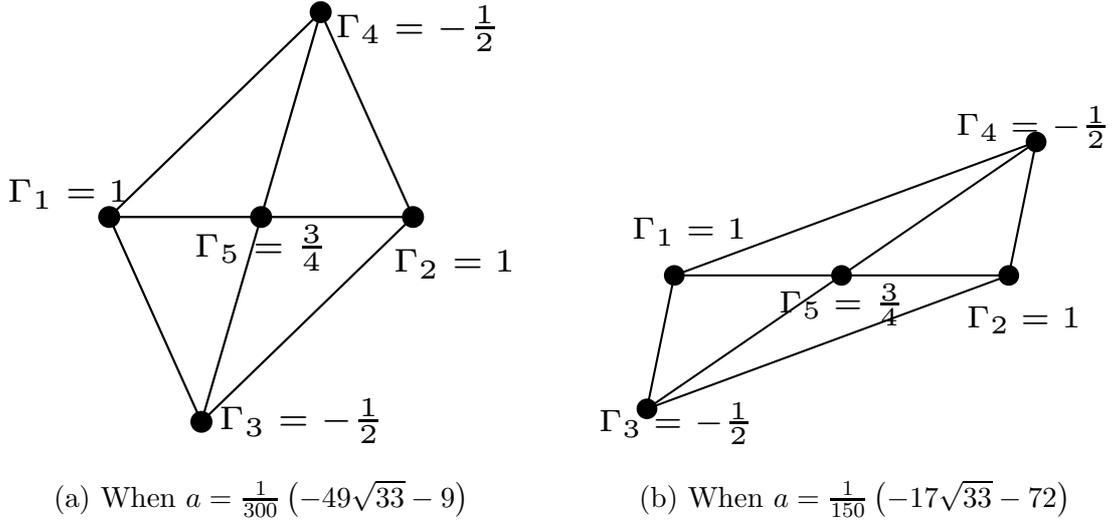
\begin{figure}[!h]
	\centering
\begin{subfigure}[b]{0.5\textwidth}
		\centering
		\resizebox{\linewidth}{!}{
	\begin{tikzpicture}
	\fill (-0.968,  0) circle (2pt);
		\draw  (-1.231,  0.15) node {\tiny$\Gamma_1=1$};

\fill (0.968,  0) circle (2pt);
		\draw  (1.231,  -0.3) node {\tiny$\Gamma_2=1$};

\fill (-0.380,  -1.316) circle (2pt);
		\draw  (0.26,  -1.310) node {\tiny$\Gamma_3=-\frac{1}{2}$};

\fill (0.380,  1.316) circle (2pt);
		\draw  (1.,  1.210) node {\tiny$\Gamma_4=-\frac{1}{2}$};

\fill (0,  0) circle (2pt);
		\draw  (0,  -0.2) node {\tiny$\Gamma_5=\frac{3}{4}$};

\draw [thin] (-0.968,  0)--(-0.380,  -1.316);
\draw [thin] (-0.968,  0)--(0.380,  1.316);
\draw [thin] (-0.380,  -1.316)--(0.968,  0);
\draw [thin] (0.380,  1.316)--(0.968,  0);
\draw [thin] (-0.968,  0)--(0.968,  0);
\draw [thin] (-0.380,  -1.316)--(0.380,  1.316);

	\end{tikzpicture}
}
	\caption{When $a=\frac{1}{300} \left(-49 \sqrt{33}-9\right)$}
\end{subfigure}\begin{subfigure}[b]{0.5\textwidth}
		\centering
		\resizebox{\linewidth}{!}{
	\begin{tikzpicture}
	\fill (-1.131,  0) circle (2pt);
		\draw  (-1.031,  0.3) node {\tiny$\Gamma_1=1$};

\fill (1.131,  0) circle (2pt);
		\draw  (1.231,  -0.3) node {\tiny$\Gamma_2=1$};

\fill (-1.316,  -0.910) circle (2pt);
		\draw  (-1.116,  -1.010) node {\tiny$\Gamma_3=-\frac{1}{2}$};

\fill (1.316,  0.910) circle (2pt);
		\draw  (1.3,  1.010) node {\tiny$\Gamma_4=-\frac{1}{2}$};

\fill (0,  0) circle (2pt);
		\draw  (0,  -0.2) node {\tiny$\Gamma_5=\frac{3}{4}$};

\draw [thin] (-1.131,  0)--(-1.316,  -0.910);
\draw [thin] (-1.131,  0)--(1.316,  0.910);
\draw [thin] (-1.316,  -0.910)--(1.131,  0);
\draw [thin] (1.316,  0.910)--(1.131,  0);
\draw [thin] (-1.131,  0)--(1.131,  0);
\draw [thin] (-1.316,  -0.910)--(1.316,  0.910);

	\end{tikzpicture}
}
	\caption{When $a=\frac{1}{150} \left(-17 \sqrt{33}-72\right)$}
\end{subfigure}
\caption{A one-parameter family of   collapse configurations}
	\label{fig:continuumcollapseconfigurations}
\end{figure}

A  straightforward computation shows that the family of   configurations above satisfies \eqref{stationaryconfiguration2} if  
\begin{center}
$ \Gamma_1=\Gamma_2=1, ~~~~~\Gamma_3=\Gamma_4=-\frac{1}{2},~~~~~ \Gamma_5=\frac{3}{4},$
\end{center}
and
\begin{equation*}
    \Lambda =\frac{-a^2+5 b^2-10 i b c-5 c^2}{2 a^2 \left(b^2-4 i b c-3 c^2\right)}.
\end{equation*}
Note that \begin{equation*}
    |\Lambda|=1 ~~~~~~~~~~\text{and}~~~~~~~~~~\Lambda\notin \mathbb{R},
\end{equation*}
thus the family of   configurations above is a continuum of  real normalized collapse configurations in  the five-vortex problem.



\bibliographystyle{plain}


\end{document}